\documentclass[twocolumn]{aa} 
\usepackage{ulem,cancel}
\usepackage{graphicx,txfonts,comment,float,amsfonts,longtable,lscape,nicefrac,mathrsfs,subfig,amsmath,relsize,bm}
\usepackage[dvipsnames]{xcolor}
\usepackage[utf8]{inputenc}
\usepackage[T1]{fontenc}
\usepackage[colorlinks=true,allcolors=blue]{hyperref}
\usepackage{placeins}
\usepackage{soul}
\newcommand{\opha}{$\rho$~Oph-A}
\newcommand{\nHii}{$n_\mathrm{H_2}$}
\newcommand{\NHii}{$N_\mathrm{H_2}$}
\newcommand{\BPOS}{$B_\mathrm{pos}$}
\newcommand{\um}{$\mu$m}
\newcommand{\Kcm}{K~cm$^{-3}$}
\newcommand{\kms}{km~s$^{-1}$}
\newcommand{\cmcube}{cm$^{-3}$}
\newcommand{\cmsq}{cm$^{-2}$}
\newcommand{\Ma}{$\mathscr{M}_\mathrm{A}$}
\begin{document} 

\title{Mapping and characterizing magnetic fields in the Rho Ophiuchus-A molecular cloud with SOFIA/HAWC$+$}
\author{Ng\^an~L\^e\inst{1,2}\thanks{Corresponding author: Ng\^an~L\^e~\newline \email{nganle191919@gmail.com}}, 
Le~Ngoc~Tram\inst{3}, 
Agata~Karska\inst{1,3}, 
Thiem~Hoang\inst{4,5}, 
Pham~Ngoc~Diep\inst{6,7},
Michał~Hanasz\inst{1},
Nguyen~Bich~Ngoc\inst{6,7},
Nguyen~Thi~Phuong\inst{6,2},
Karl~M.~Menten\inst{3}, 
Friedrich~Wyrowski\inst{3}, 
Dieu~D.~Nguyen\inst{8}, 
Thuong~Duc~Hoang\inst{9}, 
and Nguyen~Minh~Khang\inst{10}
}
\institute{$^{1}$ Institute of Astronomy, Faculty of Physics, Astronomy and Informatics, Nicolaus Copernicus University, Grudzi\k{a}dzka 5, 87-100 Toruń, Poland\\
$^{2}$ Institute For Interdisciplinary Research in Science and Education (IFIRSE), ICISE, 07 Science Avenue, Ghenh Rang Ward, 55121 Quy Nhon City, Binh Dinh Province, Vietnam\\
$^{3}$ Max-Planck-Institut für Radioastronomie, Auf dem Hügel 69, 53121, Bonn, Germany\\
$^{4}$ Korea Astronomy and Space Science Institute, 776 Daedeokdae-ro, Yuseong-gu, Daejeon 34055, Republic of Korea\\
$^{5}$ University of Science and Technology, Korea, 217 Gajeong-ro, Yuseong-gu, Daejeon 34113, Republic of Korea\\
$^{6}$ Department of Astrophysics, Vietnam National Space Center, Vietnam Academy of Science and Technology, 18 Hoang Quoc Viet, Hanoi, Vietnam\\
$^{7}$ Graduate University of Science and Technology, Vietnam Academy of Science and Technology, 18 Hoang Quoc Viet, Hanoi, Vietnam\\
$^{8}$ Université de Lyon 1, ENS de Lyon, CNRS, Centre de Recherche Astrophysique de Lyon (CRAL) UMR5574, F-69230 Saint-Genis-Laval, France\\
$^{9}$ School of Physics and Astronomy, University of Minnesota, 115 Union St SE, Minneapolis, MN 55455, USA\\
$^{10}$ Astrophysics Research Institute, Liverpool John Moores University, 146 Brownlow Hill, Liverpool L3 5RF, UK\\
} 
\date{Received September 18, 2023; accepted August 30, 2024}
\titlerunning{Magnetic fields in \opha}
\authorrunning{N.~L\^e et al., 2024}
\abstract 
{Together with gravity, turbulence, and stellar feedback, magnetic fields (B-fields) are thought to play a critical role in the evolution of molecular clouds and star formation processes. The polarization of thermal dust emission is a popular tracer of B-fields in star-forming regions.}
{We aim to map the morphology and measure 
the strength of B-fields of the nearby molecular cloud, rho Ophiuchus-A (\opha), to understand the role of B-fields in regulating star formation and in shaping the cloud.} 
{We analyzed the far-infrared (FIR) polarization of thermal dust emission observed by SOFIA/HAWC$+$ at 89 and 154~$\mu$m toward the densest part of \opha, which is irradiated by the nearby B3/4 star, Oph-S1. These FIR polarimetric maps cover an area of $\sim 4.5\arcmin \times 4.5\arcmin$ (corresponding to $0\rlap{.}''18 \times 0\rlap{.}''18$ pc$^2$) with an angular resolution of 7.8$\arcsec$ and 13.6$\arcsec$, respectively. }
{The \opha~cloud exhibits well-ordered B-fields with magnetic orientations that are mainly perpendicular to the ridge of the cloud toward the densest region. 
We obtained a map of B-field strengths in the range of 0.2-2.5~mG, using the Davis-Chandrasekhar-Fermi (DCF) method. The B-fields are strongest at the densest part of the cloud, which is associated with the starless core SM1, and then decrease toward the outskirts of the cloud. 
By calculating the map of the mass-to-flux ratio, Alfv\'en Mach number, and plasma $\beta$ parameter in \opha, we find that the cloud is predominantly magnetically sub-critical, sub-Alfv\'enic, which implies that the cloud is supported by strong B-fields that dominate over gravity, turbulence, and thermal gas energy. 
The measured B-field strengths at the two densest subsregions using other methods that account for the compressible mode are relatively lower than that measured with the DCF method. However, these results do not significantly change our conclusions on the roles of B-fields relative to gravity and turbulence on star formation. 
Our virial analysis suggests that the cloud is gravitationally unbound, which is consistent with the previous detection of numerous starless cores in the cloud. By comparing the magnetic pressure with the radiation pressure from the Oph-S1 star, we find that B-fields are sufficiently strong to support the cloud against radiative feedback and to regulate the shape of the cloud. } 
{}
\maketitle
% ------------------------------------------------
\section{Introduction}\label{sec:intro}
Stars like our Sun form in the cold and dense parts of molecular clouds, which consist of filaments, clumps, and dense cores \citep{shu87,Lada03,Kennicutt12}. The formation of stars is regulated not only by gravity, but also by magnetic fields (B-fields), turbulence, and radiative and mechanical feedback from nearby stars \citep{McKee07,Hennebelle12,hen19rev}. 
Turbulence counteracts gravity and prevents the collapse of the dense core, but it can also generate shocks that compress the gas and trigger the formation of new stars. 

B-fields are thought to play a critical role in the star formation process by supporting the material in the cloud against self-gravitational collapse \citep[see, e.g.,][]{mouscho01,Crutcher12,Pat19,pattle23}. The collapse can occur when the dense core is magnetically super-critical, with the ratio of mass to magnetic flux (known as mass-to-flux ratio) greater than 1; that is, when B-fields are not strong enough to prevent the gravitational collapse. The collapse of super-critical cores will lead to the formation of protostars. Conversely, when the core is sub-critical, strong B-fields can act against gravitational collapse. However, several mechanisms might enhance the mass-to-flux ratio and cause the core to collapse, including ambipolar diffusion \citep[see e.g.,][]{Mestel1956, Mouscho06, Liu22rev} and magnetic reconnection diffusion \citep{Lazarian99,santos2012role}. 

Feedback from massive stars (O- or B-type) can influence the properties of the parental molecular cloud and the star formation processes \citep[e.g.,][]{krum14,Pabst19, Pabst20}. Three-dimensional magnetohydrodynamic (3D MHD) simulations have suggested that B-fields also influence the structure of clouds along with radiation pressure feedback from massive stars \citep[e.g.,][]{henney09, mackey&lim11}. It has been shown that when the B-field orientations are parallel to the ionizing radiation field, the cloud will tend to have a flattened shape, while the cloud will have a column-like or pillar structure if the B-fields are perpendicular to the radiation field \citep{mackey&lim11_b}. These findings from simulations have been in agreement with observations and could explain various globular and elongated substructures in the IC~1396 region \citep{Soam18}, the famous pillar structure in the M16 region \citep{pattle18}, and the Horsehead Nebula \citep{Hwang23}.

Interstellar dust grains are known to have aspherical shapes and can be systematically aligned with a fixed direction in space, for instance, interstellar B-fields \citep[][]{hall49,hilt49}. The alignment of interstellar dust grains with B-fields involves the internal alignment of the grain axis of the maximum inertia moment (also the shortest axis) with its angular momentum (i.e., internal alignment) and the alignment of the angular momentum with B-fields (external alignment). Internal alignment is caused by internal relaxation processes, including Barnett relaxation and inelastic relaxation \citep{pur79}. The external alignment is caused by the radiative torque (RAT) alignment mechanism \citep{draine97, Lazarian07} and by the more comprehensive magnetically enhanced RAT mechanism \citep[MRAT,][]{mrat16}. The internal and external alignment processes establish the alignment of the grain with its minor axis parallel to the B-fields \citep[see, e.g.,][]{thiem22}. Dust grains absorb or scatter starlight in the ultraviolet (UV), optical, or near-infrared (NIR) range. However, aligned dust grains tend to absorb and scatter the electric field component of the starlight along their major axis more efficiently. This results in the polarization of background starlight with the polarization vector parallel to the grain's minor axis. Thus, the polarization of the starlight seen in the UV to optical or NIR wavelength has the polarization orientation parallel to the B-field orientation. Aligned dust grains also re-emit the energy absorbed from starlight in the longer wavelength from far-infrared (FIR) to sub-millimeter (sub-mm), the so-called thermal dust emission. As the electric field component of starlight is absorbed more along the grain's major axis, the dust grain will re-emit more energy with the electric field component along this major axis than along the minor axis. This causes the thermal emission of aligned dust grains to be polarized with the polarization orientation along the grain's major axis. Because dust grains have their minor axes aligned with the B-field orientation, the polarization orientation of thermal emission is perpendicular to the B-field orientation. Thus, if the position angle of the polarized thermal dust emission is known, one can rotate the polarization angle by 90\degr~to infer the position angle of the B-fields in the medium. Linear polarized thermal dust emission has been routinely used to trace B-fields in various environments of molecular clouds in the Milky Way \citep[see, e.g.,][]{Ward-th17,chuss19,pattle22} and extragalactic regions \citep[see, e.g.,][]{lopez21,lopez22,Borl23}. The Davis-Chandrasekhar-Fermi method \citep[DCF;][]{Davis51, CF53} has been widely used to measure the strengths of the B-field projected on the plane of the sky, \BPOS, based on the assumption that the kinetic energy of turbulent gas balances to that of turbulent B-fields. 

Adopting the mean value of B-field strengths along the line of sight ($B_\mathrm{los}$) measured using the Zeeman technique over different molecular clouds, \cite{Crutcher10} derived a relationship between the B-field strengths and gas volume densities ($n_\mathrm{H}$). For low densities of $n_\mathrm{H}\leq$300~\cmcube, B-fields were found not to considerably change with densities. However, for high densities above $n_\mathrm{H}\sim$300~\cmcube, the B-field strengths increase with densities as $n_\mathrm{H}^{k}$, where the power-law index $k$ is $\sim$0.65. 
\cite{pattle23} compiled a large number of measurements of the B-field strength projected on the plane of the sky (\BPOS) estimated using the DCF method. These authors showed that these \BPOS~generally follow the results from \cite{Crutcher10} until $n_H \ge 10^7$~\cmcube. Beyond this $n_\mathrm{H}$~value, the \BPOS~distribution is more sparse \citep[see Fig.~2a~in][]{pattle23}. 
Theoretical studies have suggested that the power-law index varies between $k\sim$2/3 for weak B-fields \citep{Mestel1966} and $k\la$0.5 for strong B-fields \citep{Mouscho99}. Therefore, the maps of the morphology and strength of the B-fields within a specific cloud are essential to understanding the role of the local B-fields on star formation within the cloud. 
Previous authors have attempted to derive the map of B-fields toward some well-known star-forming regions, for instance, OMC-1 \citep{Guerra21,Hwang21}, 30~Doradus  \citep[30~Dor,][]{tram23}, and the G11.11-0.12 filament \citep{ngoc23}. These studies revealed significant variations in the B-field morphology and strength across the clouds, demonstrating that B-fields strongly influence the cloud's evolution and, thus, star formation activities. 

\begin{figure}[tbp]\centering
\includegraphics[width=0.49\textwidth]{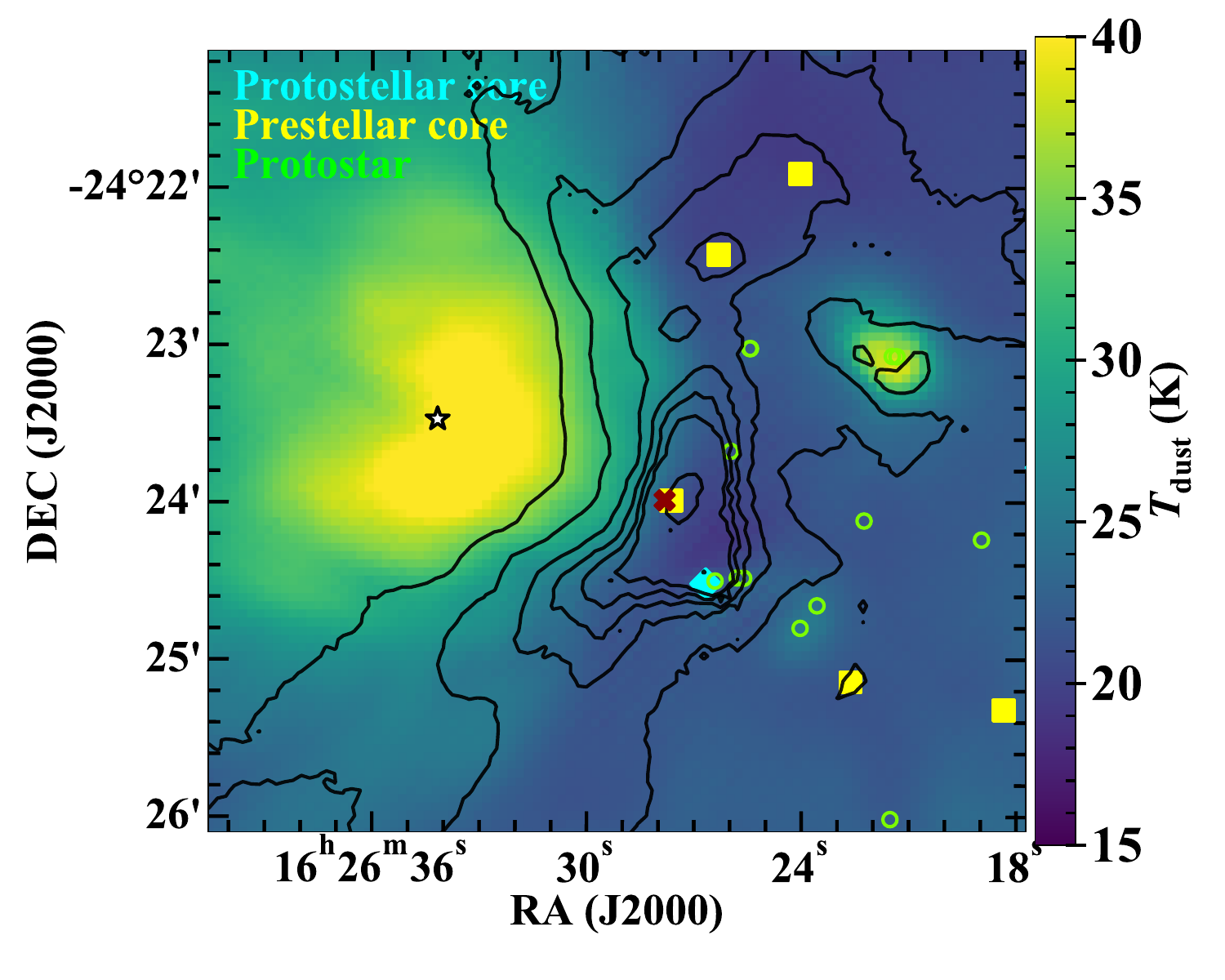}
\caption{Distribution of dust temperatures (colors) and H$_2$ column densities (contours) in \opha~\citep{Santos19}. Contour levels of $N_\mathrm{H_2}$ are shown at (1, 2, 4, 6, 8, and 10) $\times10^{22}$ \cmsq. The star symbol shows the position of Oph-S1. Small circles show the positions of YSOs \citep{Enoch2009,evans09,con10}. Yellow squares and cyan diamonds indicate the positions of prestellar and protostellar cores, respectively, identified as part of the \textit{Herschel} Gould Belt Survey \citep{Ladjelate2020}. The \lq$\times$' symbol shows the position of the SM1 core.}
\label{fig:tdust}
\end{figure}

In this work, we focus on a specific system of the $\rho$ Ophiuchus-A (hereafter~\opha) molecular cloud to study the effects of stellar feedback and B-fields on the evolution of the cloud and star formation. 
The \opha~molecular cloud is located in the northern part of the L1688 active low-mass star-forming region in the $\rho$~Ophiuchus dark cloud complex \citep{loren90,wilking08,esplin20}. The \opha~cloud was initially identified as a dense gas clump based on DCO$^+$ observations performed by \cite{loren90}, along with other clumps named $\rho$~Oph-B to $\rho$~Oph-F. This region has high extinctions ($A_V\ga$20~mag) corresponding to the high gas column densities compared to other regions, which are typically more quiescent (see Fig.~\ref{fig:tdust}). Being one of the closest low-mass star-forming regions in our Galaxy, located at a distance of $\sim$137~pc \citep{ortiz17}, the gas and dust content in the Oph--A region has been studied extensively at almost every wavelength by both ground-based and space instruments for many years \citep[see e.g.,][for a review]{wilking08}. The \opha~region hosts numerous prestellar and protostellar cores identified by the Ground Belt Survey \citep{Ladjelate2020} and protostars in the early stages \citep[i.e., Class 0/I Young Stellar Objects (YSOs),][]{Enoch2009,evans09,con10}. The \opha~cloud is mostly influenced by a nearby embedded early high-mass B3/4 star \citep{brown75, Mookerjea18}, Oph-S1, which has a mass of $\sim$8~$M_\odot$ \citep{Hamaguchi2003}. The Oph-S1 star is surrounded by a compact \ion{H}{ii} region \citep{andre88} and is believed to trigger the star formation activity in Oph--A, especially along the ridge of the dense cloud with indications of protostellar clumps \citep[see Fig.~2a in][]{motte98}. The star formation activity in the \opha~is also believed to be affected by the Sco~OB2 association located in the western part of the cloud at a distance of $\sim$145~pc \citep{tim99}. 

We investigated B-fields in \opha\  using data observed in various wavelengths:  NIR (\textit{JHK} bands) using imaging polarimeter SIRPOL \citep{Kandori06} mounted on the 1.4-m Infrared Survey Facility (IRSF) telescope \citep{Kwon15}, FIR by the High-resolution Airborne Wideband Camera-plus \citep[HAWC$+$;][]{harper18} on board the 2.7-m Stratospheric Observatory For Infrared Astronomy (SOFIA) \citep{Santos19}, and sub-mm using the linear polarimeter POL-2 placed in front of the 
Submillimetre Common-User Bolometer Array 2 \citep[SCUBA-2;][]{scuba2} camera mounted on the 15-m \textit{James Clerk Maxwell} Telescope (JCMT) \citep{Kwon18}. \cite{Santos19} performed the analysis on polarimetric data from SOFIA/HAWC$+$ observations at 89 and 154~\um~to study the change in the slope of the polarization spectra with respect to the column densities and dust temperatures in \opha. Later, \cite{tram21} used a similar SOFIA/HAWC$+$ dataset to investigate the relation between the polarization fraction of thermal dust emission versus dust temperatures in the cloud and to understand the mechanism of dust grain alignment and disruption induced by RATs. \citet{Kwon18} estimated the $B_\mathrm{pos}$ values in \opha,~using the JCMT/POL-2 data at 850~$\mu$m, toward a few of the densest parts of the cloud, yielding  $B_\mathrm{pos}$ values in the range of $\sim$0.5--5~mG. The highest strength was measured toward the densest part of the cloud associated with the starless core SM1 (see Fig.~\ref{fig:tdust}). These \BPOS~strengths in \opha~are comparable to values found in the OMC$-1$ region \citep{Hwang21}. \cite{Kwon18} also found that the B-field direction is generally perpendicular to the main structure of the cloud, which is generally seen in the dense cores or high-density structures of molecular clouds. The \opha~region provides an ideal isolated system for studying the impact of the B-fields in the star-forming region and investigating the interplay between B-fields and the stellar feedback into the molecular cloud. In this paper, we aim to: 1) determine the B-field morphology and strengths in the entire \opha~region using the SOFIA/HAWC$+$ dust polarization data; 2) investigate the role of B-fields relative to gravity and turbulence in the star formation activity in the cloud; and 3) quantify the combined effect of B-fields and radiation pressure from the nearby Oph-S1 star on the shape of the \opha~cloud. 

The paper is organized as follows. We describe a set of archival observational data in Sect.~\ref{sec:data}. We present the results of mapping the B-field morphology and strength using various techniques in Sect.~\ref{sec:Bfields} and discuss the roles of B-fields relative to gravity, turbulence, and radiation pressure feedback from the Oph-S1 star in Sect.~\ref{sec:dis}. Finally, we summarize our results in Sect.~\ref{sec:sum}. 
% =====================
%=====================
% ---------------------------------
\section{Observational data} \label{sec:data} 
\subsection{Dust polarization data from SOFIA/HAWC$+$}\label{ssec:hawc-obs}
We use the FIR polarimetric observations from the onboard SOFIA/HAWC$+$ toward \opha. The observations were done on 2017 May 17, under the proposal ID $\mathrm{70_{-}0511}$ (PI: Darren Dowell), using the nod-match-chop mode, with 154\degr~chop angle, 480\arcsec~chop throw, and 4-points dithers. Here, we use the SOFIA/HAWC$+$ Level-4 data for two bands C and D, centered at 89 and 154~$\mu$m, respectively, which is fully calibrated and can be directly downloaded from the SOFIA archive\footnote{\url{https://irsa.ipac.caltech.edu/Missions/sofia.html}}. The final polarimetric maps cover an area of $\sim4.5\arcmin\times4.5\arcmin$ (corresponding to $0\rlap{.}''18 \times 0\rlap{.}''18$ pc$^2$) toward the central region in \opha~with the angular resolutions of 7.8$\arcsec$ and 13.6$\arcsec$ full-width-half-maximum (FWHM) at 89 and 154~\um, respectively. The pixel size of HAWC$+$ maps is $\sim$2.0\arcsec~at 89~\um~and 3.4\arcsec~at 154~\um. The observations and data reduction are described in detail in \cite{Santos19}. 
%=====================
\begin{figure*}[ht]\centering
\includegraphics[width=\textwidth]{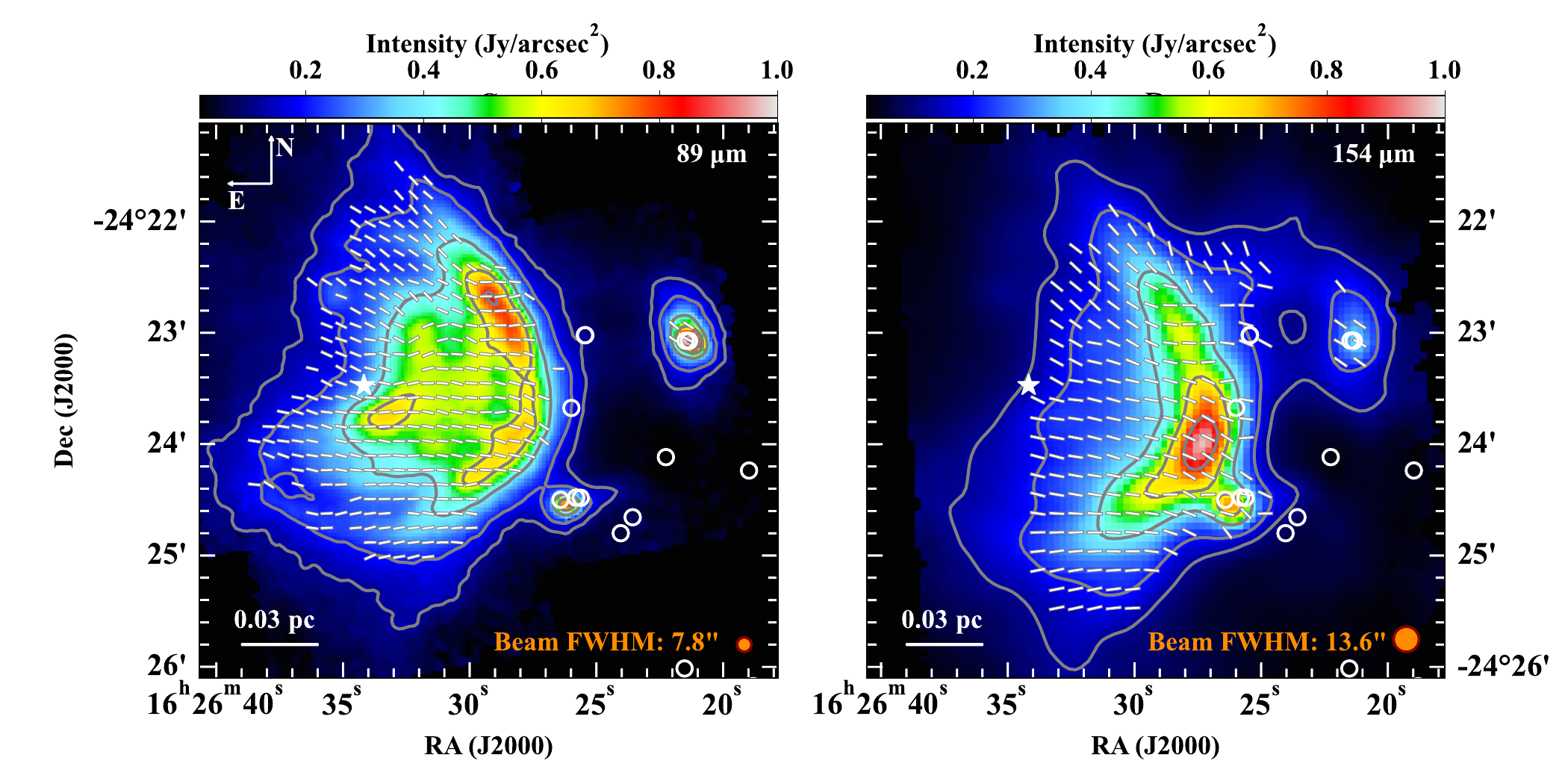}
\caption{Maps of the B-field orientation (white segments) on top of the total intensity (Stokes-$I$, color maps) toward \opha, observed with SOFIA/HAWC$+$ in band C (left) and band D (right), centered at 89 and 154~$\mu$m, respectively. The gray contours in each map show the Stokes-$I$ intensity at the corresponding wavelength, with contour levels at 0.12, 0.2, 0.4, 0.6, and 0.8~Jy/arcsec$^2$. The star symbol marks the position of Oph-S1 star. Small white circle symbols indicate the positions of YSOs. The filled orange circles show the beam size of each color map. }
\label{fig:Bfield_orientation}
\end{figure*} 
%=================================================

From the maps of the Stokes-$I$, Stokes-$Q$, and Stokes-$U$ (along with their uncertainties $\sigma_{I}$, $\sigma_{Q}$, $\sigma_{U}$), the maps of polarization fraction ($p_\mathrm{0}$) and its corresponding uncertainty ($\sigma_{p}$) were calculated. The de-biased polarization fraction, $p$, is calculated as: $p^2=p_\mathrm{0}^2-\sigma_\mathrm{p}^2$ \citep{Vaillancourt06}. The maps of polarization angle ($\theta$) and its corresponding uncertainty ($\sigma_\theta$) in degrees can be derived from Stokes-$Q$ and Stokes-$U$. We refer  to \cite{Gordon18} for a full description of estimates of the polarization fraction, polarization angle, and their corresponding uncertainties. To enhance the quality in measuring the maps of B-field orientation and strength in \opha, we apply some critical conditions in choosing vectors (pixels) in SOFIA/HAWC$+$ maps. We mask out from SOFIA/HAWC$+$ maps the pixels for which $I/\sigma_I$<54 for band C and $I/\sigma_I$<214 for band D. In both bands, we further removed the pixels for which we have $p/\sigma_p$<3 and $p$>50\% from the HAWC$+$ maps (see Appendix~\ref{app:histogram}). 

%=================================================
\begin{figure*}[ht]\centering
\includegraphics[width=\textwidth]{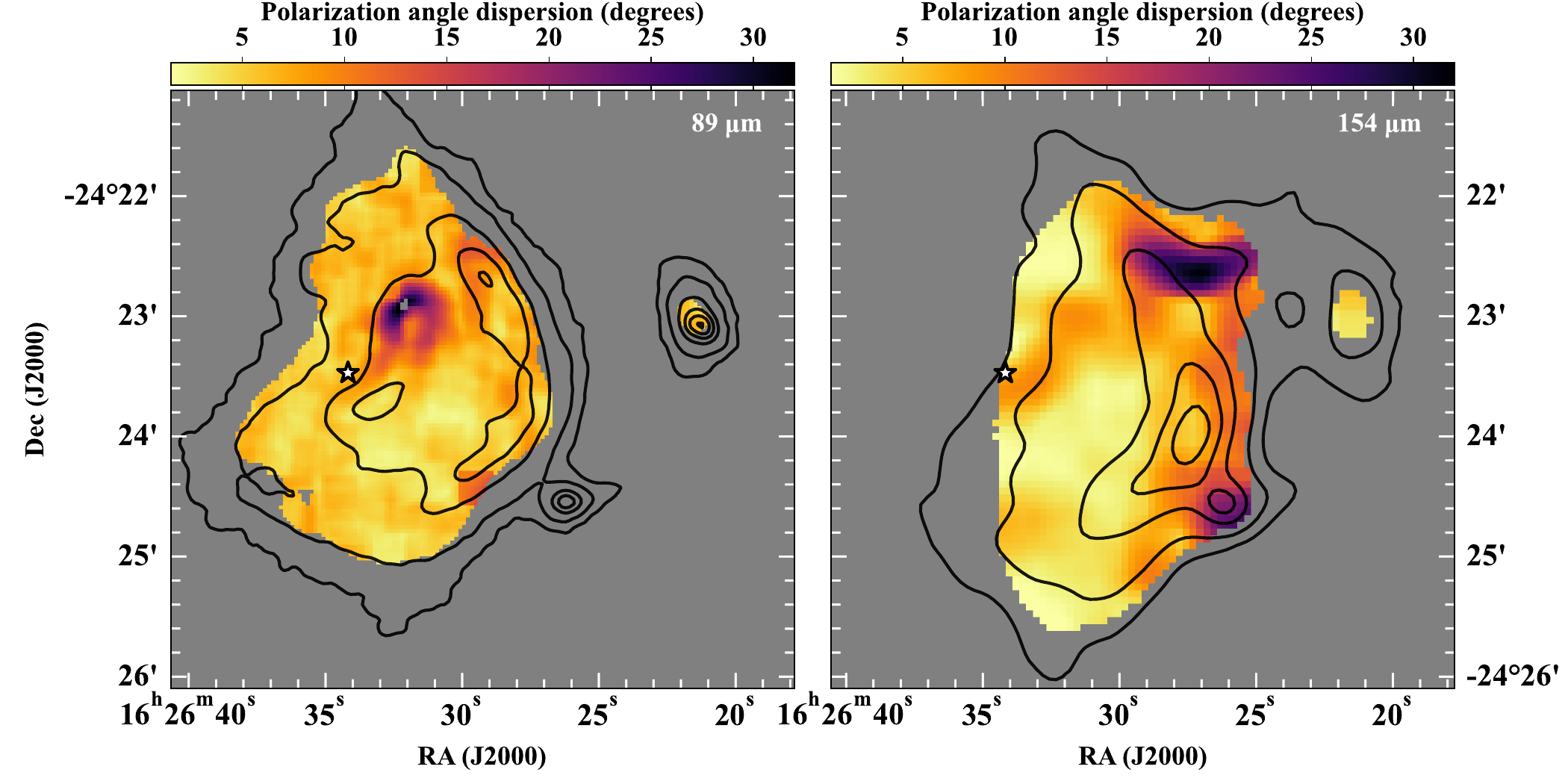}
\caption{Polarization angle dispersion maps using data in band C (left panel) and band D (right panel). The black contours are from the total intensity map at the corresponding wavelength, with the contour levels at 0.12, 0.2, 0.4, 0.6, and 0.8 Jy/arcsec$^2$. The star symbol in each map indicates the position of the Oph-S1 star. }\label{fig:disp_ang_maps} 
\end{figure*} 
%================================================= 

\subsection{Spectroscopic data}\label{ssec:line-obs}

We used the spectroscopic observations of HCO$^+$(4$-$3) line at $\sim$356.734~GHz obtained from the Heterodyne Array Receiver Program \citep[HARP;][]{Buckle09} mounted on JCMT. The observations were carried out during June and July 2011 (Proposal ID: M11AU13), covering the \opha~region with a beam size (or FWHM) of $\sim$14$\arcsec$. We used the reduced spectral map of HCO$^+$(4$-$3) line directly downloaded from the JCMT science archive\footnote{\url{https://www.cadc-ccda.hia-iha.nrc-cnrc.gc.ca/en/jcmt/}}. The data cube had an original spectral resolution of 0.025~\kms~and was smoothed to have a final spectral resolution of 0.5~\kms. For comparison to the HCO$^{+}$(4$-$3) data, we also used observations of N$_2$H$^{+}$(3$-$2) line at $\sim$279.515~GHz obtained from the APEX-2 instrument on the 12-m Atacama Pathfinder EXperiment telescope \citep[APEX;][]{gusten2006}. The data cube has an original spectral resolution of 0.082~\kms. Here, we used the back-end data cube with a spectral resolution of 0.25~\kms, covering the \opha~region with the beam size of $\sim$22$\arcsec$ \citep[for details, see][]{liseau15}. 

\subsection{Dust temperature and column density maps}\label{ssec:dust-gas-obs}

We adopted the maps of dust temperature ($T_\mathrm{dust}$) and gas column density ($N_\mathrm{H_2}$) from \cite{Santos19}. These maps were obtained from the fitting of a modified black body to the spectral energy distribution (SED) using the continuum flux maps from \textit{Herschel}/PACS \citep{pacs} at 70, 100, and 160~$\mu$m, with an FWHM of 11.4$\arcsec$ \citep[for details, see Sect.~3.2 in][]{Santos19}. Figure~\ref{fig:tdust} shows the spatial distribution $T_\mathrm{dust}$ toward \opha, overlaid with contours of column density, $N_\mathrm{H_2}$. In this study, we use the map of $T_\mathrm{dust}$ as the map of the gas kinetic temperature, assuming that the gas is in the local thermodynamic equilibrium (LTE) and well-coupled with the dust. 

% ------------------------------------------------
\section{Magnetic fields in \opha}\label{sec:Bfields}
\subsection{Magnetic field morphology}\label{ssec:Bfields_orientation}

We construct the B-field morphology in \opha~by rotating polarization angles by 90\degr. Figure~\ref{fig:Bfield_orientation} shows the maps of B-field orientation (half-vectors) inferred from maps of the polarization angle at 89 and 154~\um~. The background color maps show the Stokes-$I$ intensity at the respective wavelength. Here, the position angle of B-fields ($\theta_B$) is measured east of north, and in the range between 0$\degr$ and 180$\degr$. The polarimetric data at both wavelengths can well reveal B-fields toward the dense region in \opha,~where Stokes-$I$>0.12~Jy/arcsec$^2$ (the area inside the lowest contour in each map). The data at 89~\um~can further resolve the B-fields morphology extended toward the warmer dust region located at the east side of the Oph-S1 star. 

The B-field vectors in \opha~are generally well-ordered and largely follow the material structure in low- to intermediate-densities regimes (see Fig.~\ref{fig:Bfield_orientation}). B-field orientations seem to be bent toward the higher density region in the center of the map and are mostly perpendicular to the ridge of the cloud. On the eastern side of the map, the orientation of B-field vectors tends to vary in the counter-clockwise direction from north to south; thus, the position angle of B-fields increases. Particularly, in the northeast part of the map, B-field vectors run along the northeast to southwest direction with a median $\theta_B$ value of $\sim$56\degr~and $\sim$49\degr~at 89 and 154~\um, respectively. This feature is also seen in a small area in the northwest part of the map. 
Toward the east side, B-field vectors orient nearly to the east-west direction, with median $\theta_B$ values of $\sim$83\degr~at 89~\um~and $\sim$75\degr~at 154~\um. On the other hand, B-field vectors in the southeast region tend to orient consistently to the southeast to northwest directions, with median $\theta_B$ values of $\sim$92\degr~and $\sim$85\degr~at 89 and 154~\um, respectively. 
Interestingly, the data at 154~\um~shows B-field vectors running nearly vertically, that is, along the  north-south direction toward a small area at the north part of the map, which is not seen in the data at 89~\um~(see Fig.~\ref{fig:Bfield_orientation} and also Fig.~\ref{fig:orient_compare}). 

In addition, there are two outstanding components of B-fields toward the regions characterized by the highest H$_2$ column densities: one component associated with the peak of continuum emission at 154~\um~runs along the northeast to southwest direction with the median B-field orientation of $\sim$70$\degr$; the other is located at the north of the first component and associated with the peak of continuum emission at 89~\um, which exhibits the B-field vectors along the east-west direction with a median position angle of $\sim$88\degr~and $\sim$81\degr~at 89 and 154~\um, respectively (see Fig.~\ref{fig:orient_compare}). We will further discuss the B-field morphology together with B-field strengths in Sect.~\ref{ssec:dis_Bfields}.

\subsection{Measurement of magnetic field strength using the DCF method}\label{ssec:Bfields_strength}
We first used the classical~DCF method \citep{Davis51,CF53} (hereafter we refer to as the DCF method) to calculate the strengths of the B-field projected in the plane-of-the-sky ($B_\mathrm{pos}^\mathrm{DCF}$) in the entire region of the \opha~cloud. We  compared our obtained results with those obtained using the more advanced techniques described in Sect. \ref{ssec:B_other_med}.

According to the DCF method, the B-field strength can be estimated by \cite{Ostriker2001}:
\begin{equation}\label{eq:dcf}
B_\mathrm{pos}^\mathrm{DCF}=f\sqrt{4\pi\rho}\frac{\sigma_\mathrm{NT}}{\delta\theta}, 
\end{equation}
where $f$ is the correction factor, $\rho = \mu m_\mathrm{H}n_\mathrm{H_2}$ is the gas mass density in units of g~cm$^{-3}$, $\mu$=2.8 is the mean molecular weight per hydrogen molecule \citep{Kauffmann08}, and $m_\mathrm{H}$ is the mass of the hydrogen atom, $n_\mathrm{H_2}$ is the gas volume density in units of cm$^{-3}$. Then, $\sigma_\mathrm{NT}$ is the one-dimensional (1D) non-thermal velocity dispersion along the line of sight (LOS) of the ideal gas species that traces the dense region, similarly to the dust-polarized emission in units of km~s$^{-1}$, and $\delta\theta$ is the polarization angle dispersion in degrees. The correction factor, $f$, is uncertain, and numerical simulations by \cite{Ostriker2001} show $f=0.5$~when the polarization angle dispersion does not exceed 25$\degr$. However, varying values of $f$ will result in different \BPOS~strengths \citep[see e.g., ][for more detailed discussions]{skalidis21,liu21}.  

Here, we use the formula given in \cite{crutcher04}:
\begin{equation}\label{eq:Bpos}
B_\mathrm{pos}\approx9.3\sqrt{n_\mathrm{H_{2}}}\frac{\Delta\varv_\mathrm{NT}}{\delta\theta}~\left(\mu\mathrm{G} \right),
\end{equation}
where $\Delta \varv_\mathrm{NT} =\sigma_\mathrm{NT}\sqrt{8\ln2}$ is non-thermal FWHM velocity dispersion measured in units of km~s$^{-1}$, and the correction factor $f=0.5$ has been used. This formula is widely used to estimate $B_\mathrm{pos}$ via the DCF technique. 

\subsubsection{Polarization angle dispersion maps}\label{ssec:angular_dispersion}
It is essential to see how the B-field orientation changes locally within the molecular cloud. To do so, we calculated $\delta\mathrm{\theta}$ across the \opha~region. We used the method presented in \cite{Hwang21}, for which $\delta\mathrm{\theta}$ is calculated pixel-by-pixel throughout the map of \opha. We summarize the procedure of calculating $\delta\mathrm{\theta}$ toward the pixel $i^\mathrm{th}$ in the map as follows:
\begin{enumerate}
\item We used a box covering the pixel $i^\mathrm{th}$ with a size of 9$\times$9 pixels, corresponding to two beam sizes of the SOFIA/HAWC$+$ observations (i.e., 15.6$\arcsec$ and 36.4$\arcsec$ in bands C and D, respectively).
\item We calculated the mean value of polarization angles within the small box, $\bar{\theta_\mathrm{i}}$, which implies the local mean polarization angle toward the  $i^\mathrm{th}$ pixel.
\item For each pixel within the small box, we calculated the difference between the observed polarization angle and the mean value. We computed the root mean square (RMS) of the distribution of angle differences in the box. This RMS value indicates the polarization angle dispersion calculated toward the  $i^\mathrm{th}$ pixel.
\item The uncertainty of the polarization angle dispersion toward the pixel $i^\mathrm{th}$ is propagated from the uncertainty of polarization angles in the small box and determined as the mean value of polarization angle uncertainties within the small box. 
\end{enumerate}
By repeating the process for all pixels in the entire region, we obtained the map of $\delta\theta$ in \opha. We note that if there are more than 50$\%$ of pixels undefined (i.e., having NaN value of $\theta$) in the computing box, the result of polarization angle dispersion calculation toward pixel $i^\mathrm{th}$ is ignored. We applied this criterion to keep the high quality of the $\delta\theta$ estimate. It is also necessary to de-bias the polarization angle dispersion due to the uncertainty of the observed polarized angles. 

Figure~\ref{fig:disp_ang_maps} shows the maps of polarization angle dispersion toward \opha~using HAWC$+$ data in two bands C (left panel) and D (right panel), respectively. The polarization angle dispersion calculated using data in bands C and D cover ranges of 2.3$\degr$--32.2$\degr$ and 1.3$\degr$--31.5$\degr$, which correspond to the median values of 5.7\degr~and 5.8\degr, respectively. 
The differences of $\delta\theta$ in two wavelengths imply that the data at the shorter wavelength tends to trace warm dust surrounding the high-mass star S1 and toward the outer layers; whereas the data at the longer wavelength probes the colder dust in the denser layers of the cloud. 
We note, however, that in maps of $\delta\theta$ in both bands, there are a few pixels where $\delta\theta>25\degr$ (27 pixels in band C and 45 pixels in band D). Thus, we do not use the $\delta\theta$ value in these pixels to estimate \BPOS~in Sect.~\ref{ssec:Bpos}. 

\subsubsection{Velocity dispersion map}\label{ssec:velocity-dispersion}

We used the HCO$^{+}$(4$-$3)~molecular data observed by JCMT/HARP to estimate the total velocity dispersion, $\sigma_\varv$. This molecular line has a critical density of $\sim$10$^{6}$~\cmcube~\citep{Greve09} and thus is considered a good tracer for probing the dense region in the \opha~cloud. The HCO$^{+}$(4$-$3) emission is well-represented by a single Gaussian profile. Thus, in each pixel, we fit the line with a single Gaussian. The central velocity ($\varv_0$) across \opha~covers the range from 2.25 to 3.76~km~s$^{-1}$~with a median value of 3.12 \kms, which is in agreement with the systemic velocity of \opha~of $\sim$3.44~km~s$^{-1}$ \citep{andre07}. We used the line width of the Gaussian as the measure of total velocity dispersion toward each pixel. 

To construct the map of the 1D non-thermal (turbulent) velocity dispersion ($\sigma_\mathrm{NT}$), we subtracted the thermal component ($\sigma_{\varv,\mathrm{th}}$) from the map of the total velocity dispersion, given by:
\begin{equation}
\sigma_\mathrm{NT}=\sqrt{\sigma_\varv^2-
\sigma^2_{\varv,\mathrm{th}}}=\sqrt{\sigma_\varv^2- \frac{k_\mathrm{B}T_\mathrm{gas}}{m_\mathrm{mol}}}
~\left(\mathrm{km~s}^{-1}\right),\label{eq:NT}
\end{equation}
where $m_\mathrm{mol}$=29~amu is the mass of the HCO$^{+}$ molecule, $k_\mathrm{B}$ is the Boltzmann constant, and $T_\mathrm{gas}$ is the kinetic temperature of the gas. Here, we assume that the gas is in LTE condition and well-coupled with the dust in the cloud. Thus, the gas kinetic temperature is equal to the excitation temperature in all levels ($T_\mathrm{ex}$) and equal to the dust temperature: $T_\mathrm{gas}=T_\mathrm{ex}=T_\mathrm{dust}$. We convolved the map of $T_\mathrm{dust}$ to the lower resolution of the molecular line map (14$\arcsec$) to obtain the same pixel size on both maps. We converted the velocity dispersion of the non-thermal component into the FWHM velocity dispersion, given as $\Delta\varv_\mathrm{NT}=\sigma_\mathrm{NT}\sqrt{8\ln{2}}$. The uncertainty of non-thermal FWHMs is propagated from the uncertainty of the line width obtained from the Gaussian fit. 

\begin{figure}[tbp]\centering
\includegraphics[width=0.49\textwidth]{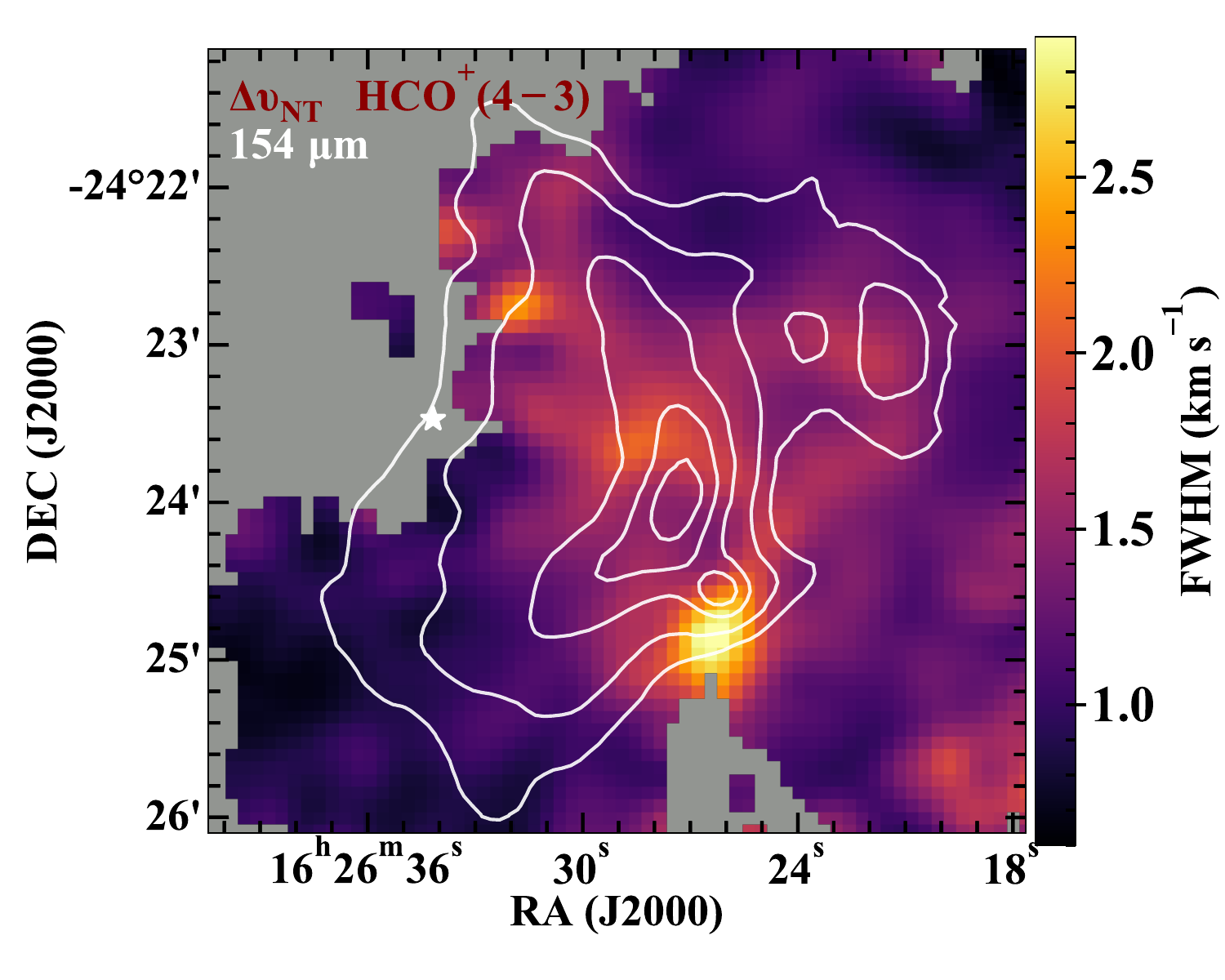}
\caption{Distribution of the non-thermal FWHM of the HCO$^{+}$(4$-$3) molecular line toward \opha. The star symbol marks the position of the Oph-S1 star. White contours show the continuum emission (Stokes-$I$) at 154~\um, with contours levels at 0.12, 0.2, 0.4, 0.6, and 0.8 Jy/arcsec$^{2}$. }\label{fig:delta_velo}
\end{figure}

Figure~\ref{fig:delta_velo} shows the map of the 1D non-thermal FWHM of the HCO$^{+}$ molecular line toward \opha. The $\Delta\varv_\mathrm{NT}$ values are in the range from 0.62 to 2.98~km~s$^{-1}$ with a median value of 1.19~km~s$^{-1}$. We find a good agreement between our result obtained with HCO$^{+}$(4$-$3) and those from other line tracers, namely, 0.95~\kms~using N$_2$H$^+$(3$-$2) from APEX and 1.5~\kms~using C$^{18}$O(3$-$2) from JCMT/HARP (see Appendix~\ref{app:velo_disper_comparison} for details). Thus, it is plausible to use results from the HCO$^{+}$(4$-$3) line in the following analysis.
%=================================================
%=================================================

\subsubsection{Gas volume density map}\label{ssec:gas_density}
%=================================================
\begin{figure*}[tbp]\centering
\includegraphics[width = 0.95\textwidth]{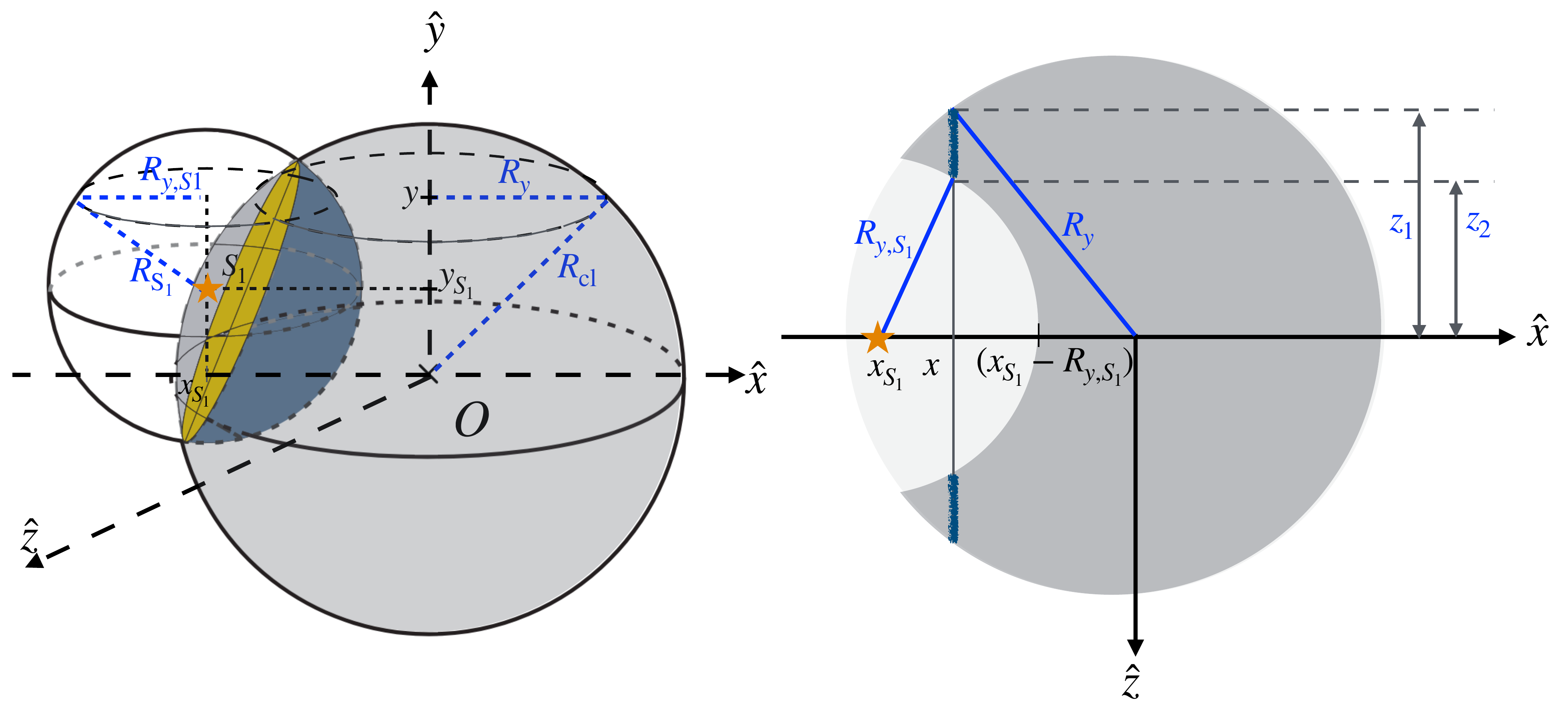}\\
\includegraphics[width = 0.485\textwidth]{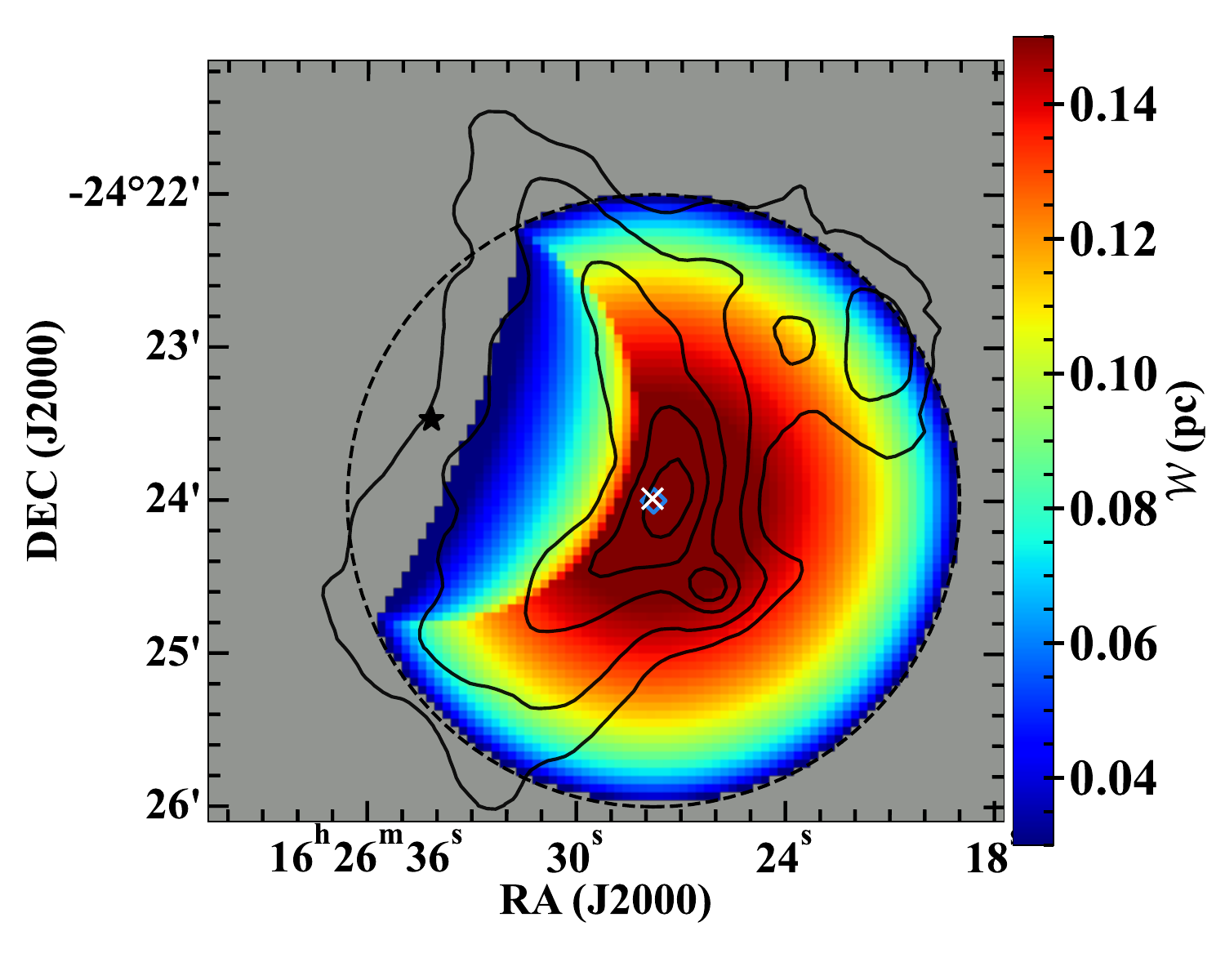}
\includegraphics[width = 0.485\textwidth]{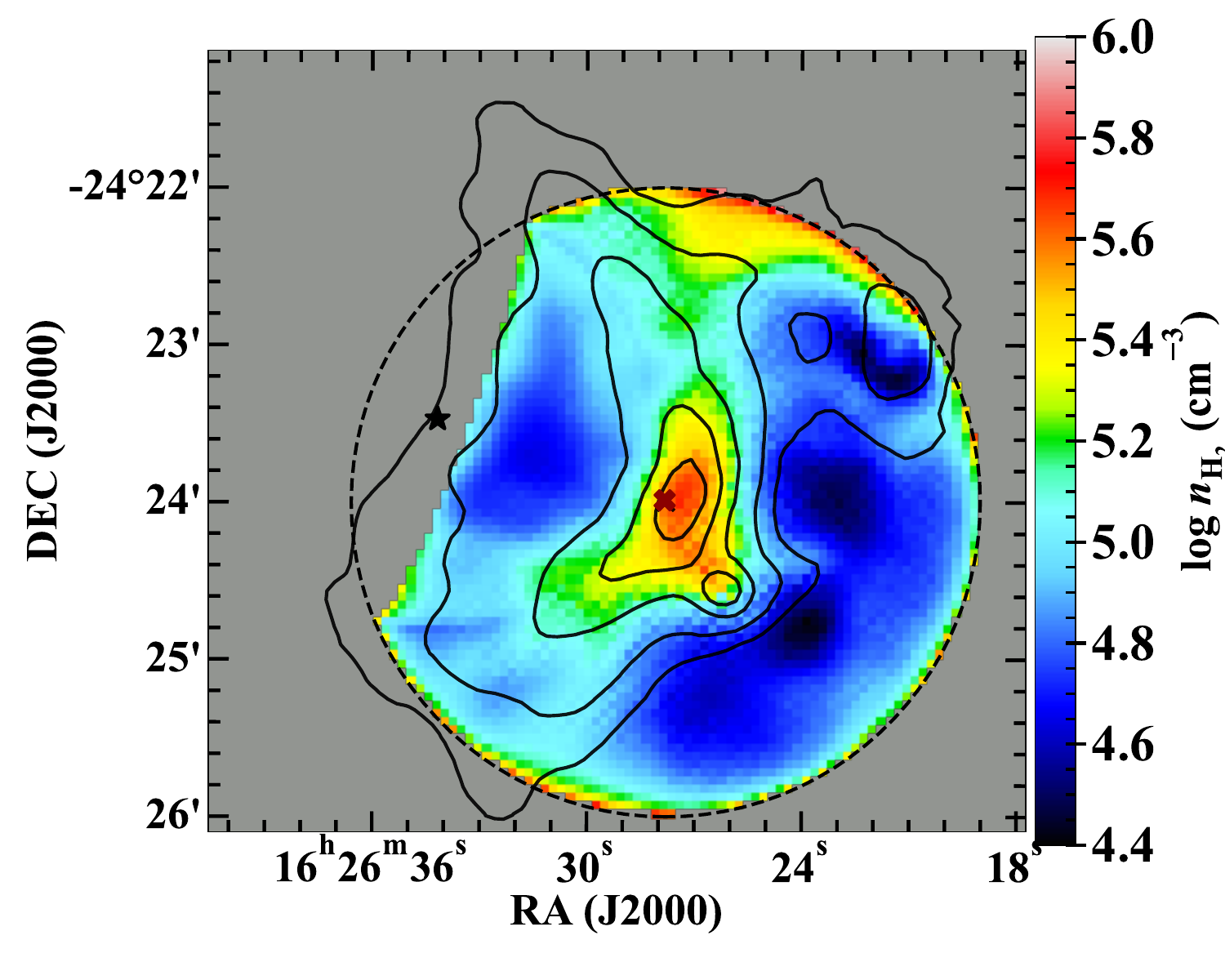}
\caption{Cartoon of 3D structure of the \opha~cloud where we assume that the cloud has a spherical shape but is pushed by radiation feedback from a nearby high-mass star Oph-S1, shown at the top (left panel, see text) and a plane of the constant coordinate $y$ with the LOS $\hat{z}$ (right panel). Bottom row displays the map of the depth of the cloud, $\mathcal{W}$ (left panel), and the map of the gas volume density (\nHii) inferred from \NHii (right panel). In each panel, the "$\times$" and star symbols mark the peak position of \NHii~and location of Oph-S1 star, respectively. Toward maps of $\mathcal{W}$ and \nHii~in bottom panels, the black dashed circle has a radius $R_\mathrm{cl}$=120\arcsec~and black solid contours show the continuum emission at 154~\um~with levels similar as shown in Fig.~\ref{fig:delta_velo}. }
\label{fig:volume_map}
\end{figure*}

We estimated the gas volume, \nHii, across the dense cloud from \NHii. Here, we assume that the cloud has a spherical Plummer-like shape \citep{liseau15,Santos19}, which is frequently used to describe the globular dense cores \citep{Plummer1911}. We consider a sphere centered at the peak position of \NHii, located at (RA, Dec)= ($16^{h}26^{m}27\fs8$, $-24\degr24\arcmin00\farcs0$), which is also associated with the position of the starless core SM1 (the separation is much less than 1\arcsec). We took the radius of the cloud $R\rm{_{cl}}\approx120\arcsec$~\citep[$\sim$0.074~pc,][]{Santos19}. To describe the effect of stellar feedback from the high-mass Oph-S1 star, we further add a spherical shell with a radius of $R_{\rm S_1}\approx80\arcsec$ surrounding this star. We also assume that the positions of peak column density and the S1~star are located in the same plane perpendicular to the LOS. The top panels in Fig.~\ref{fig:volume_map} show the 3D schematic view of the cloud (left) and a cross-section on the $xz$-plane at a given $y$ position (right). The $x$-~and $y$-axes perpendicular to each other are in the plane of the sky and the $z-$axis is along the LOS. The center of the sphere is located at $(x_{\rm 0},~y_{\rm 0},~z_{\rm 0})=(0,~0,~0)$. The depth of the cloud at each position ($\nolinebreak x,~y$) along $z-$axis is calculated by: 
\begin{equation}\label{eq:depth}
\mathcal{W}(x,y) =2\left( z_1 - z_2\right)=2\left(\sqrt{R_y^2 - x^2} - \sqrt{R_{y,S_1}^2 - (x-x_{\rm S_1})^2}\right), 
\end{equation}
where $R_{y}= \sqrt{R_{\mathrm{cl}}^2 - y^2}$ and $R_{y,S1}= \sqrt{R_{\rm{S1}}^2 - (y-y_{S1})^2}$ are the radius of horizontal circles perpendicular to the plane of sky at $y$ position (see top right panel in Fig.~\ref{fig:volume_map}). We obtain the map of $\mathcal{W}$, shown in the bottom left panel in Fig.~\ref{fig:volume_map}. The values of $\mathcal{W}$ are in the range of $\sim 20\arcsec - 222\arcsec$. We then infer \nHii~from \NHii~using the formula:
\begin{equation}
n_\mathrm{H_2}=\frac{N_\mathrm{H_2}}{\mathcal{W}}. 
\end{equation}

The bottom right panel in Fig.~\ref{fig:volume_map} shows the spatial extent of \nHii~across \opha, overlaid by the contours representing the continuum emission in band D (Stokes-$I$). The \nHii~map has an angular resolution of 11.4\arcsec and a pixel size of $\sim$3.2$\arcsec$, similar to that of the \NHii~map. The gas density spans from $\sim $$10^{4.4}-10^6$~\cmcube~and is highest at the center of the cloud. The pattern of \nHii~distribution well follows the intensity map in band D, revealing the ridge shape of the dense cloud. Some pixels located at the circular boundary in the \nHii~map show anomalously high densities; this is due to the very low depths caused by the spherical model, and thus we exclude these pixels from our \BPOS~strength measurement.  

%=================================================
\begin{figure*}[h!]\centering
\includegraphics[width=\linewidth]{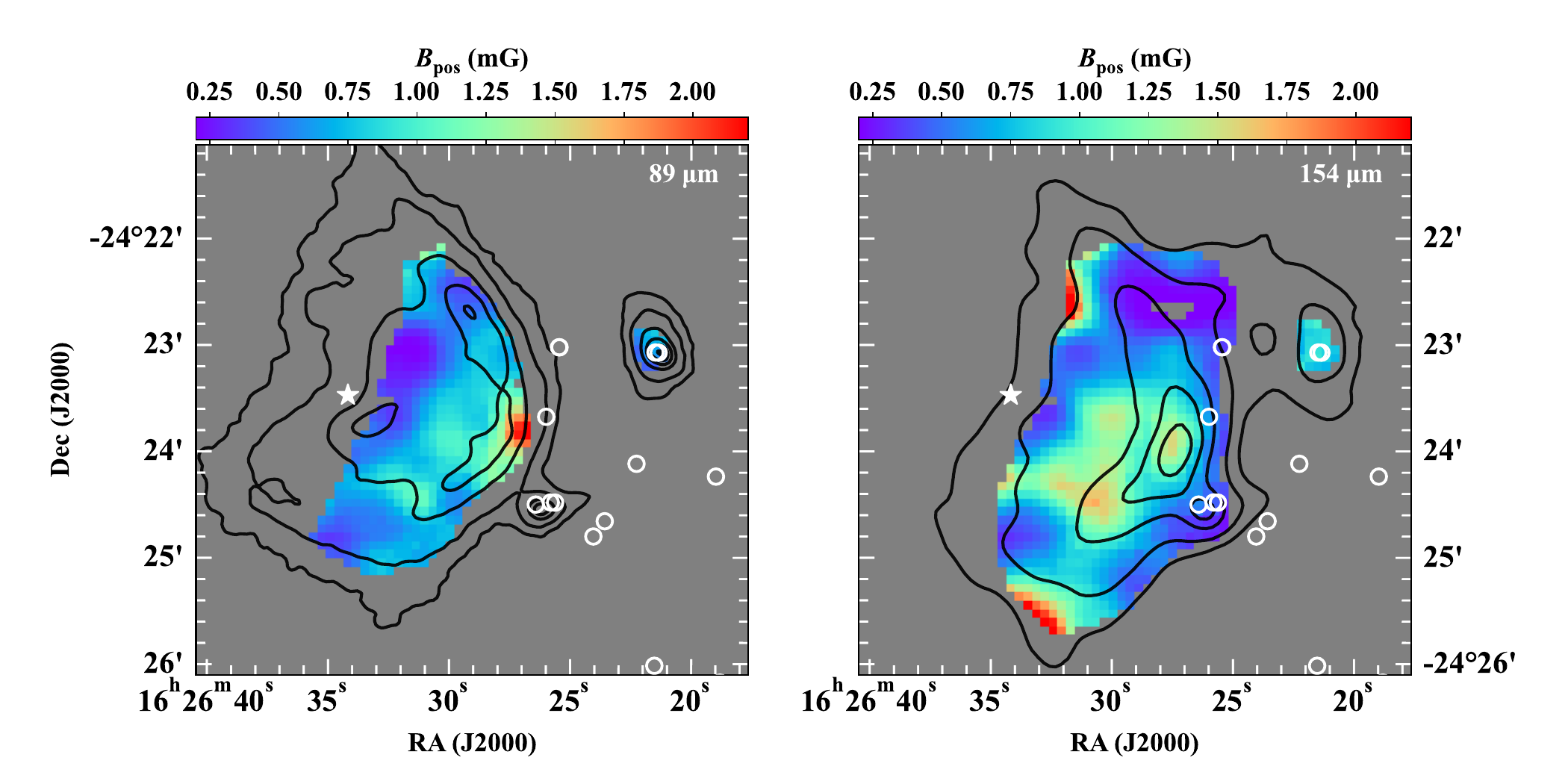} 
\caption{Maps of B-field strengths projected in the plane-of-the-sky (\BPOS) in \opha~using data in band C (left panel) and band D (right panel), respectively. The contours in each map are similar as in Fig.~\ref{fig:Bfield_orientation}. The star symbol marks the position of the Oph-S1 star and small circle symbols show the position of YSOs. }
\label{fig:Bpos_map}
\end{figure*}
%=================================================

\subsubsection{Magnetic field strength maps}\label{ssec:Bpos}
We used the maps of $\delta\theta$, $\Delta\varv_\mathrm{NT}$, and \nHii~to measure $B_\mathrm{pos}^\mathrm{DCF}$ in the entire region of \opha. All input maps were smoothed to the angular resolution of 14$\arcsec$, corresponding to that of the $\Delta\varv_\mathrm{NT}$ map. We then reconstructed the maps of $B_\mathrm{pos}$ at 89 and 154~\um~toward \opha~using Eq.~\ref{eq:Bpos}, which are shown in Fig.~\ref{fig:Bpos_map}. 

The $B_\mathrm{pos}^\mathrm{DCF}$~values vary from $0.2-2.3$~mG and from $0.2-2.5$~mG for data at 89 and 154~\um, respectively. The highest strength is found at the center region of the cloud, associated with the emission peak at 154~\um~and also the position of the SM1 core. The median $B_\mathrm{pos}^\mathrm{DCF}$~values are $0.7\pm0.3$ and $0.7\pm0.4$~mG\footnote{The uncertainties of these median values are solely of the estimated $B_\mathrm{pos}^\mathrm{DCF}$~distribution across the cloud. }, as measured using 89 and 154~\um~data, respectively. 
The uncertainty of \BPOS~is propagated based on the uncertainty of the velocity dispersion of the gas and polarization angle dispersion. Since \nHii~map is inferred from \NHii~resulted in SED fitting models \citep[see][]{Santos19}. In addition, the estimated depth of the cloud is based on the 3D geometrical shape of the cloud and also is uncertain. 
Thus, we did not take into account the uncertainty of the gas volume density in the total uncertainty of the B-field strengths. For this reason, the uncertainty of $B_\mathrm{pos}$ might be underestimated. 

\subsection{Measurement of magnetic field strength using other methods}\label{ssec:B_other_med}

Several new techniques have been proposed to improve the DCF method for the last decades. For instance, \cite{hi09} introduced the angle dispersion function (ADF) method where the dispersion in polarization angles $\delta\theta$ in Eq.~\ref{eq:Bpos} is replaced by the square root of the structure function of polarization angles. This provides an improved measurement of B-field strength but also gives us the ratio of the turbulent to large-scale B-field strengths. Alternatively, \cite{ST21} introduced a modification of the DCF method (hereafter ST), assuming that the mean (ordered) and fluctuation components of the total B-field energy, $B_0$ and $B_\mathrm{t}$, respectively, are not independent. 
More recently, the differential measure approach \citep[DMA;][]{laz22} is considered promising to enhance the accuracy of B-field strength measurements at scales of $\ell$ smaller than the turbulence injection scale and takes into account the anisotropic properties of MHD turbulence (given by factor $f^\prime$). 
For the sake of cross-checking, the B-field strength measurement in \opha~using the DCF method, we re-calculated the mean B-field strengths using the ADF, ST, and DMA methods at the two subregions (a) and (e) associated with the densest parts of the cloud (these subregions are named as in \cite{Kwon18}, see also in Fig.~\ref{fig:orient_compare}). 

% ================
\begin{figure*}
\sidecaption\centering 
\includegraphics[width = 0.7\textwidth]{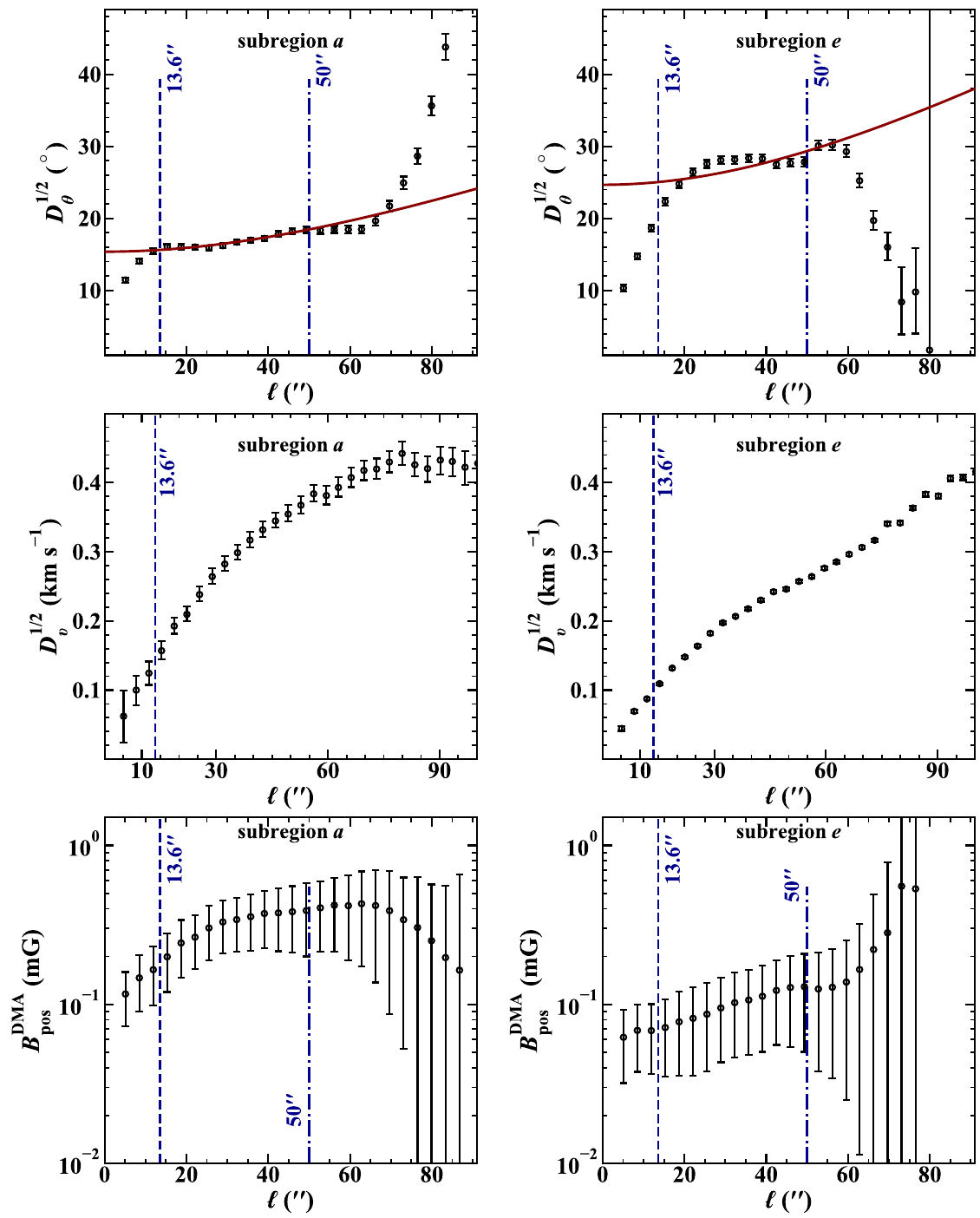}
\caption{Structure function of polarization angles (top panels) and of central velocity (middle panels) measured in the subregions (a) (left) and (e) (right). The bottom panels show the resulting $B_\mathrm{pos}^\mathrm{DMA}$ strength estimated via Eq.~\ref{eq:dma} in the two subregions (a) and (e). The HAWC$+$ data at 154~\um~is used to calculate the structure-function (data points). The red solid line in the top panels shows the fitted line of the angular structure-function to the quadratic function, where we chose data points with 13.6\arcsec$<\ell\leq$50\arcsec. The blue dash-dotted line in each panel indicates the upper $\ell$-limit of 50\arcsec~used for fitting the angular structure-function and for estimating the mean value of $B_\mathrm{pos}^\mathrm{DMA}$ in the two subregions (a) and (e) (see texts).   The blue dashed line in each panel indicates the beam size of HAWC$+$ observation at 154~\um. }\label{fig:fit_ADF}
\end{figure*}

\subsubsection{ADF method} 

According to the ADF method, the $B_\mathrm{pos}^\mathrm{ADF}$~can be estimated using Eq.~\ref{eq:Bpos} but the polarization angle dispersion $\delta\theta$ is replaced by the structure function of the polarization angles, so-called angular structure function \citep{hi09}. The angular structure function, $D^{1/2}_\theta$, at an angular scale $\ell$ is estimated as:
\begin{equation}\label{eq:SF}
D^{1/2}_\theta(\ell) \equiv \left< \Delta \theta^2 (\ell) \right> ^{1/2}
= \left\{ \frac{1}{N(\ell)} \sum_{i=1}^{N(\ell)} \Bigl[ \theta(x) - \theta(x+\ell)\Bigr]^2\right\}^{1/2}, 
\end{equation} 
where $\left< ... \right>$ denotes an average, $\Delta\theta(\ell) \equiv\theta(x) - \theta(x+\ell)$ is the polarization angle difference between individual pairs of polarization vectors at position $x$ and $x+\ell$, where $\theta(x)$ and $\theta(x+\ell)$ are the corresponding polarization angles at position $x$ and $x+\ell$, respectively, and $N(\ell)$ is number of pairs of vectors with a displacement of $\ell$. 
At scales $\ell$ much smaller than the scale for a variation of the large-scale B-field structure (typically of $\sim$5$\arcmin$ seen by \textit{Planck} observations), the total angular dispersion function can be expressed as:
\begin{equation}\label{eq:quadratic_func}
D_\theta(\ell) \simeq b^2 + m^2\ell^2 + \sigma_\mathrm{M}^2(\ell), 
\end{equation}
where $b$ represents the turbulent component of B-fields ($B_\mathrm{t}$), $m\ell$ describes the large-scale structure of B-fields ($B$), and $\sigma_\mathrm{M}(\ell)$ is subject to measurement uncertainties. Also, the ratio between the turbulent and large-scale B-field strength indicates the angular dispersion, given as \citep{hi09}:
\begin{equation}
\delta\theta\simeq \frac{\left<B^2_\mathrm{t}\right>^{1/2}}{B} =\frac{b}{\sqrt{2-b^2}}.
\end{equation}
When $B_\mathrm{t} \ll B $, i.e., $b\ll1$~rad, we can obtain $\delta\theta \sim b/\sqrt{2}$. 

For each subregion (a) and (e) in \opha, we calculate the angular structure function $D_\theta(\ell)$ using HAWC$+$ data at 154~\um~via Eq.~\ref{eq:SF} and then perform the fit to the quadratic function described in Eq.~\ref{eq:quadratic_func} using the Python {\tt curve$_{-}$fit} routine. 
We select data points that 13.6\arcsec$<\ell\leq$50\arcsec~to avoid fitting to structures higher than the beam size and to better fit to smaller-scale structures. Top panels in Fig.~\ref{fig:fit_ADF} show structure functions of polarization angles as a function of the distance of displacement $\ell$ toward subregions (a) and (e). From the fitted $b$ value of the fits, we derive $\delta\theta$ values of $10.9\degr\pm 0.2\degr$ and $17.5\degr\pm 0.2\degr$ toward subregions (a) and (e), respectively. We find these values in good agreement with mean polarization angle dispersion values estimated using our approach described in Sect.~\ref{ssec:angular_dispersion}, which are $\sim$9.2\degr~and $\sim$14.3\degr~in subregions (a) and (e), respectively. Toward subregions (a) and (e), mean gas volume densities are $2.1\times10^5$~and $1.1\times10^5$~\cmcube~respectively, and their mean non-thermal FWHMs are 1.74 and 1.45~km~s$^{-1}$, respectively. We substitute new $\delta\theta$ values together with the above mean values of \nHii~and $\Delta\varv_\mathrm{NT}$ into Eq.~\ref{eq:Bpos} to calculate $B_\mathrm{pos}^\mathrm{ADF}$~strengths in subregions (a) and (e), which are $0.69$~mG and $0.26$~mG, respectively. These values are in good agreement with $B_\mathrm{pos}^\mathrm{DCF}$~strengths obtained using the DCF method, where $B_\mathrm{pos}^\mathrm{DCF}$~values are in ranges of 0.32--1.55~mG and 0.15--0.97~mG toward subregions (a) and (e), respectively, with corresponding mean $B_\mathrm{pos}^\mathrm{DCF}$~values of $\sim$0.84 and $\sim$0.37~mG. 

\subsubsection{ST method} \label{ssec:st}
The ST method is applicable for the compressible B-fields mode and thus is expected to provide a more accurate estimate of B-field strengths \citep{skalidis21}. 
Another advantage of the ST method is that this method accounts for the correlation between the fluctuation and mean components of the total B-field energy not to be zeros. 
Here, we measure maps of \BPOS~strength ($B_\mathrm{pos}^\mathrm{ST}$) in band C and band D using the formula as appeared in \cite{ST21}: 
\begin{equation}\label{eq:ST_method}
B_\mathrm{pos}^\mathrm{ST} = \sqrt{4\pi\rho} \frac{\Delta\varv_\mathrm{NT}}{\sqrt{2\delta\theta}} = \frac{1}{\sqrt{2\delta\theta}}B_\mathrm{pos}^\mathrm{DCF}. 
\end{equation}

We show the maps of $B_\mathrm{pos}^\mathrm{ST}$ across \opha~in band C and band D in Fig.~\ref{fig:Bpos_map_ST}. We find that $B_\mathrm{pos}^\mathrm{ST}$ value ranges from $0.03 -0.7$~mG at 89~\um~and $0.02-1.3$~mG at 154~\um, appeared to be lower than $B_\mathrm{pos}^\mathrm{DCF}$. A median $B_\mathrm{pos}^\mathrm{ST}$ value of $\sim$$0.2\pm0.1$~mG is seen in both bands. For a direct comparison with those using other methods, we estimate the mean value of $B_\mathrm{pos}^\mathrm{ST}$~at 154~\um~toward two subregions (a) and (e), which are $\sim$0.19 and $\sim$0.07~mG, respectively. These values are about 4.5 times lower than those estimated using the DCF method. 

\subsubsection{DMA method} 
We also calculate \BPOS~trengths in two subregions (a) and (e) using the DMA method,  $B_\mathrm{pos}^\mathrm{DMA}$,  using Eq.~\ref{eq:dma} following \cite{laz22}: 
\begin{equation}\label{eq:dma}
B_\mathrm{pos} = f^\prime\sqrt{4\pi\rho}\frac{D^{1/2}_\varv(\ell)}{D^{1/2}_\theta(\ell)}, 
\end{equation}
where $D^{1/2}_\theta(\ell)$ and $D^{1/2}_\varv(\ell)$ are the second order structure function of polarization angles and central velocity at an angular scale $\ell$, $f^\prime$ is the correction factor depending on the anisotropic properties of MHD turbulence. Here, we adopt $f^\prime$=1 related to the naive formalism as suggested in \cite{laz22}. 

We calculate the structure function of the central velocity of HCO$^+$(4$-$3) line toward subregions (a) and (e), using the similar form in Eq.~\ref{eq:SF} but for the central velocity $\varv_0$, shown in middle panels in Fig.~\ref{fig:fit_ADF}. We then measure $B_\mathrm{pos}^\mathrm{DMA}$~strengths in two subregions (a) and (e) via Eq.~\ref{eq:dma}, using the information from the $D^{1/2}_\theta(\ell)$ together with $D^{1/2}_\varv(\ell)$ and the mean values of \nHii~in the subregions. The $B_\mathrm{pos}^\mathrm{DMA}$ as a function of displacement $\ell$ is shown in bottom panels of Fig.~\ref{fig:fit_ADF} for subregions (a) and (e). It can be seen that the resulting $B_\mathrm{pos}^\mathrm{DMA}$ in the range $13.6\arcsec <\ell <50\arcsec$ is likely flatten within their uncertainty. Thus, in each subregion we take the mean of $B_\mathrm{pos}^\mathrm{DMA}$ in this range to be the average value, which are $\sim$0.34~mG and $\sim$0.12~mG in subregions (a) and (e), respectively. These values are slightly higher than the resulting \BPOS~estimated using the ST method, but they are at the low-end of that measured using the DCF method (i.e., $\sim$2--3 times less than the mean values estimated using the DCF method presented in Sect.~\ref{ssec:Bfields_strength}). These differences are indeed expected since the DCF method somewhat overestimates the \BPOS~strength in the cloud. We  discuss this issue further in Sect.~\ref{ssec:dis_Bfields}. 

%=================================================
\subsection{Mass-to-flux ratio}\label{ssec:mass_to_flux}
The relative importance of B-fields compared to gravity in the cloud can be quantified by the mass-to-flux ratio, $M/\phi$, which can be estimated in units of the critical ratio, $\left(M/\phi_{B}\right)_\mathrm{crit} =1/(2\pi\sqrt{G})$ \citep{nakano78}: $\lambda\equiv\left(M/\phi \right)_\mathrm{observed}/\left(M/\phi\right)_\mathrm{crit}$. We measure $\lambda$ in the entire region of \opha, using the formula introduced by \cite{crutcher04}:
\begin{equation}\label{eq:mass_to_flux_intro}
\lambda=7.6\times10^{-21}\frac{N_\mathrm{H_2}}{B},
\end{equation}
where $N_\mathrm{H_2}$ and $B$ are in units of cm$^{-2}$ and $\mu$G, respectively. 

Due to the lack of $B_\mathrm{los}$~component measurements in the \opha~cloud, we adopted the $B_\mathrm{pos}$ component as the total B-field strengths $B$: $B\equiv B_\mathrm{tot}\approx B_\mathrm{pos}$. For the  analysis, we  employed only the $B_\mathrm{pos}^\mathrm{DCF}$ map estimated using the DCF method and inferred from the polarimetric data in band D; this is because it covers a larger area of the higher density region in \opha~than that coming from the data in band C. 

\begin{figure}[h!]\centering
\includegraphics[width=0.44\textwidth]{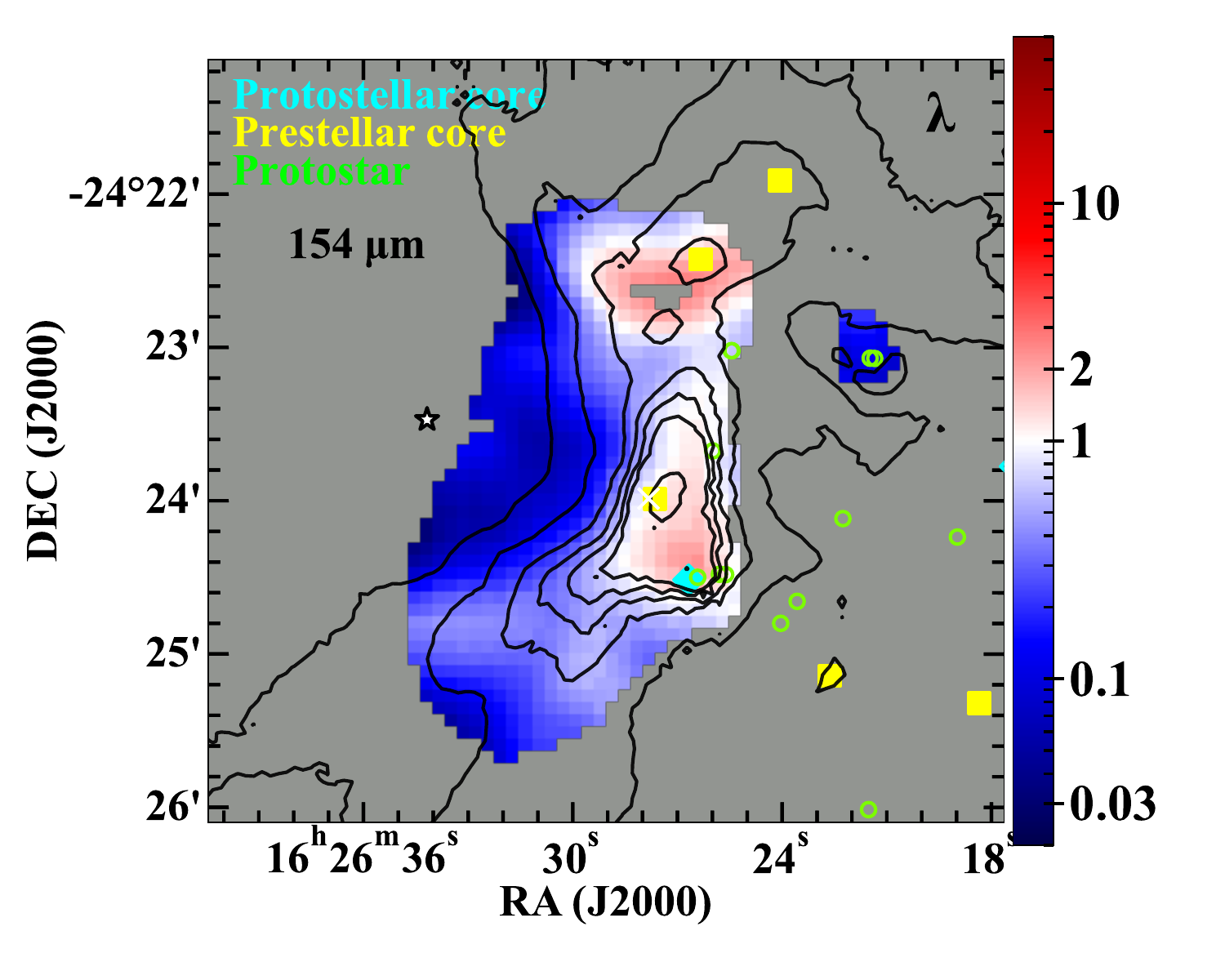}\\
\includegraphics[width=0.44\textwidth]{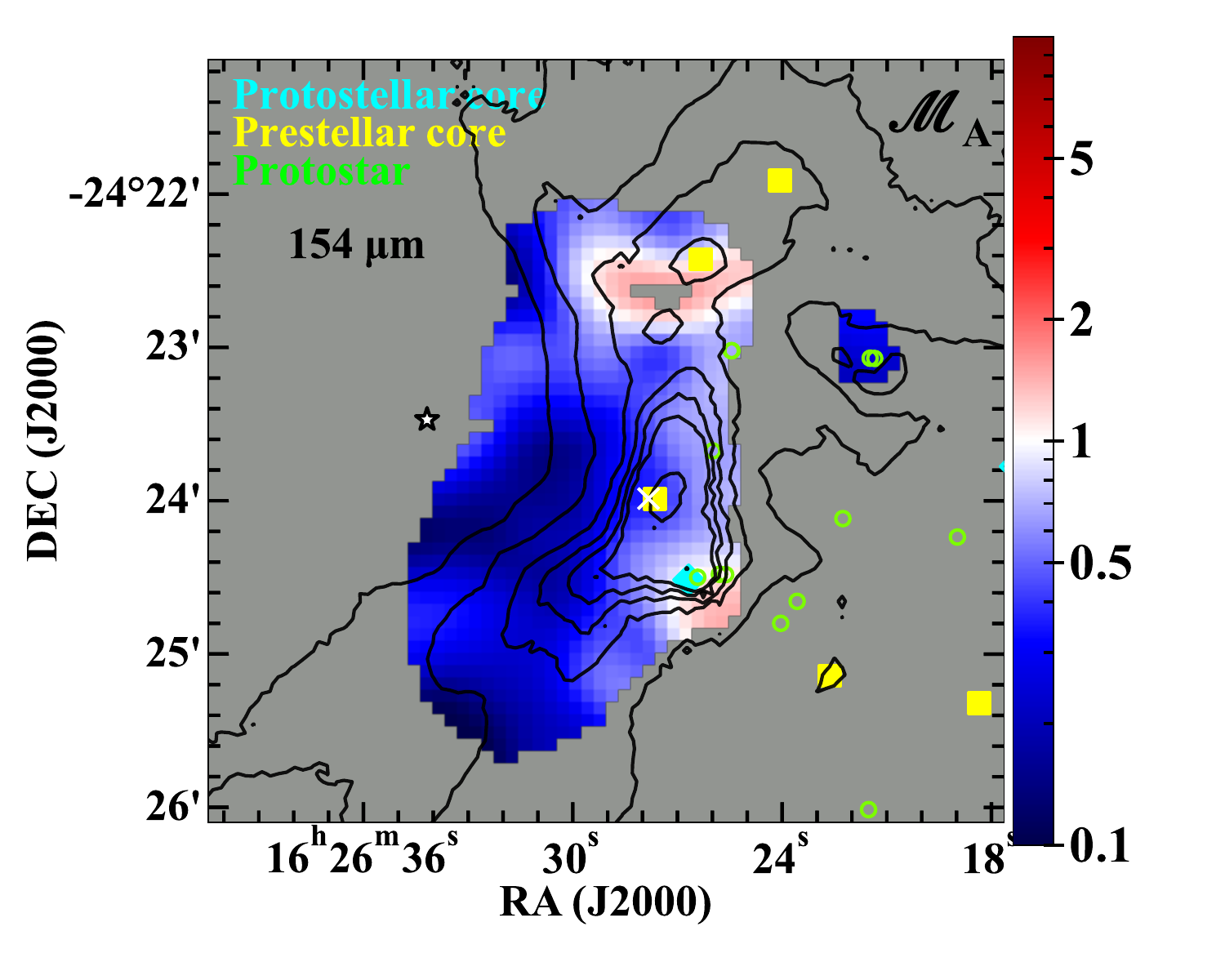}\\ 
\includegraphics[width=0.44\textwidth]{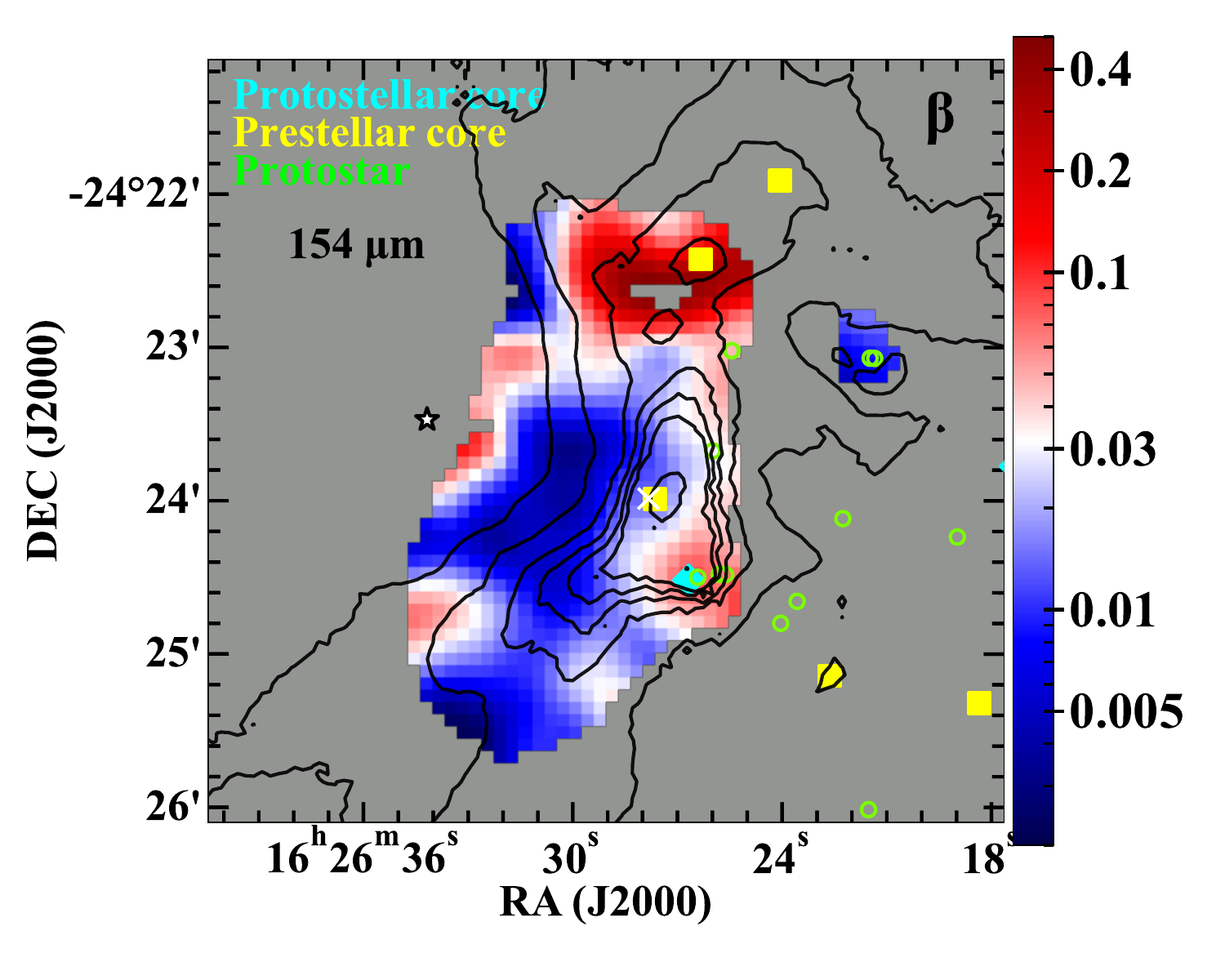} 
\caption{Maps of $\lambda$ (top panel), Alfv\'en Mach number (middle panel), and plasma $\beta$ parameter (bottom panel) in \opha~using the $B_\mathrm{pos}^\mathrm{DCF}$ in band D. In each map, yellow square and cyan diamond symbols indicate the positions of prestellar and protostellar cores, respectively. Green circles indicate the positions of YSOs \citep{Enoch2009,evans09,con10}. The star and \lq$\times$' symbols mark the positions of the Oph-S1~star and the SM1~core, respectively. Black contours show \NHii, similar as in Fig.~\ref{fig:tdust}. }
\label{fig:lambda_alfven_beta_pressure_maps}
\end{figure}

The top panel in Fig.~\ref{fig:lambda_alfven_beta_pressure_maps} shows the spatial distribution of the mass-to-flux ratio in \opha. It reveals an increasing gradient of $\lambda$ from the central and eastern parts of the clump to the outer regions, spanning from 0.03--2.56, with a median value of $\sim$0.26. The region exhibiting the lowest values of $\lambda \ll$1 is close to the Oph$-$S1 star, suggesting that the region is firmly magnetically sub-critical, namely, B-fields are strong enough to prevent the gravitational collapse. This can also be explained by the influence of the strong winds and radiation fields from the Oph$-$S1 star which tend to push the material away to to its vicinity at the west side. The mass-to-flux ratio increases toward the outskirt regions of the Oph-S1 star, especially toward regions along the ridge of the dense cloud (i.e. center, north, and southeast parts of the map) where the gas column density is $\ga 4\times10^{22}$~cm$^{-2}$. These regions are magnetically super-critical and so, the self-gravitational collapse might have occurred there, resulting in several protostars and a single protostellar core identified in the region \citep{Enoch2009,evans09,con10}.
%=================================================

\subsection{Alfv\'enic Mach number}\label{ssec:M_A}
To investigate the interplay between B-field and turbulence due to the non-thermal motion of the gas in \opha, we calculate the map of Alfv\'enic Mach number, given as:
\begin{equation}\label{eq:Ma}
\mathscr{M}_\mathrm{A} =\sqrt{3}\frac{\sigma_\mathrm{NT}}{\varv_\mathrm{A}}, 
\end{equation}
where $\sigma_\mathrm{NT}$ is given by Eq. \ref{eq:NT}, and $\varv_\mathrm{A}$ is the Alfv\'enic speed driven by the B-fields ($\varv_\mathrm{A}=B/\sqrt{4\pi\rho}$), both in units of km~s$^{-1}$. The factor $\sqrt{3}$ in Eq.~\ref{eq:Ma} assumes that turbulence is isotropic, and allows the calculation of the 3D Alfv\'enic Mach numbers. We note that adding this factor will provide upper limits to the estimated \Ma. 

The middle panel in Fig.~\ref{fig:lambda_alfven_beta_pressure_maps} shows the map of $\mathscr{M}_\mathrm{A}$ toward \opha~using results of polarimetric data at 154~\um~and HCO$^+$(4--3) as the gas tracer. The $\mathscr{M}_\mathrm{A}$ values are in the range from 0.1--1.5 across the cloud, with a median value of 0.38.  Most of \opha~shows $\mathscr{M}_\mathrm{A}$<1 (magnetically sub-Alfv\'enic), suggesting the B-field dominates turbulence in the cloud. However, two small regions at the north and southwest of the map exhibit $\mathscr{M}_\mathrm{A}\geq$1, indicating strong Alfv\'enic motions of the gas (magnetically super-Alfv\'enic). 
These regions are also associated with the positions of a few prestellar and protostellar cores, and YSO candidates \citep{Enoch2009,evans09,con10}. 

\subsection{Plasma $\beta$ parameter}\label{ssec:beta}
To compare the magnetic presure ($P_\mathrm{mag}$) and thermal pressure ($P_\mathrm{th}$) in \opha, we constructed the map of the plasma beta parameter ($\beta$), given as:
\begin{equation}\label{eq:beta}
\beta = \frac{P_\mathrm{th}}{P_\mathrm{mag}}=2\left(\frac{c_\mathrm{s}}{\varv_\mathrm{A}}\right)^{2}, 
\end{equation} 
where $c_\mathrm{s} = \sqrt{{k_\mathrm{B}T_\mathrm{gas}} /\left(\mu~m_\mathrm{H}\right)}$ is the thermal sound speed measured in units of km~s$^{-1}$, assuming $T_\mathrm{gas}$=$T_\mathrm{dust}$ (see Sect.~\ref{ssec:velocity-dispersion}).

The bottom panel in Fig.~\ref{fig:lambda_alfven_beta_pressure_maps} shows the map of the $\beta$ parameter toward \opha~in band D. We obtained $\beta$<1  across the map with a median value of 0.02. This result implies that the cloud is supported by strong magnetic pressure, which is much more prominent than that induced by the thermal motions of the gas.

\subsection{Virial mass analysis}\label{ssec:virial_mass}
To probe the dynamical state of the \opha~cloud, we calculated the virial parameter ($\alpha_\mathrm{vir}$), which is the ratio of the virial mass ($M_\mathrm{vir}$) and the isothermal mass ($M_\mathrm{iso}$) of the cloud. 
Here, we assume the cloud as a uniform density sphere with an effective radius of $R_\mathrm{eff}$ as a whole, having a mean mass density of $\left<\rho\right>$. The isothermal mass, $M_\mathrm{iso}$, considering the self-gravity of the cloud
can be estimated as:
\begin{equation}\label{eq:iso_mass}
M_\mathrm{iso}=\frac{4}{3}\pi\left<\rho\right> R_\mathrm{eff}^{3}. 
\end{equation}
The virial mass, $M_\mathrm{vir}$, determined from magnetic, thermal, and non-thermal kinetic energies can be estimated as \citep[see e.g.,][]{BM92,pillai11}: 
\begin{equation}\label{eq:vir_mass}
M_\mathrm{vir}=\frac{5R_\mathrm{eff}}{G}\left(\frac{\left< \varv_\mathrm{A}\right>^2}{6}+\left<c_\mathrm{s}\right>^2+\left<\sigma_\mathrm{NT}\right>^2\right), 
\end{equation}
where $\left< \varv_\mathrm{A}\right>$, $\left<c_\mathrm{s}\right>$, and $\left<\sigma_\mathrm{NT}\right>$ are  the mean values of the Alfv\'enic velocity, thermal sound speed, and of non-thermal velocity dispersion toward the cloud, respectively, while $G$ is the gravitational constant. 

We calculate $R_\mathrm{eff}$ by selecting an ellipse covering the area where \BPOS~was measured. 
This ellipse is centered at (RA, Dec) =  ($16^{h}26^{m}29\fs5, -24\degr 23\arcmin 50\farcs0$) with the semi-major and semi-minor axes are $a$=120$\arcsec$ and $b$=75$\arcsec$, respectively. The $R_\mathrm{eff}$ is then determined as $R_\mathrm{eff}=\sqrt{ab}\approx95\arcsec$ ($\sim$0.063~pc assuming the distance to \opha~of 137~pc).  
Within this area, the mean gas volume density is $\left< n_\mathrm{H_2}\right> = 2.2\times 10^5$~\cmcube, corresponding to the mean mass density $\left<\rho \right>= 1.01\times 10^{-18} \rm g~cm^{-3}$. The corresponding mean values of velocities are: $\left< \varv_\mathrm{A}\right>=3.36~\rm km~s^{-1}$, $\left<c_\mathrm{s}\right> = 0.09 ~\rm km ~s^{-1}$, and $\left<\sigma_\mathrm{NT}\right> = 0.59 ~\rm km ~^{-1}$. We used these mean values of the velocities, non-thermal velocity dispersion, and the gas volume density to calculate $M_\mathrm{iso}$ and $M_\mathrm{vir}$ in Eqs.~\ref{eq:iso_mass}~and \ref{eq:vir_mass}, yielding $\alpha_\mathrm{vir}=M_\mathrm{vir}/M_\mathrm{iso}\approx$15. This shows that the cloud is gravitationally unbound and that B-fields and turbulence can support the cloud against self-gravity. This high value of $\alpha$~obtained in the cloud overalls could explain the low efficiency of star formation in \opha; however, in a few regions, the density is very high (up to 10$^5$ or 10$^6$~\cmcube). This is consistent with the lack of prestellar cores and protostellar sources in the central region, which can be seen in the lower density parts of the cloud only (see e.g., Fig.~\ref{fig:lambda_alfven_beta_pressure_maps}). To study smaller scales across the clump, polarimetric observations with a higher spatial resolution would be required.

% ------------------------------------------------
\section{Discussion}\label{sec:dis}
\subsection{Morphology and strengths of magnetic field in \opha~}\label{ssec:dis_Bfields}
The B-field orientation in \opha~is well-ordered and predominantly perpendicular to the ridge of the dense cloud (Sect.~\ref{ssec:Bfields_orientation}). The feature is illustrated most clearly by the data at the longer wavelength (154~\um). The data at 89~\um~better traces warmer dust in the lower density region of \opha, which surrounds the Oph-S1 star. 

The B-field properties measured using SOFIA/HAWC$+$ polarization data are qualitatively consistent with previous measurements of B-field orientations in \opha~obtained at 850~\um~using SCUBA-2/POL-2 \citep{Kwon18} and SCUBA/SCUPOL \citep{mat09} on JCMT and at NIR wavelengths (\textit{JHK} bands) using IRSF/SIRPOL \citep{Kwon15}. In the POL-2 dataset, \opha~was divided into ten subregions, identified based on their B-field position angle and the polarization fraction, covering the entire extent of HAWC$+$ maps (see Fig.~\ref{fig:orient_compare}). The highest density region is characterized by an average B-field position angle of 50\degr, which is also consistent with the NIR data toward the L1688 main cloud \citep{Kwon15}. Table~\ref{tab:compare_sofia_jcmt} shows the median B-field position angle ($\theta_B$) in each subregion, which has been covered by the FIR and sub-mm observations a well. The median B-field position angles from the 89, 154, and 850~\um~datasets are in rather good agreement, with the difference in $\theta_B$ values ranging from 3$\degr$ to 19$\degr$ ($\leq$30\%). These differences could be explained by the fact that the sub-mm data probe colder dust layers of the cloud, while the FIR data trace warmer layers. Indeed, such foreground layers were identified through the detection of the self-absorption features in the spectra of the [\ion{O}{i}] line at 63~\um~\citep{Liseau06} and the [\ion{C}{ii}] line at 158~\um~\citep{Mookerjea18}. 

%=================================================
\begin{table*}[h!]\centering
\caption{B-field position angle ($\theta_B$) and strength (\BPOS) in subregions for the dataset at 89, 154, and 850~\um.}\label{tab:compare_sofia_jcmt}
\begin{tabular}{c|ccc|cc|ccc}
\hline\hline
Subregion$^{a}$&\multicolumn{1}{c}{$\theta_{B,89}$}&\multicolumn{1}{c}{$\theta_{B,154}$}&{$\theta_{B,850}$}&$\theta_{B,89-850}$ Diff.$^{b}$ &$\theta_{B,154-850}$ Diff.$^{b}$ & \multicolumn{1}{c}{$B_\mathrm{pos,89}$} & \multicolumn{1}{c}{$B_\mathrm{pos,154}$}& \multicolumn{1}{c}{$B_\mathrm{pos,850}$}\\
~& \multicolumn{3}{c}{(deg)} & (\%) & (\%) &\multicolumn{3}{c}{(mG)}\\
\hline
(a) & $73.9\pm3.1$ & $70.8\pm3.2$ & $54.4 \pm 1.5$& 36& 30 &    4.5     &       2.1     &       5.0\\
(b) & $56.3\pm2.7$ & $48.7\pm0.6$ & $39.7 \pm 4.4$& 42& 23 &    0.1     &       0.2     &       --\\
(e) & $87.9\pm2.4$ & $80.6\pm3.4$ & $99.6 \pm 2.7$& 12& 19 & 0.7 &      0.3     &       0.8\\
(f) & $60.1\pm1.5$ & $55.0\pm1.2$ & $76.2 \pm 4.7$& 21& 28 &    0.2     &       0.3     &       --\\
(g) & $82.9\pm4.2$ & $75.4\pm1.4$ & $66.3 \pm 3.3$& 25& 14      &       0.5     &       0.7     &       --\\
(i) & $92.0\pm4.3$ & $84.9\pm1.6$ & $75.0 \pm 3.0$& 23& 13 &    0.4     &       0.5     &       --\\
\hline
\end{tabular}
\begin{flushleft}
\textbf{Notes}: $^{(a)}$ The subregions are named as in \cite{Kwon18}. 
$^{(b)}$ The relative difference in $\theta_{B,\lambda-850}$ is taken as $|1-\theta_{B,\lambda}/\theta_{B,850}|\times100$, where $\theta_{B,\lambda}$ is the B-field position angle at wavelength $\lambda$=[89, 154] (\um). 
\end{flushleft}
\end{table*}
%=================================================

\cite{Kwon18} calculated $B_\mathrm{pos}$ in three subregions associated with the densest parts of \opha~using non-thermal velocity dispersion from the N$_2$H$^+$(1--0) line \citep{andre07} and the sub-mm data at 850~\um. In our study, we calculated the pixel-by-pixel $B_\mathrm{pos}$ toward the entire \opha~region. To facilitate comparisons with results obtained from the sub-mm data, we calculated the mean value of $B_\mathrm{pos}$ in two out of these three subregions, covered by the HAWC$+$ data (see Table~\ref{tab:compare_sofia_jcmt}). Mean values of $B_\mathrm{pos}$ using FIR data are lower than that obtained with the sub-mm data, which traces colder dust layers in the cloud characterized also by the higher densities. The HAWC$+$ data, on the other hand, essentially traces the warmer layers in the lower density region. In addition, \cite{Kwon18} adopted values of the polarization angle dispersion as the standard deviation of the polarization angles, which are much smaller than those using the spread of the dust polarization angles (see Sect.~\ref{ssec:angular_dispersion}). As a consequence, the $B_\mathrm{pos}$ obtained with sub-mm data is expected to be stronger than the one obtained with FIR data. We find that our \BPOS~estimate for \opha~is typically higher than those in other molecular clouds, ranging from 50 to 400 $\mu$G \citep[for e.g.,][]{coude19,ngoc21,Pattle21,Ward-Thompson23}, but similar to those found in the DR21 (OH) region \citep{poidevin13} or the OMC$-1$ cloud in Orion A region \citep{Pattle17,Hwang21} where \BPOS~of a few mG was measured. 

In this study, we adopt \BPOS~component as the total B-field strength. If we consider the contribution of the LOS component, the total B-field strength will be stronger than what we measured. The mass-to-flux ratio, Mach Alfv\'enic number, and plasma $\beta$~parameter will decrease, but this will not significantly change our conclusion that B-fields are generally dominant over gravity, turbulent, and gas kinetic energies, especially toward the center of the cloud. 
Furthermore, as presented in Sect.~\ref{ssec:B_other_med}, we test that the \BPOS~strengths obtained with the DCF method in subregions (a) and (e) are rather in good agreement with values obtained with the ADF method but higher than those using the ST and DMA methods (see Table~\ref{tab:compare_Bstrengths}).  We use new \BPOS~values estimated with ADF, ST, and DMA methods toward two subregions (a) and (e) to estimate the mass-to-flux ratio, Mach Alfv\'enic number, and $\beta$ plasma, which are all higher than that measured with the DCF method (see Table~\ref{tab:compare_Bstrengths}). Subregion (a) changes from magnetically sub/trans-critical to super-critical but remains to be super-Alfv\'enic, and has $\beta$<1 in overall. A similar behavior is seen in subregion (e) except that this subregion changes from trans- to super-Alfv\'enic (see Table~\ref{tab:compare_Bstrengths}). 

We go on to assume that the B-field strength in the entire cloud decreases by a factor of $\sim$4.5, as measured with the ST method. In this case, the majority of the cloud will become magnetically trans- or super-critical ($\lambda \ga$1), except for the regions in the vicinity of the Oph--S1 star (see the top panel of Fig.~\ref{fig:lambda_alfven_beta_pressure_maps_ST}). However, this trend of $\lambda$ remains similar as obtained using $B_\mathrm{pos}^{\mathrm{DCF}}$. The cloud will become mostly super-Afv\'enic with \Ma>1, except for the southeast region of the cloud (see the middle panel of Fig.~\ref{fig:lambda_alfven_beta_pressure_maps_ST}). This assumption will also result in the Alfv\'enic velocity decreasing by a factor of $\sim$4.5 and thus, $M_\mathrm{vir} \approx 2 M_\mathrm{iso}$. This implies that the cloud is in the magneto hydrostatic equilibrium and may undergo gravitational collapse due to turbulent fluctuations on smaller scales. 

\begin{table*}[h!]\centering\tiny
\caption{\label{tab:compare_Bstrengths} Comparison of \BPOS~strength, mass-to-flux ratio ($\lambda$), Afv\'enic Mach number ($\mathscr{M}_\mathrm{A}$), and $\beta$ plasma toward subregions (a) and (e) measured using different methods.}
\begin{tabular}{c|cccc|cccc|cccc|cccc}
\hline\hline
~ &\multicolumn{4}{c|}{$B_\mathrm{pos}$~(mG)}& \multicolumn{4}{c|}{$\lambda$}& \multicolumn{4}{c|}{$\mathscr{M}_\mathrm{A}$}& \multicolumn{4}{c}{$\beta$}\\
Subregion &DCF & ADF & ST & DMA& DCF & ADF & ST & DMA&DCF & ADF & ST & DMA&DCF & ADF & ST & DMA\\
\hline 
a &0.84 & 0.69&0.19& 0.34 & 0.91 &1.12&4.03& 2.49 & 1.57 & 1.92&6.93&3.88 & 0.02 &0.03&0.37&0.12\\
e &0.37 & 0.26&0.07&0.12 & 0.84 & 1.22&4.20&1.99 & 0.62 & 0.90&3.09&1.96& <0.01 &<0.01&0.11& 0.04\\
\hline
\end{tabular}
\end{table*}

\subsection{Magnetic field orientation versus density structure of \opha}
To investigate what is the relative orientation of the B-fields with respect to the gas structure in \opha, we quantify the projected offset angle between the orientations of B-field vectors projected in the plane-of-the-sky and the main structure associated with the ridge of the cloud, given as $\Delta \theta=\theta_B-\theta_\mathrm{ridge}$, where $\theta_\mathrm{ridge}$ is the position angle of the ridge. We identified the shape of the ridge using the Python package {\tt RadFil}\footnote{\url{https://github.com/catherinezucker/radfil}} \citep[][see Appendix~\ref{app:ridge}]{Zucker18}. 

Figure~\ref{fig:map_offset_angle} shows the map of offset angles ($\Delta \theta$) in \opha, overlaid by contours of continuum emission at 154~\um. Clearly, the offset angle appears to increase when moving from the lower density regions to the denser regions, indicating that the relative orientation of B-fields with respect to the main structure of the cloud turns from parallel to perpendicular toward the high-density region. This is in agreement with \textit{Planck} polarization observations at 353~GHz (850~\um) toward the entire L1688 region where the relative orientation of B-fields, with respect to the gas structure changes gradually. It runs parallel in the low-density region to perpendicular toward the higher density region, with the transition occurring at $N_\mathrm{H_2}\sim 3\times10^{21}$~\cmsq~and corresponding to the visual extinction $A_V$ of $\sim$3.3~mag\footnote{We convert $A_V$ from $N_\mathrm{H_2}$ using the relation: $N_\mathrm{H_2}/A_V= 0.94\times10^{21}$ molecules~cm$^{-2}$~mag$^{-1}$~with a \lq\lq standard\rq\rq~assumption of the total-to-selective extinction ratio, $R_V=A_V/E(B-V)$, of 3.1 \citep{Frer82}.} \citep{soler19}. This transition has also been observed in other nearby molecular clouds in the Gould Belt using \textit{Planck} 353~GHz polarization observations, typically occurring at $N_\mathrm{H_2} \sim 2.5 \times 10^{21}$~\cmsq~or $A_V\sim$2.7~mag \citep[see e.g.,][]{planck16, soler17}. This phenomenon can be explained by the motion of the high-density material during the self-gravitational collapse. When the contraction occurs, the material becomes denser and compressed, which could cause the B-field lines to be re-configured. Eventually, the B-field lines are
dragged and reshaped to become more uniform and perpendicular to the high-density structure of the field.

%=================================================
\begin{figure}\centering
\includegraphics[width=0.46\textwidth]{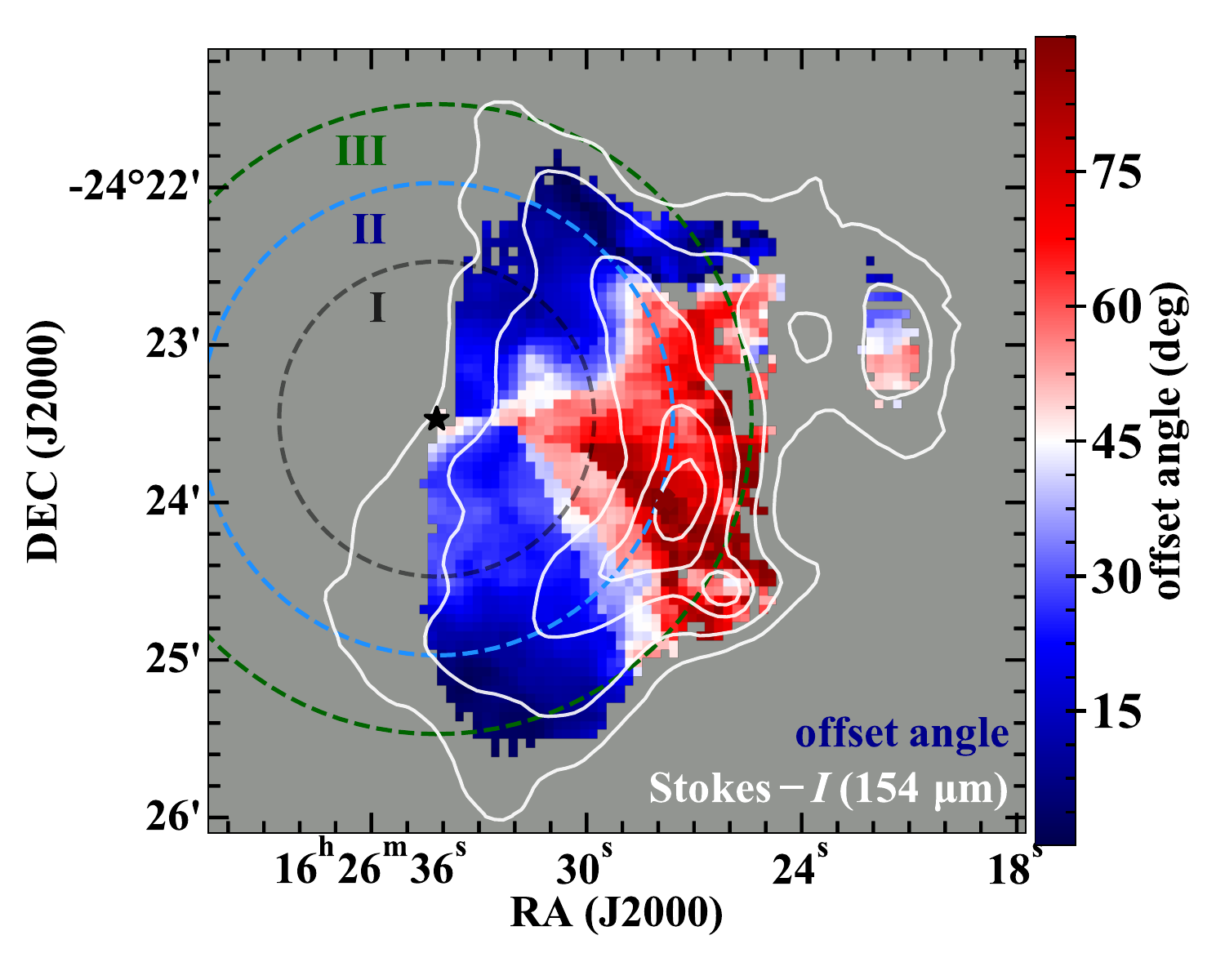}
\caption{Map of the offset angle between B-field orientations and main cloud structure in \opha. White contours show the continuum emission at 154~\um~with similar levels as in Fig.~\ref{fig:delta_velo}. The dashed circles centered at the position of the Oph-S1 star have a radius of 60$\arcsec$ (black), 90$\arcsec$ (blue), and 120$\arcsec$ (green), corresponding to the levels of $A_V$ of $\sim$6, 30, and 200~mag. The star symbol marks the position of the Oph-S1 star. }
\label{fig:map_offset_angle}
\end{figure}
%=================================================

%=================================================
\begin{figure}[h!]\centering
\includegraphics[width=0.46\textwidth]{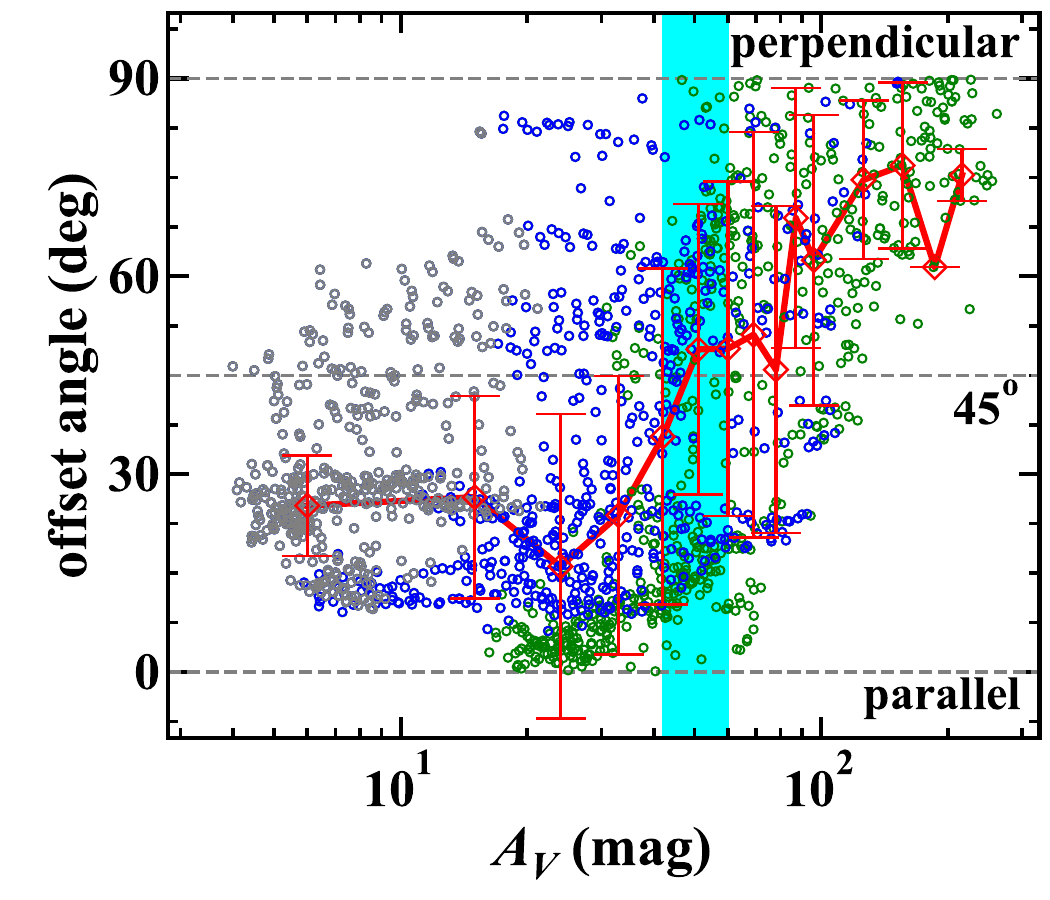}
\caption{Distribution of the offset angles between the orientations of B-fields and the ridge in \opha, as a function of $A_V$ (circle symbols). The data in gray color represent the offset angle collected within the black circle "I" centered at the position of the Oph-S1 star with a radius of 60$\arcsec$ in Fig.~\ref{fig:map_offset_angle}. The data in blue and green additionally represent the offset angle collected within the two rings "II" and "III" in Fig.~\ref{fig:map_offset_angle}, respectively. The red diamonds represent the median value of the offset angles in each interval of $A_V$=3~mag. The number of red diamonds is reduced to improve the clarity of the figure. The red curve connecting all the red diamonds shows the trend of the offset angles vs. $A_V$ in \opha. The cyan box indicates the range of $A_V$=42--60~mag where relative orientation between B-fields and the ridge changes from parallel to perpendicular. The gray horizontal dashed lines indicate perpendicular, 45\degr, and parallel relative orientations of B-fields with respect to the ridge.}\label{fig:offset_angle_Av}
\end{figure}
%=================================================
The HAWC$+$ data traces B-fields in the dense region of \opha~with $N_\mathrm{H_2}\sim (3-240)\times 10^{21}$~\cmsq~(see Fig.~\ref{fig:tdust}), which corresponds to $A_V$ in range of $\sim$3.2--256~mag. Thus, we further investigate in detail the relation between the relative orientation of B-fields with respect to the cloud structure and the $A_V$ within \opha, as shown in Fig.~\ref{fig:offset_angle_Av}. Clearly, in the regime where $A_V\la$30~mag the offset angle ($\Delta\theta$) decreases with $A_V$ and drops from $\sim$60\degr~at $A_V$=4~mag down to 0\degr~at $A_V\sim$20~mag. This indicates that the relative orientation of B-fields changes from perpendicular back to parallel at $A_V\sim$18--33~mag. Similar trends of this transition were found in the dense region of the Serpens South filament \citep[][]{Pillai20} and in the Serpens Main dense cloud \citep[][]{Kwon22}, occurring at $A_V\sim$21~mag and $\sim$50~mag, respectively. This is likely because the B-fields are already dragged into the flow of very dense material during the gravitational contraction. Interestingly, we observe that the offset angle then progressively increases with $A_V\ga$40~mag in our region, indicating the relative orientation of B-field with respect to the ridge changes back from parallel to perpendicular (with the transition at $A_V\sim$42--60~mag). We note, however, that this feature was not seen in the Serpens South filamentary cloud, whose maximum is $A_V\sim$60~mag (see Fig.~3 in \cite{Pillai20}); however, it was seen in the Serpens Main cloud region, where $A_V \ga$170~mag (see Fig.~6 in \cite{Kwon22}). %\cite{Pillai20}). 

\subsection{Magnetic field versus gas density}\label{ssec:Bfields_nH}
The B-field strength ($B_\mathrm{tot}$ or $B$) can be expressed as a function of $n_\mathrm{H}$ ($n_\mathrm{H} = 2 \,n_\mathrm{H_2} $): $B \propto n_\mathrm{H}^{k}$. Theoretical models predict $k\sim$2/3 in the weak B-field regions \citep{Mestel1966} and $k\la$0.5 in the strong B-field regions, where magnetic energy dominates over the gravity \citep{Mouscho99}. Consequently, in the latter case, it suggests that $n_\mathrm{H}$ would increase faster than $B$ if there is any collapse in the dense cloud. 
%=================================================
\begin{figure}[t!]\centering
\includegraphics[width = 0.475\textwidth]{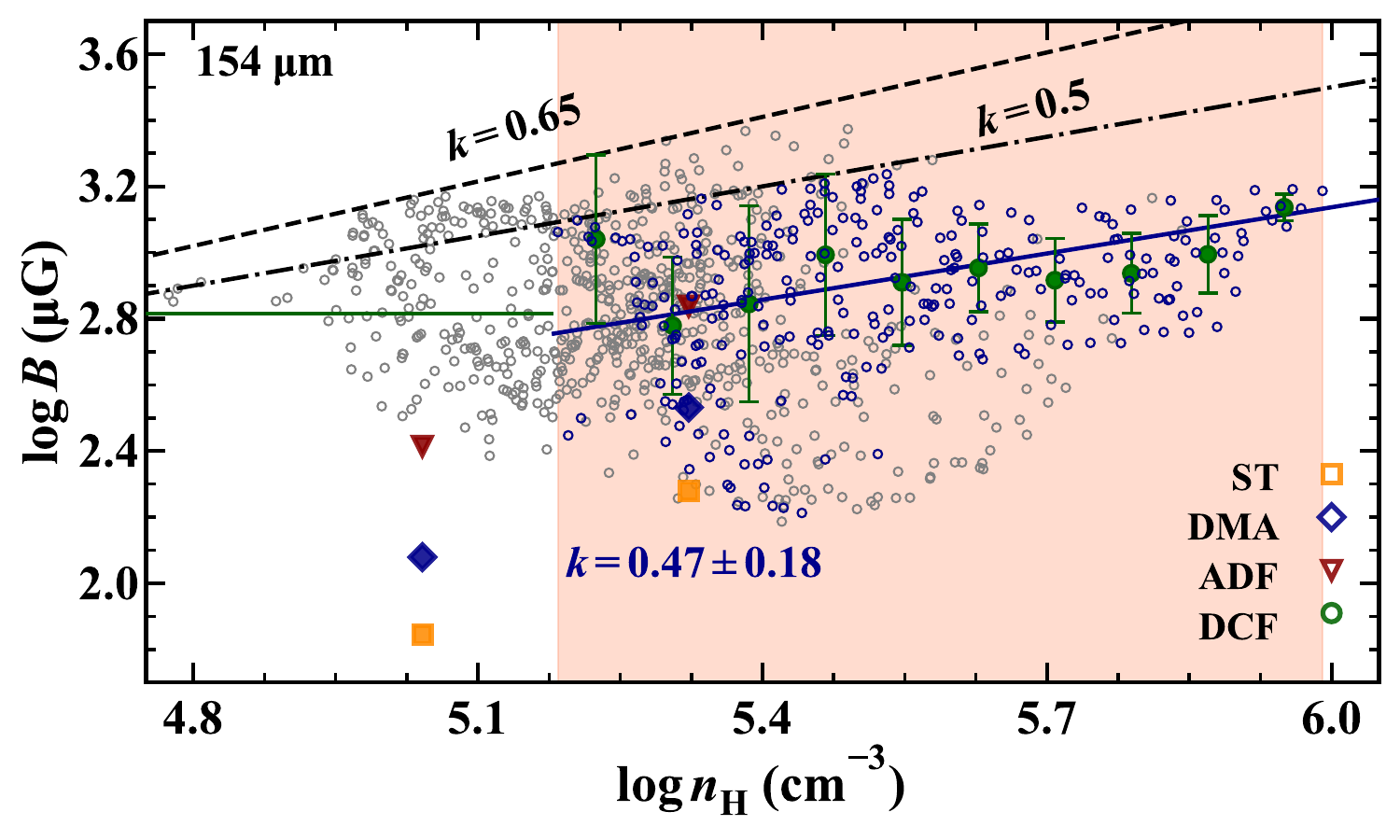}
\caption{Correlation between $\log B$ and $\log n_\mathrm{H}$ in \opha. Small circles represent $B$ strength estimated with the DCF method at band D. Small circles in blue further represent data within the region where Stokes$-I \ge 0.4 ~\mathrm{Jy/arcsec^2}$, corresponding to the densities $\log n_\mathrm{H} \ge 5.2$. The bigger circles in green represent median values of $B$ binned in each interval of $\log n\mathrm{_H}$ of $\sim$0.08 (see texts). Mean values of $B$ estimated with ADF (red triangles), DMA (blue diamonds), and ST (orange squares) methods toward subregion (a) and (e) are overplotted. The solid green line represents the median value $\log B \approx2.81$ at low densities. The solid blue line shows the power-law fit toward the binned data of $B$ using the DCF method toward more denser region (covered by the orange shaded box), resulting $k=0.47\pm0.18$. The black dash-dotted line indicates the critical power-law $B\propto n_\mathrm{H}^{0.5}$ for strong B-fields, predicted from the theoretical models \citep{Mouscho99}. The black dashed line indicates the relation found by \cite{Crutcher10} using the Zeeman observational data. }\label{fig:Bpos_vsnH2}
\end{figure}
%=================================================

Here, we show in Fig.~\ref{fig:Bpos_vsnH2} the relation $\log B- \log n_\mathrm{H}$ toward the \opha~cloud, using B-fields measured with the DCF method in band D. We find that for densities $\nolinebreak \la 10^{5.2}~\mathrm{cm^{-3}}$, the B-fields distribution tends to be flattened. In a regime with densities higher than $ \nolinebreak 10^{5.2}~\mathrm{cm^{-3}}$ we start seeing the increasing trend of B-fields with respect to gas densities. Thus, we decided to test the $\log B- \log n_\mathrm{H}$ relation at the densest part of the cloud only. We chose this region as the area within the third contour of Stokes-$I$ at band D (see in the bottom right panel of Fig.~\ref{fig:volume_map}), where Stokes$-I \ge0.4~\rm Jy/arcsec^2$. 
Within this region, it reveals a relatively weak correlation between the B-fields and densities (with the Pearson coefficient of the correlation is found $r\sim \rm 0.32$, corresponding to a significance of $\sim 5\sigma$). We performed the power-law fit using the Python {\tt curve$_{-}$fit} routine ($\log B=k\times \log (n_\mathrm{H})+b$), using mean values of $B$ estimated within every interval of $\log n_\mathrm{H}$=0.08\footnote{The bin width is chosen by using the Freedman-Diaconis rule. }. This gives a power-law index of $k=0.47\pm0.18$. In Fig.~\ref{fig:Bpos_vsnH2}, we also overplot the results of  \BPOS using other methods in subregions (a) and (e). We find that the fitted line well covers the data point representing the resulting \BPOS~value using ADF method and shows good agreement with those estimated with ST and DMA in subregion (a) within and uncertainty of $1\sigma$. The resulting $k=0.47\pm0.18$ is found consistent with theoretical predictions for strong B-fields \citep{Mouscho99}. This is also consistent with $\beta\ll 1$, representing strong B-fields in \opha~(see bottom panel in Fig.~\ref{fig:lambda_alfven_beta_pressure_maps} and Sect.~\ref{ssec:beta}). 
We then performed the fit to $B_\mathrm{pos}^\mathrm{ST}$~values measured with the ST method from Sect.~\ref{ssec:st} and find $k = 0.49\pm0.27$. We found a similar power-law index $k$ for the dataset at band C, namely, $0.43\pm0.12$ and $0.39\pm0.08,$ using result of $B_\mathrm{pos}^\mathrm{DCF}$ and $B_\mathrm{pos}^{\mathrm{ST}}$, respectively (see Appendix~\ref{app:velo_disper_comparison}). 

Our resulting $k$ value is smaller than $k \sim$~0.65 found by \cite{Crutcher10}. However, their result was based on the mean values of $B$ and $n_\mathrm{H}$ in different molecular clouds, which contain significant observational uncertainty and has been questioned by later studies \citep[for e.g.,][]{Tritsis15, Li21}.

\subsection{Magnetic fields versus stellar feedback}\label{ssec:Star-feedback}
\begin{figure}[tbp]
\centering
\includegraphics[width = 0.47\textwidth]{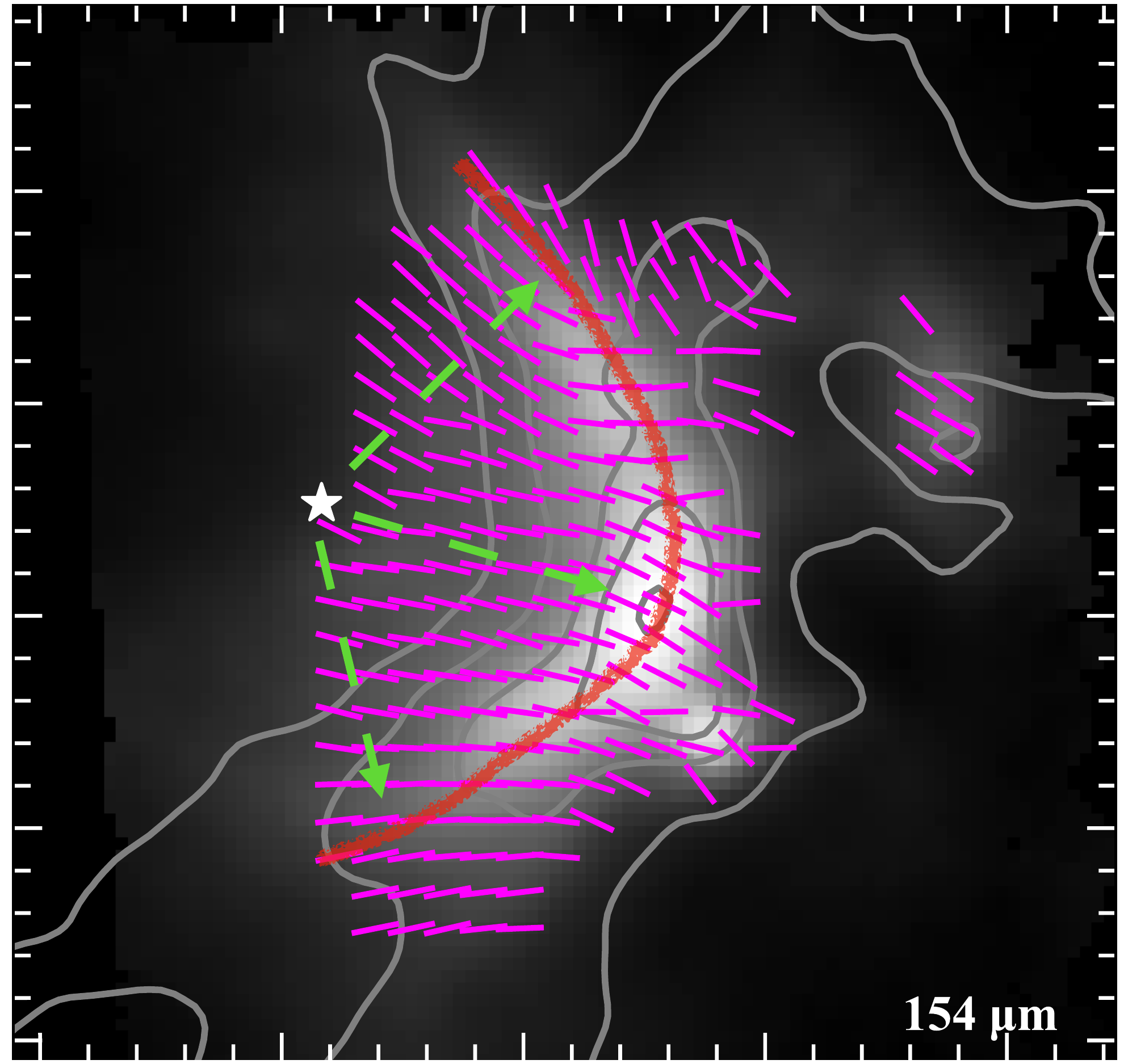}
\caption{Image of the continuum emission at 154~\um~toward Oph$-$A. B-field vectors obtained from SOFIA/HAWC$+$ at 154~\um~are shown in magenta solid segments. The green dashed arrows indicate the radiation directions from the Oph-S1 star toward the cloud. The white star marks the position of the Oph-S1 star. The red curve illustrates the elongated shape of the ridge in the dense cloud. The gray contours are the H$_2$ column density, with contour levels at (1, 2, 5, 10, 20)$\times$10$^{22}$~cm$^{-2}$.}
\label{fig:sketch_radiation}
\end{figure}

High-mass stars irradiate, heat up, and ionize the surrounding gas and create the \ion{H}{ii} region \citep[see e.g.,][]{zinn07}. The ionized, hot gas also expands and interacts with the vicinity environment. The radiation pressure acting on both dust and gas (assuming dust is well-coupled to the gas in the cloud) can drive the material away into the denser region. The lower density matter in the outer regions of \opha~is more exposed to the ionization front and pushed away faster than the higher density matter in the inner regions. This results in dust and gas being compressed and forming the dense shell structure at the edge of the cloud. Here, we investigate a possible impact of the radiation field from the high-mass star Oph-S1 on the \opha~cloud. The \opha~molecular cloud is directly exposed to the strong UV radiation field emitted by the Oph-S1 star from the east. Figure~\ref{fig:sketch_radiation} shows the map of the continuum emission at 154~\um~from SOFIA/HAWC+, superimposed by the white contours showing the H$_2$ column density. 

We calculated the external radiation pressure of the Oph-S1 star acting on dust and gas in the densest part of the cloud, given by:
\begin{equation}\label{eq:p_rad}
P_\mathrm{ext,rad} = \frac{L_\mathrm{bol}}{4\pi\mathcal{R}^2c},
\end{equation}
where $L_\mathrm{bol}$=1100~$L_\odot$ is the bolometric luminosity of the Oph-S1 star adopted from \cite{lada84} and $\mathcal{R}\approx$ 0.06~pc is the projected distance from the cloud to the Oph-S1 star. We obtained $P_\mathrm{ext,rad}/k_\mathrm{B}\approx 2.4\times 10^{6}$~K~cm$^{-3}$, which is in agreement with the result calculated from the ionizing photons emitted by the Oph-S1 star \citep[$\sim$2$\times 10^6$~\Kcm,][]{Pattle15}. Here, we omit the fact that the UV radiation from the Oph-S1 star can be absorbed by intervening dust in between the star and the \opha~region. Thus, the actual $P_\mathrm{ext,rad}$ acting on dust and gas in \opha~is overestimated in the Eq.~\ref{eq:p_rad}. To make a comparison with the external radiation pressure, we calculated the internal thermal pressure due to the hot gas in \opha, $P_\mathrm{th} = n_\mathrm{H_2} k_\mathrm{B}T_\mathrm{gas}$. We considered the hot gas within the densest part of \opha,~where Stokes-$I \ge$0.4~Jy/arcsec$^2$ at 154~\um~only, which corresponds to the area inside the third contour shown in bottom right panel of Fig.~\ref{fig:volume_map}. Within this region, we adopted $n_\mathrm{H_2}$ in the range of 10$^4$--10$^6$~cm$^{-3}$ and the average temperature of the hot gas $T_\mathrm{gas}\approx T_\mathrm{dust}\approx$20~K, which yields $P_\mathrm{th}/k_\mathrm{B}$ in the range of (0.2--20)$\times 10^6$~K~cm$^{-3}$. With this assumption, we note that $P_\mathrm{th}/k_\mathrm{B}$ could be underestimated, since there is a gas heating process in the cloud dominated by the UV radiation from the Oph-S1 star and by the photoelectric emission from dust exposed to the UV radiation (i.e., $T_\mathrm{gas} \ga T_\mathrm{dust}$). The highest internal thermal pressure of $\sim$20$\times$10$^6$~K~cm$^{-3}$ is seen toward the densest part of the cloud, associated with the position of the SM1 core. This suggests that the internal thermal pressure is strong enough to help the material refrain from external pressure. We note that, however, in the vicinity region away from the center (e.g., the tails of the ridge shown in Fig.~\ref{fig:sketch_radiation}), the internal thermal pressure is weaker, at least one order of magnitude less than the external one. In this case, the radiation pressure from the Oph-S1 star on the east side is dominant and able to push away all the matter into the surrounding environment toward the west side. However, it is clear that the material in the low-density region still survives, despite owning a low internal thermal pressure (see Fig.~\ref{fig:sketch_radiation}). This indicates that there must be another mechanism at play, for example, the presence of strong B-fields of a few mG in the cloud. We roughly estimated  the magnetic pressure ($P_\mathrm{mag}=B^2/(8\pi)$) to be 10$^{7}$--10$^{8}$~K~cm$^{-3}$ for B-field strengths between 1--2~mG. This indicates that the magnetic pressure is at least two orders of magnitude higher than the external pressure of $\sim$10$^{6}$~K~cm$^{-3}$, sufficient to support the matter against the external pressure along with the gas thermal pressure. 
% =================================================
% =================================================

We further investigated the interplay between B-fields and radiation directions on the evolution of the cloud. We determined the main radiation direction from the Oph-S1 star toward the densest region associated with the position of the starless core SM1 as the sky-projected direction from the Oph-S1 star and SM1 core. We found that the main radiation direction is $\sim$70\degr, measured east of north. The main B-field orientation toward the densest region of \opha~of $\sim$74\degr\ is almost parallel to the radiation field. The strong radiation pressure from the Oph-S1 star pushes all the material away and creates a dense shell toward the west side. To prevent dust and gas compression in the dense region, the B-fields exert a magnetic pressure perpendicular to the B-field orientation, seen perpendicular to the direction of the radiation pressure. This suggests that the magnetic pressure does not help the matter resist the external pressure, but tends to stretch the structure out along the northwest-southeast direction, resulting in an elongated shape of the dense shell. In the west side, the external thermal pressure is blocked by the gas thermal pressure due to the high-density material, resulting in the ridge shape in the cloud with very high densities. These results are consistent with results from 3D MHD simulation, which predicted that if the B-field orientation is parallel or along with the radiation direction, the cloud will tend to have a flattened shape \cite{henney09}. However, in the northeast and southeast regions, the offset angles between the B-field vectors and radiation field are larger; namely, the B-field orientation becomes perpendicular to the radiation direction. In this case, the magnetic pressure and tension induced by B-fields run parallel with the radiation pressure, suggesting magnetic pressure can support the matter to confront the radiative compression and prevent the  dust and gas from being swept away. This results in bending the material structure and forming a dense ridge with an umbrella-like shape. Interestingly, in the case study of \opha~the ionization source is very close to the cloud (<0.1~pc) and the offset position angle between the radiation and B-fields is not homogeneous; instead, it varies from parallel to orthogonal. This allows us to closely examine how the interplay between the B-fields and radiation fields could influence the material structure of the cloud. 

% ------------------------------------------------
\section{Summary}\label{sec:sum}

In this paper, we present our analysis  of the Ophiuchus dark cloud complex, using the SOFIA/HAWC$+$ polarimetric
observations toward the \opha~molecular cloud in the L1688 active star-forming
region. Our main results are as follows: 
\begin{itemize}
\item Using the dust polarization maps at 89 and 154~\um, we derived the map of B-field morphology in \opha. We found that B-fields are generally well-ordered and primarily perpendicular to the ridge of the cloud toward the higher density regime. We estimated $B_\mathrm{pos}$ in the entire region of \opha~using the DCF method, ranging from 0.2 to 2.3~mG at 89 and to 2.5~mG at 154~\um. We found that $B_\mathrm{pos}$ is prominent toward the densest region of the cloud associated with the starless core SM1 and decreases toward the outskirts. This is the first time we have successfully estimated the distribution of the B-field strengths in \opha, which is essential to understanding how B-fields vary in the molecular cloud globally. Our derived B-field morphology and strengths well agree with previous measurements using the sub-mm dust polarization \citep{Kwon18}. 
We also estimated the \BPOS~strength using more advanced techniques, including the ADF, ST, and DMA, toward two subregions associated with the densest part of the cloud.  While the \BPOS~strengths estimated with the ADF method were in agreement with those from DCF, the ST and DMA methods provided lower \BPOS~values of $\sim$4.5 and 2-3 times, respectively. 

\item We studied the role of B-fields relative to gravity and turbulence by calculating the maps of the mass-to-flux ratio, Alfv\'enic Mach number, and plasma $\beta$ parameter toward \opha,~using the inferred $B_\mathrm{pos}^\mathrm{DCF}$ from SOFIA/HAWC$+$ data at 154~\um. We found that toward the central dense parts the cloud is magnetically sub-critical and sub-Alfv\'enic where B-fields dominate over the gravity and turbulence. We obtained a result of $\beta \ll 1$ in \opha, suggesting the cloud is supported by the strong magnetic pressure and much greater than the thermal gas pressure. The lower \BPOS~strengths obtained with the ST and DMA methods do not significantly change our conclusions on the roles of B-fields relative to gravity and turbulence on star formation.  
Our virial analysis in the entire cloud showed the virial parameter $\alpha_\mathrm{vir}\sim$15, suggesting that the cloud is gravitationally unbound and able to against the self-gravitational collapse. 
The cloud may be in the magneto hydrostatic equilibrium and may undergo gravitational collapse due to turbulent fluctuations on smaller scales. 
We should notice that our rough estimation of $\alpha_\mathrm{vir}$ is a global parameter, applied for the entire cloud. In contrast, \opha~is known as one of the densest parts of the active star-forming region L1688 hosting a few pre- and proto-stellar cores and embedded protostars. Higher spatial resolution observations toward the smaller scale region in the cloud will be necessary to investigate the dynamical state of the cloud in greater detail. 

\item We examined the role of B-fields in supporting the cloud against the radiation pressure from the nearby high-mass star Oph-S1. We found that at the center part of the cloud, the magnetic pressure indeed does not play a role in helping the material to avoid the onset of radiation pressure. However, toward the northeast and southeast regions of the cloud, the magnetic pressure positively supports the cloud to prevent both the effects of radiative pressure and gas thermal pressure. 

\end{itemize}

\begin{acknowledgements}
We thank our anonymous referee for their careful reading of our manuscript and valuable suggestions, which improved the quality of this work, particularly with regard to the realistic estimate of the gas volume density in the \opha~cloud. 
We thank Bhaswati Mookerjea for the valuable comments and discussion at the early stage of the project and for kindly sharing the final data cube of the HCO$^+$(4--3) transition line. We thank Fabio Santos for sharing the maps of dust temperatures and gas column density. We also thank Alex Lazarian for the constructive discussions concerning the DCF and DMA methods. N.L. thanks Nguyen Tat Thang for his valuable comments as part of the 3D structure of the \opha. 
N.L. and A.K. acknowledge support from the First TEAM grant of the Foundation for Polish Science No. POIR.04.04.00-00-5D21/18-00 (PI: A. Karska). A.K. acknowledges also support from the Polish National Agency for Academic Exchange grant No. BPN/BEK/2021/1/00319/DEC/1. N.B.N. was funded by the Master, Ph.D. Scholarship Programme of Vingroup Innovation Foundation (VINIF), code VINIF.2023.TS.077. M.H. acknowledges support by the National Science Centre through the OPUS grant no. 2015/19/B/ST9/02959. 
N.L. also acknowledges support by the grant from the Simons Foundation to IFIRSE, ICISE (916424, PI: N.H.). P.N.D. and N.T.P. were funded by the Vietnam Academy of Science and Technology under project code THTETN.03/24-25. 
This work is based on observations made with the NASA/DLR Stratospheric Observatory for Infrared Astronomy (SOFIA). SOFIA is jointly operated by the Universities Space Research Association, Inc. (USRA), under NASA contract NNA17BF53C, and the Deutsches SOFIA Institut (DSI) under DLR contract 50 OK 2002 to the University of Stuttgart. This work made use of data from the JCMT/HARP and APEX. JCMT is operated by the East Asian Observatory on behalf of National Astronomical Observatory of Japan; the UK STFC under the auspices of grant number ST/N005856/1; Academia Sinica Institute of Astronomy and Astrophysics; the Korea Astronomy and Space Science Institute; the Operation, Maintenance and Upgrading Fund for Astronomical Telescopes and Facility Instruments, budgeted from the Ministry of Finance of China. This work used the facilities of the Canadian Astronomy Data Centre (CADC) operated by the National Research Council of Canada. 
\end{acknowledgements}
\bibliographystyle{aa} 
\bibliography{bib} 
% ------------------------------------------------
\begin{appendix} 
% ------------------------------------------------
\section{Histogram of SOFIA/HAWC$+$ dataset}\label{app:histogram}

\begin{figure}[h!]
\centering
\includegraphics[scale = 0.25]{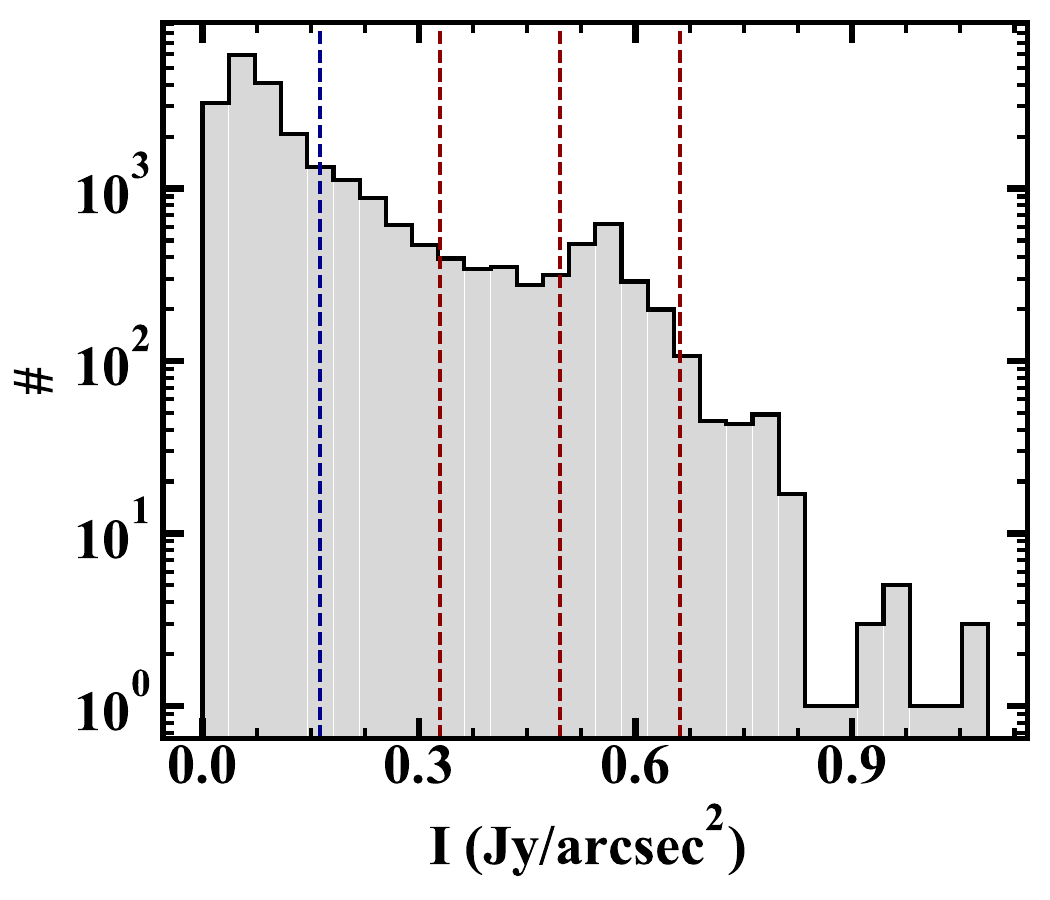}
\includegraphics[scale = 0.25]{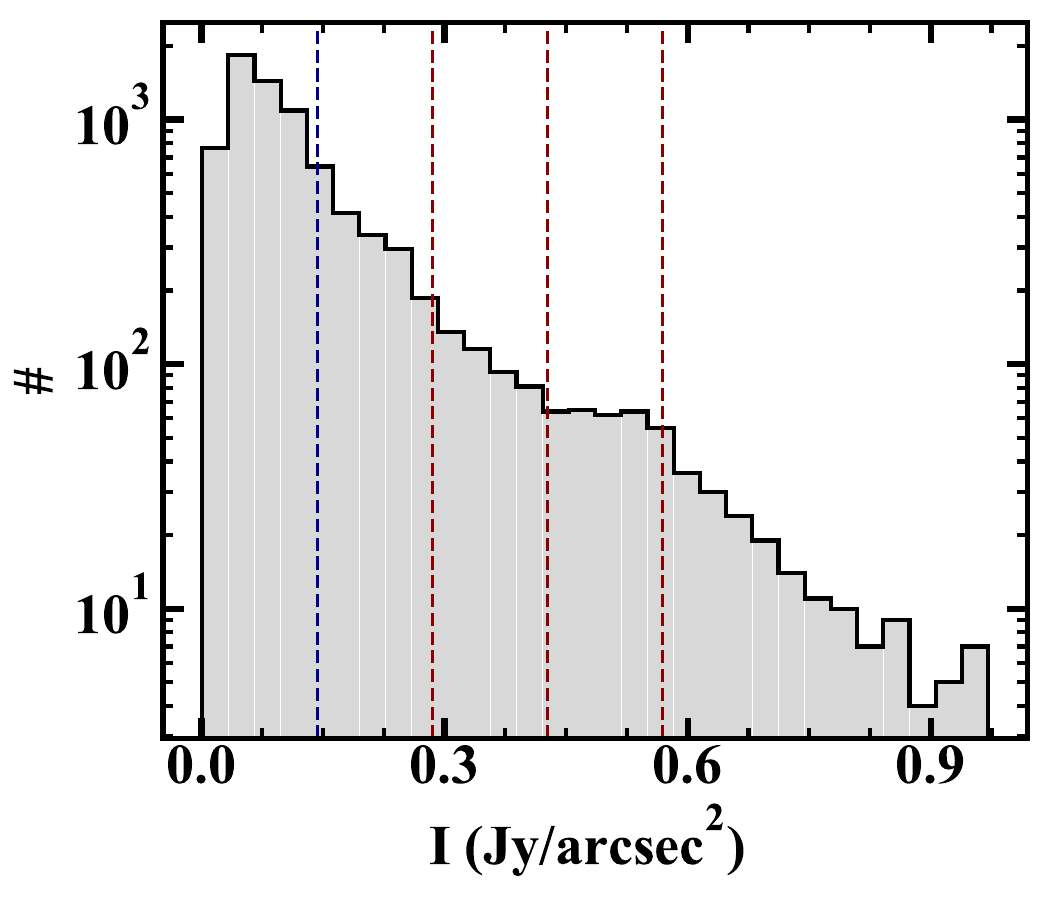}\\
\includegraphics[scale = 0.25]{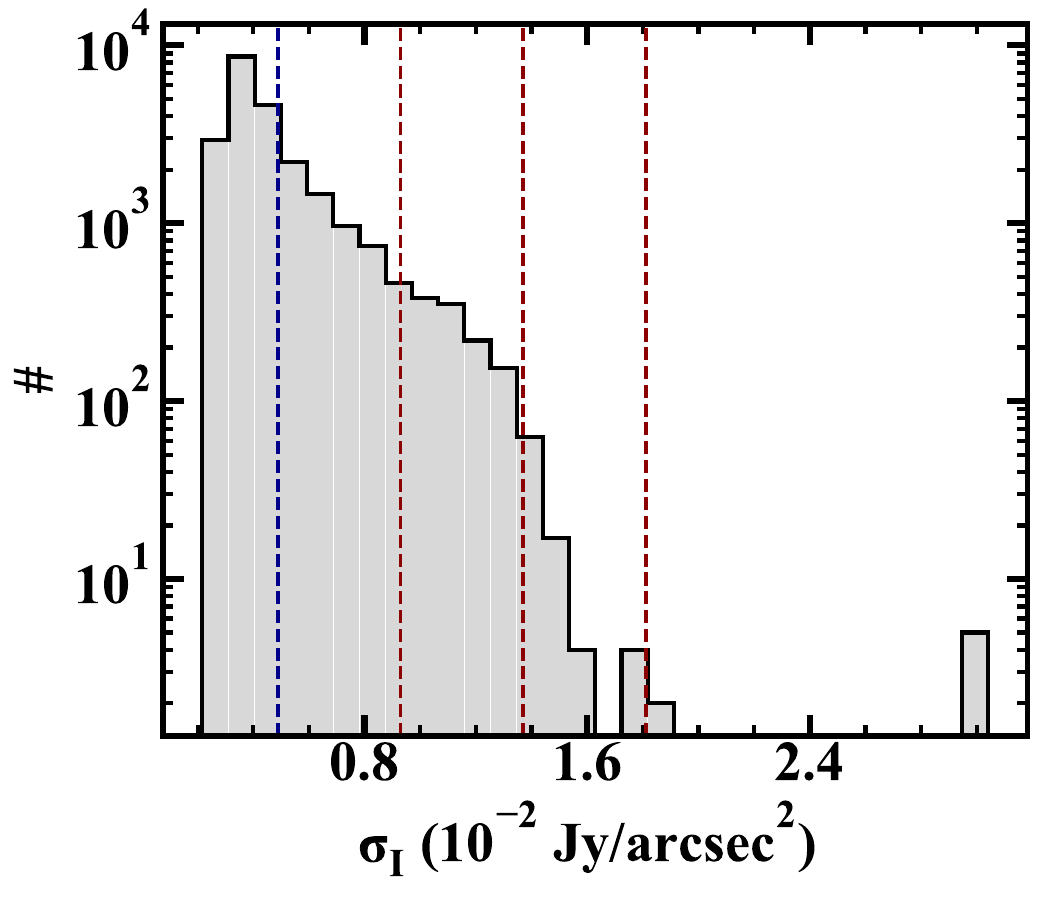}
\includegraphics[scale = 0.25]{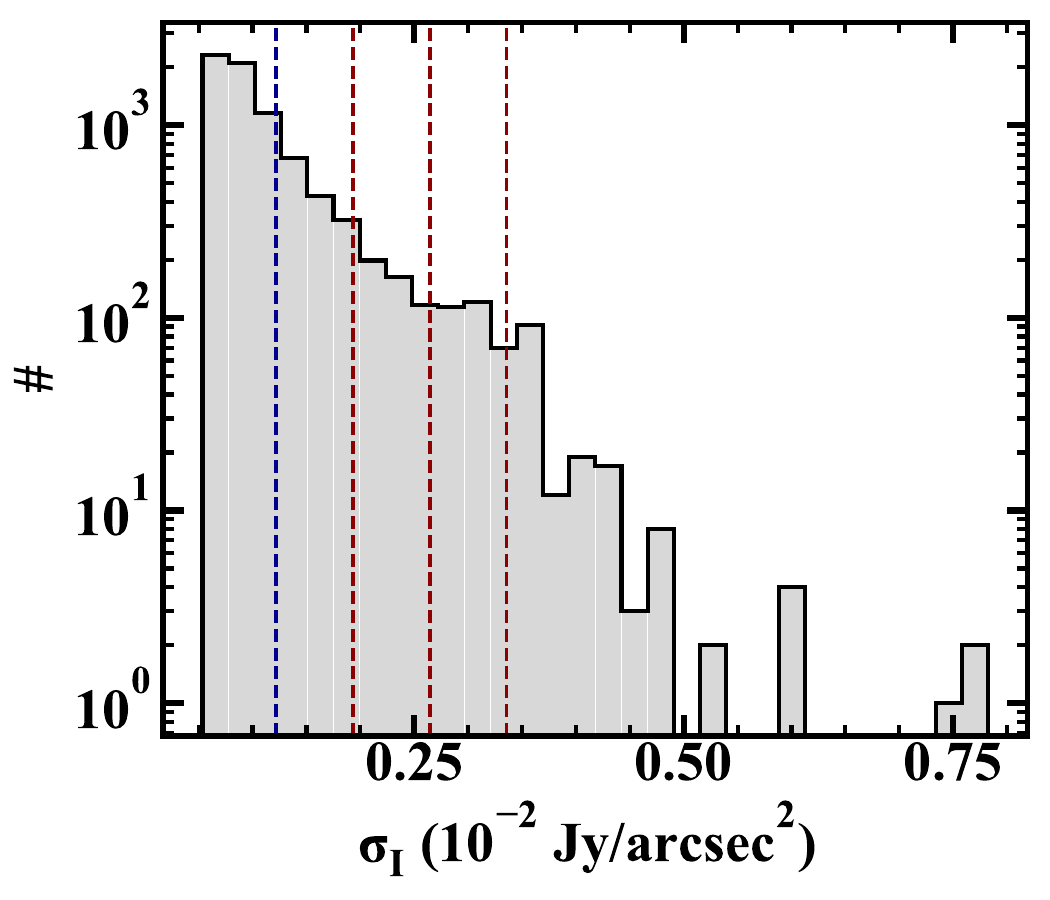}\\
\includegraphics[scale = 0.25]{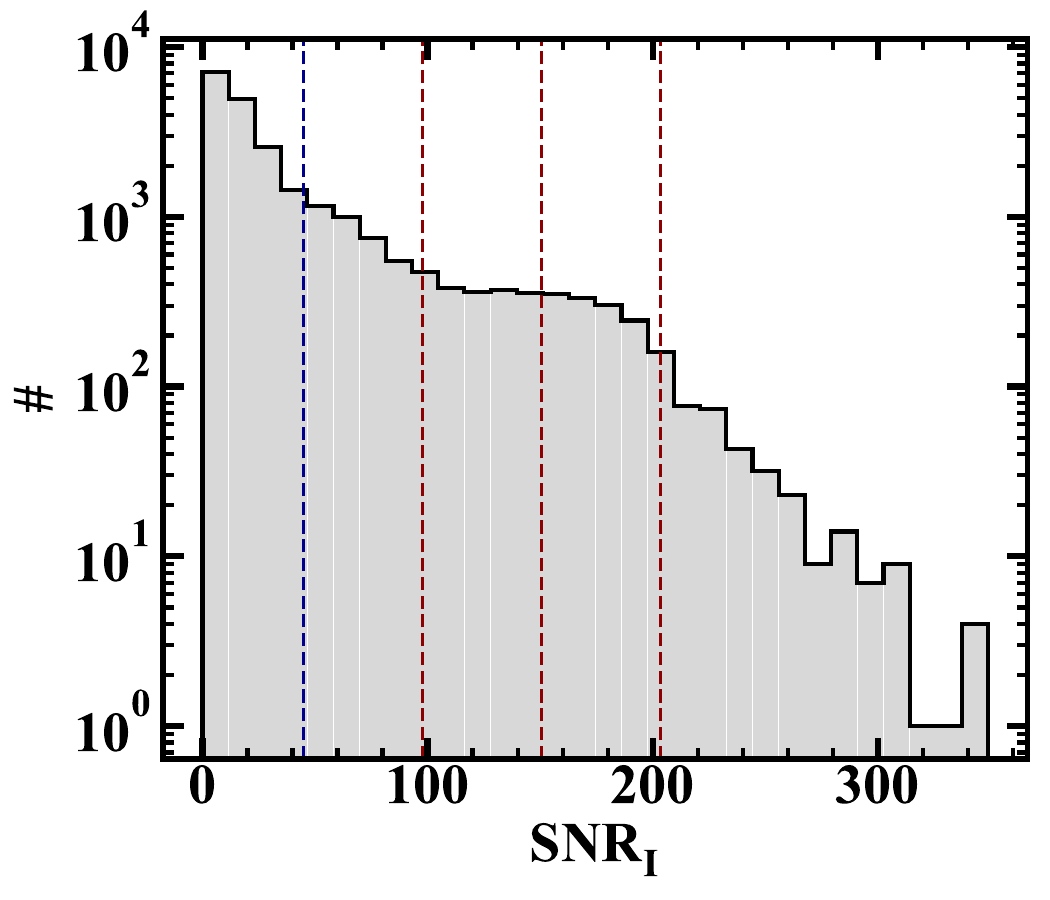}
\includegraphics[scale = 0.25]{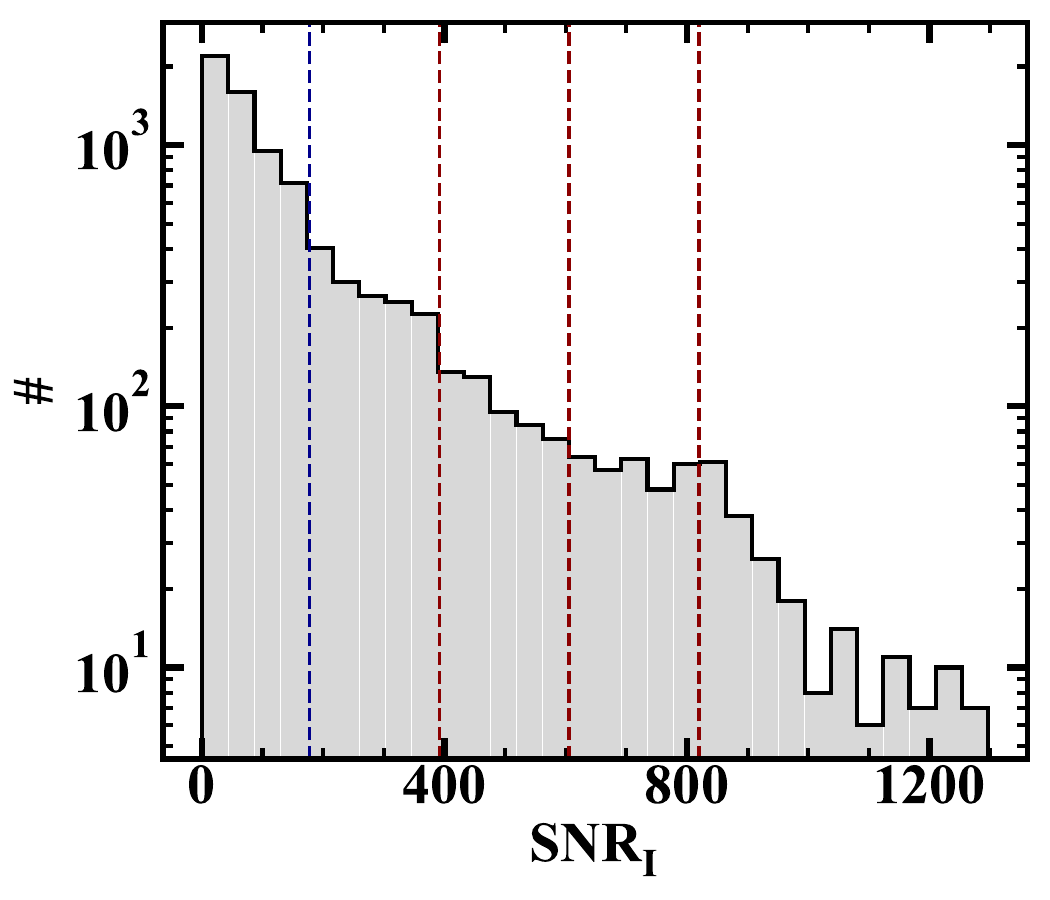}
\caption{Histograms of Stokes-$I$ (top panels), $\sigma_I$ (middle panels), and SNR$_I$ (bottom panels) of the SOFIA/HAWC$+$ data. Left panels represent data in band C at 89~\um~and right panels represent data in band D at 154~\um. The blue vertical dashed lines in each panel represent the mean value of the distribution. The red vertical dashed lines show the position of mean value plus 1$\sigma$, 2$\sigma$, and 3$\sigma$ of the distribution from left to right, respectively. }
\label{fig:histogram_I}
\end{figure}

\begin{figure}[h!]
\centering
\includegraphics[scale = 0.25]{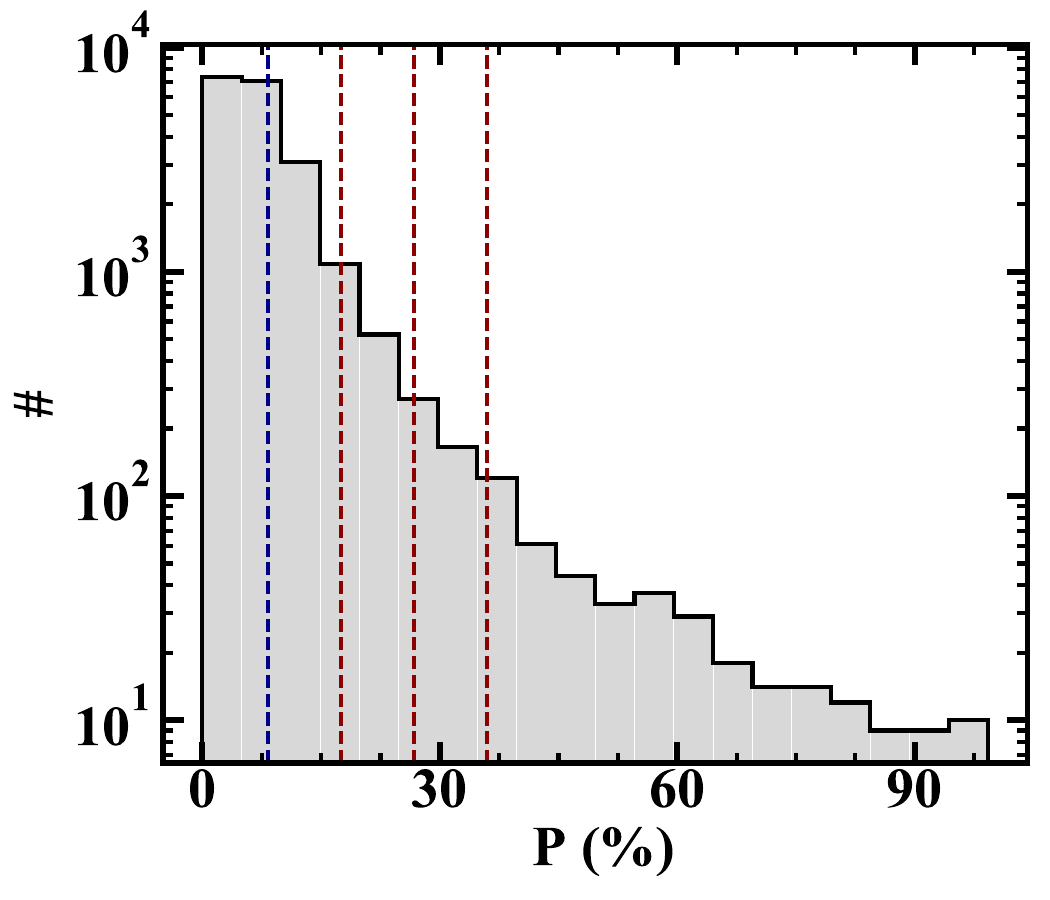}
\includegraphics[scale = 0.25]{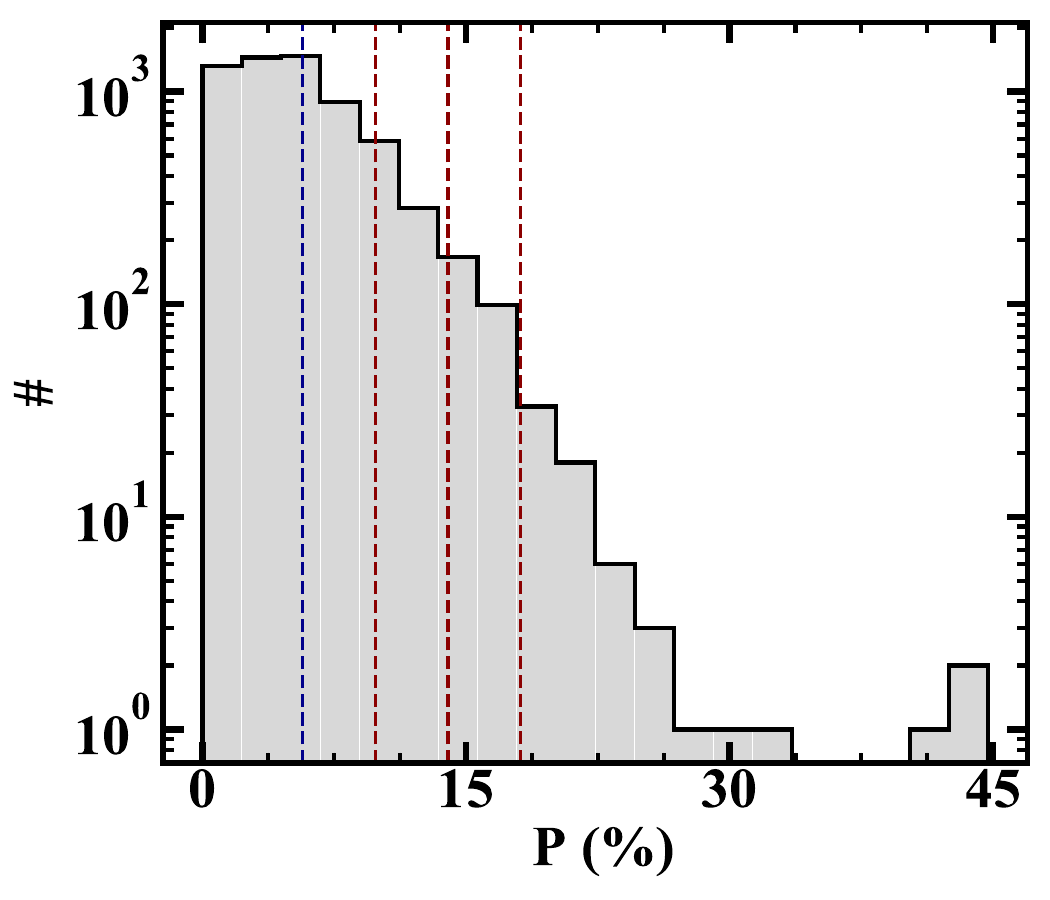}\\
\includegraphics[scale = 0.25]{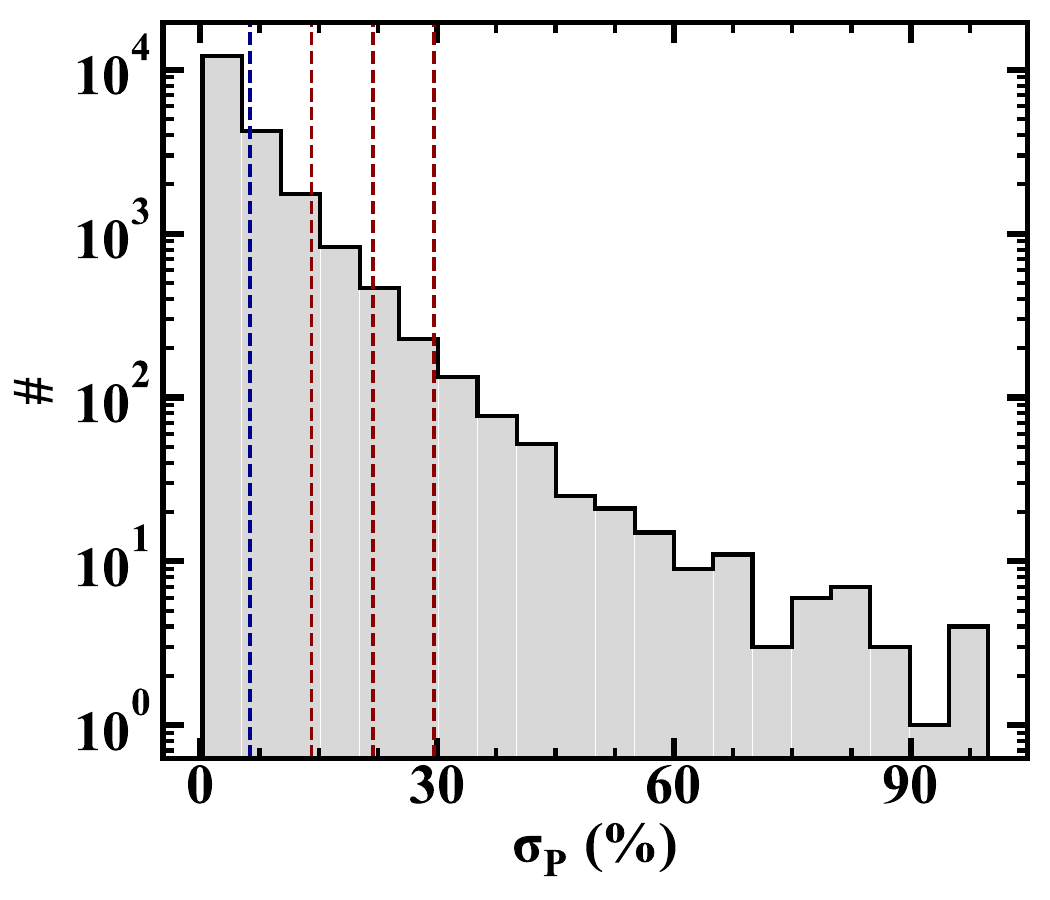}
\includegraphics[scale = 0.25]{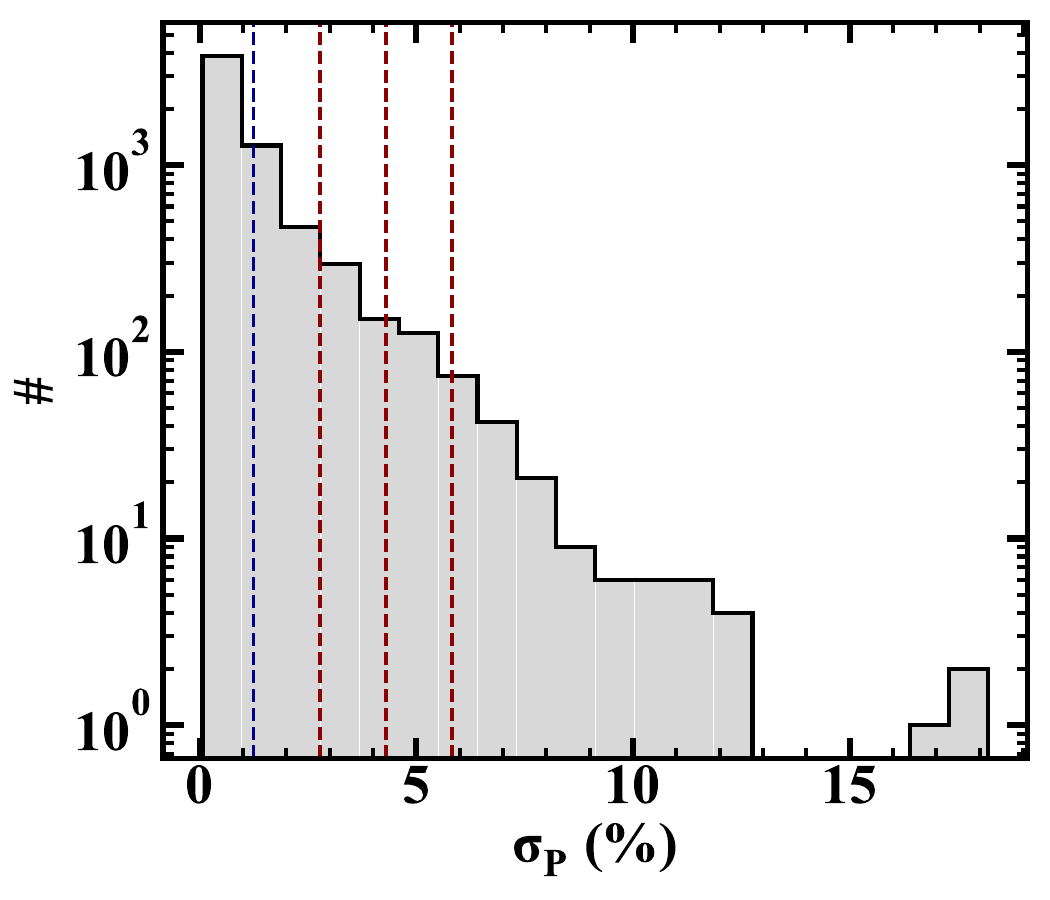}\\
\includegraphics[scale = 0.25]{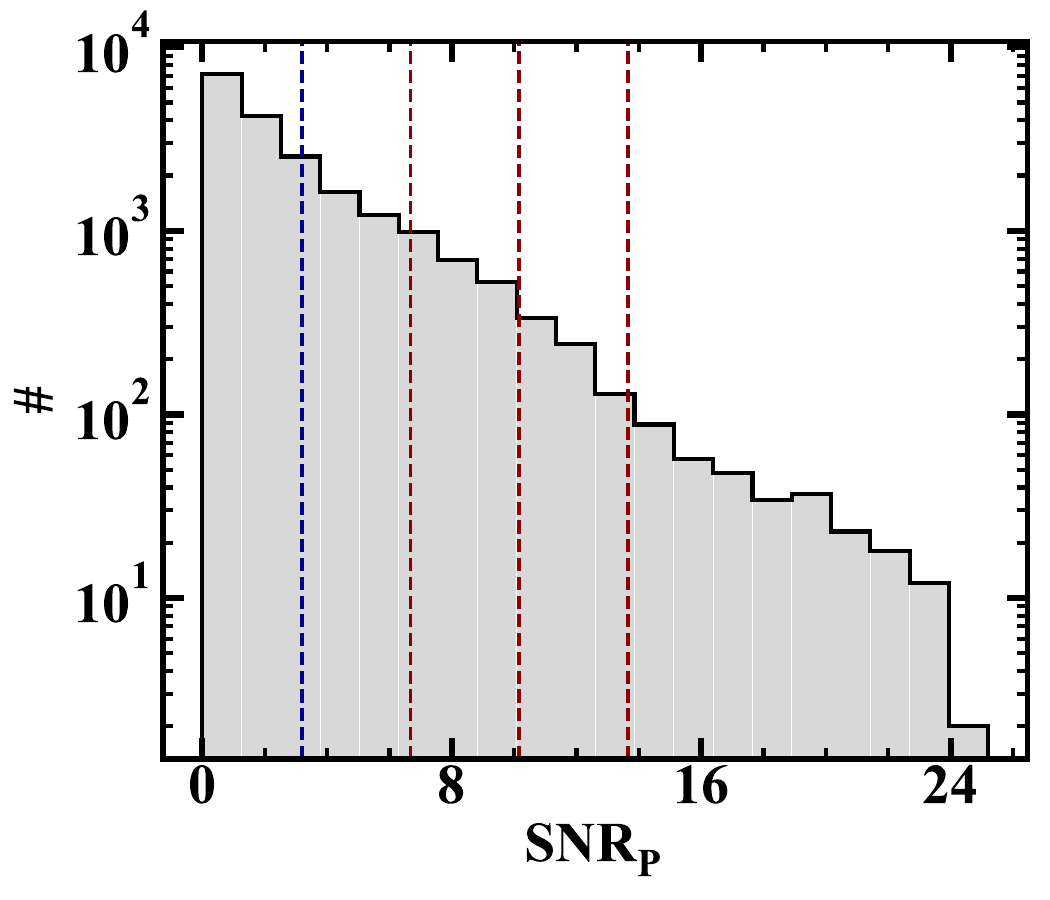}
\includegraphics[scale = 0.25]{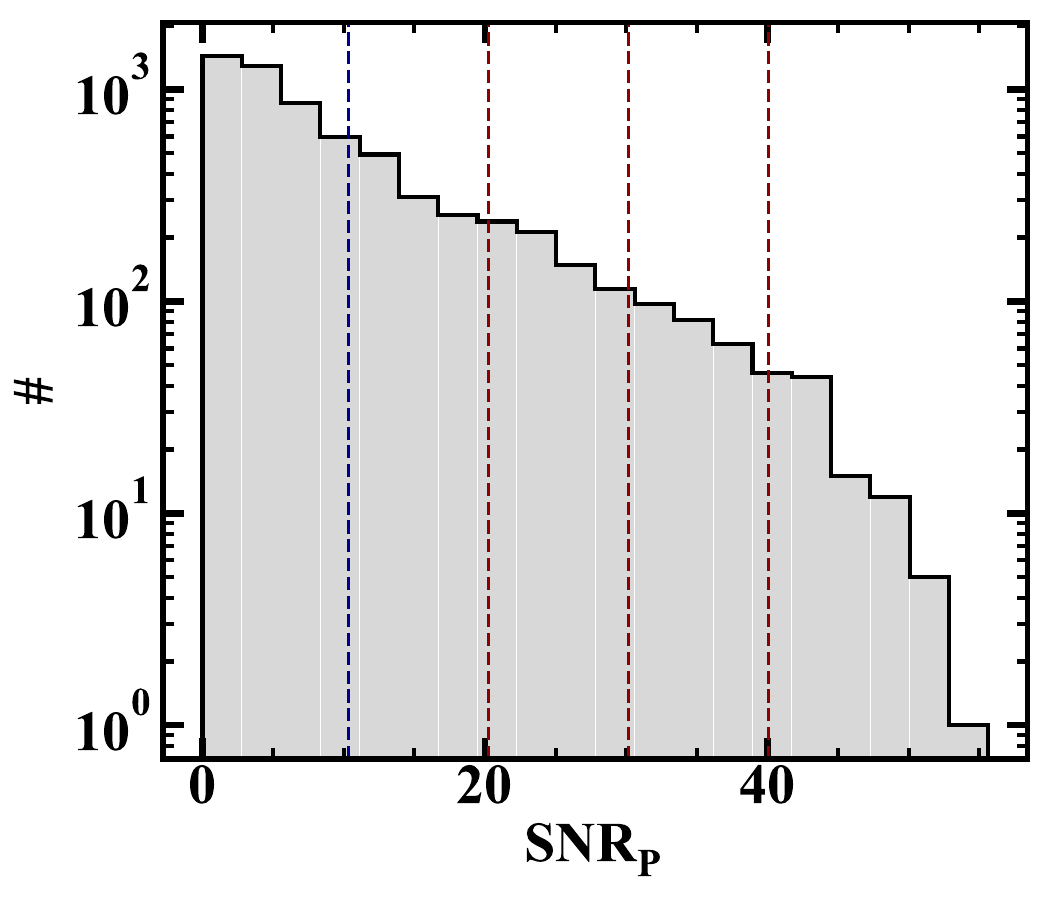}
\caption{Histograms of $p$ (top panels), $\sigma_p$ (middle panels), and SNR$_p$ (bottoms panels) of the SOFIA/HAWC$+$ data. Left panels represent data in band C at 89~\um~and right panels represent data in band D at 154~\um. The blue vertical dashed lines in each panel represent the mean value of the distribution. The red vertical dashed lines show the position of mean value plus 1$\sigma$, 2$\sigma$, and 3$\sigma$ of the distribution from left to right, respectively. }
\label{fig:histogram_P}
\end{figure}

Figures~\ref{fig:histogram_I} and \ref{fig:histogram_P} show histograms of Stokes-$I$, $\sigma_I$, SNR$_I$, $p$, $\sigma_p$, and SNR$_p$ of the SOFIA/HAWC$+$ data in band C and band D. Table~\ref{tab:sum_I_P} shows the summary of the mean and RMS of the distribution of Stokes-$I$, $p$, their uncertainties and SNRs. 

%=================================================

%=================================================
%=================================================
\begin{table*}[h!]\small 
\caption{Mean and RMS of Stokes-$I$, $p$, and their SNR in bands C and D. } \label{tab:sum_I_P} 
\centering 
\begin{tabular}{l| rr rr rr | cc rr rr} 
\hline\hline 
Wavelength & \multicolumn{2}{c}{$I$~(Jy/arcsec$^2$)} & \multicolumn{2}{c}{$\sigma_I$ (Jy/arcsec$^2$)}& \multicolumn{2}{c}{ SNR$_I$}& \multicolumn{2}{c}{$p$~($\%$) }& \multicolumn{2}{c}{ $\sigma_P$ ($\%$) }& \multicolumn{2}{c}{ SNR$_P$} \\
&       Mean    &       RMS     &       Mean    &       RMS     & Mean  &         RMS     &       Mean    &       RMS     & Mean  &       RMS     &         Mean    &       RMS     \\
\hline
89~\um  &       0.163   &       0.166   &       0.005   &       0.004   &         44.8    &       52.9    &       8.3     &       9.2     &       6.3     &         7.8     &       3.2     &       3.5\\
154~\um &       0.143   &       0.142   &       0.001   &       0.001   &         177.5   &       214.1   &       5.7     &       4.1     &       1.2     &         1.5     &       10.3    &       9.9\\
\hline
\end{tabular}
\end{table*}
%=================================================
% ------------------------------------------------
\section{Magnetic field orientations and strengths at different wavelengths}\label{app:compa_B}
\begin{figure}[h!]\centering
\includegraphics[width = 0.43\textwidth]{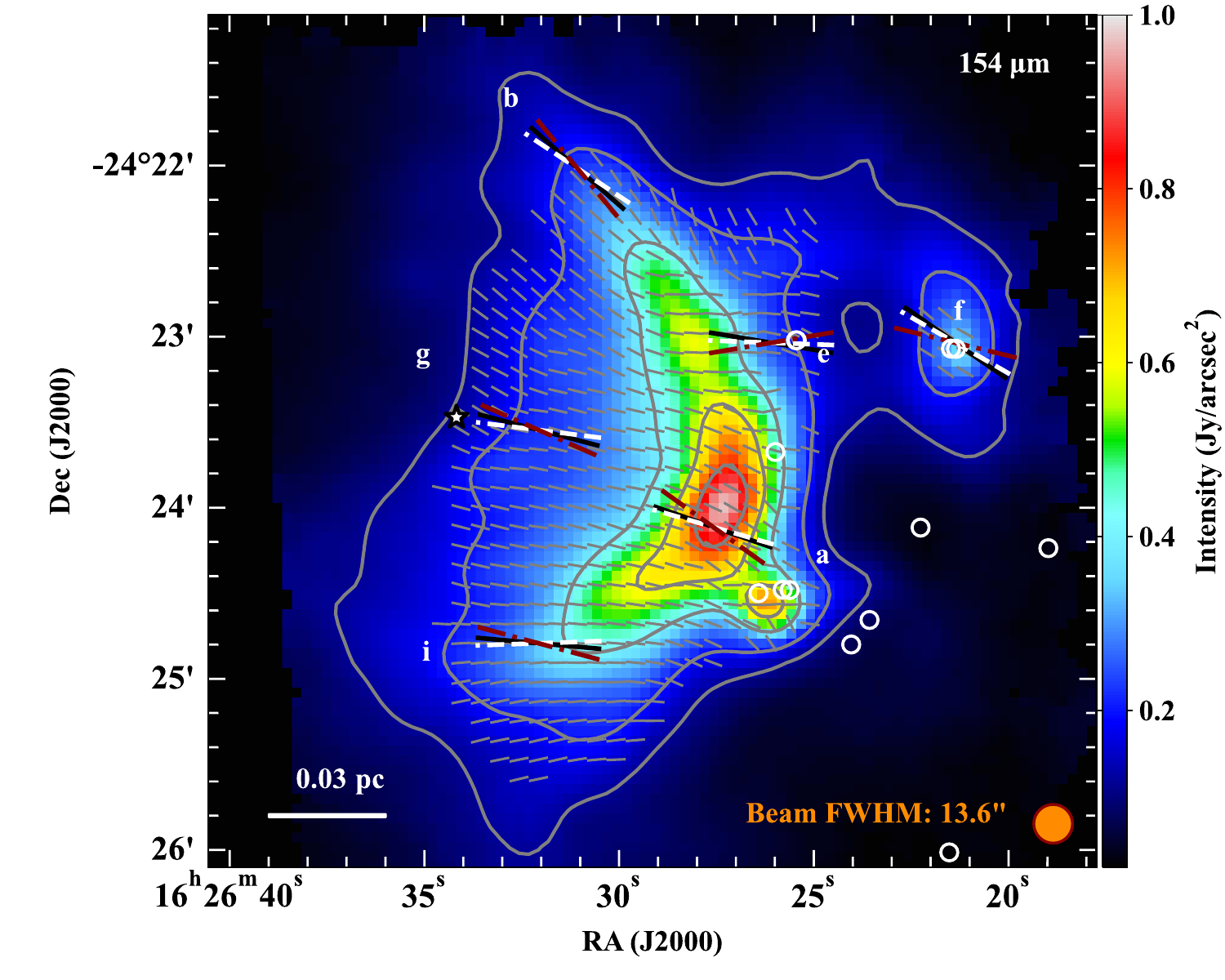}
\caption{Same as the right panel in Fig.~\ref{fig:Bfield_orientation} but with B-field vectors from POL-2 data added \citep[red dash-dotted vectors,][]{Kwon18}. These vectors represent the mean B-field position angles in subregions (see texts). White and black dashed vectors show the corresponding mean B-field position angles at 89 and 154~\um.}\label{fig:orient_compare}
\end{figure}

In Fig.~\ref{fig:orient_compare}, we compare the B-fields vectors obtained from HAWC$+$ data at 89 and 154~\um~(this study), and from POL-2 data at 850~\um~\citep{Kwon18}. Vectors show the mean B-field position angles at the subregions in \opha. 

\begin{figure}[ht!]\centering
\includegraphics[width =0.95\linewidth]{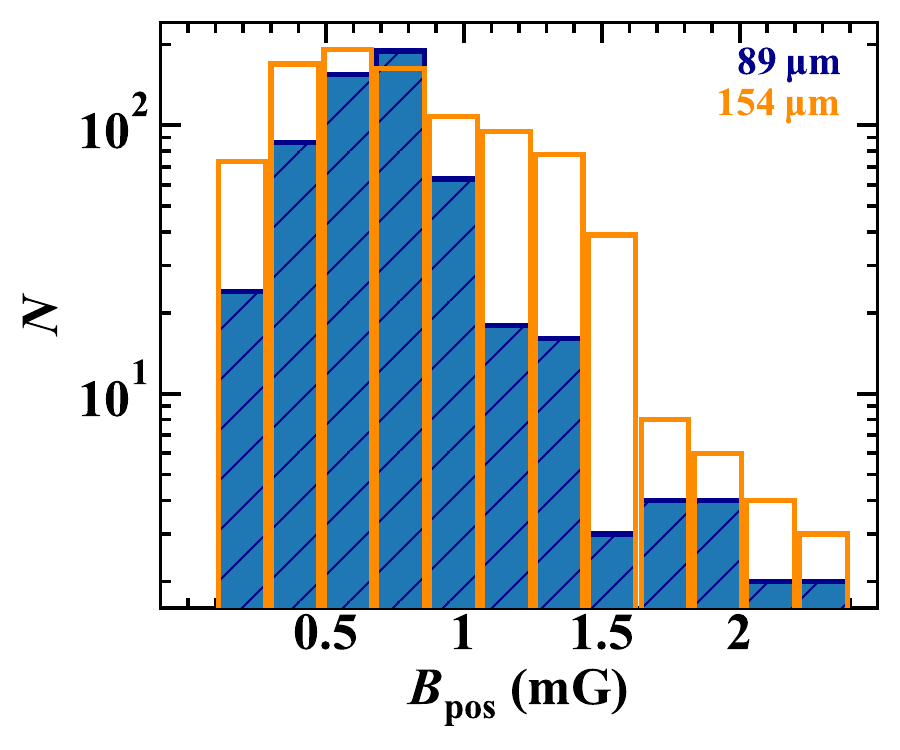}
\caption{Histogram distribution of the estimated \BPOS~using the DCF method, obtained with HAWC$+$ data at 89~\um~(in blue color with hatches) and 154 (in orange color)~\um. }\label{fig:bpos_compare}
\end{figure}

\begin{figure*}\centering
\includegraphics[width=\linewidth]{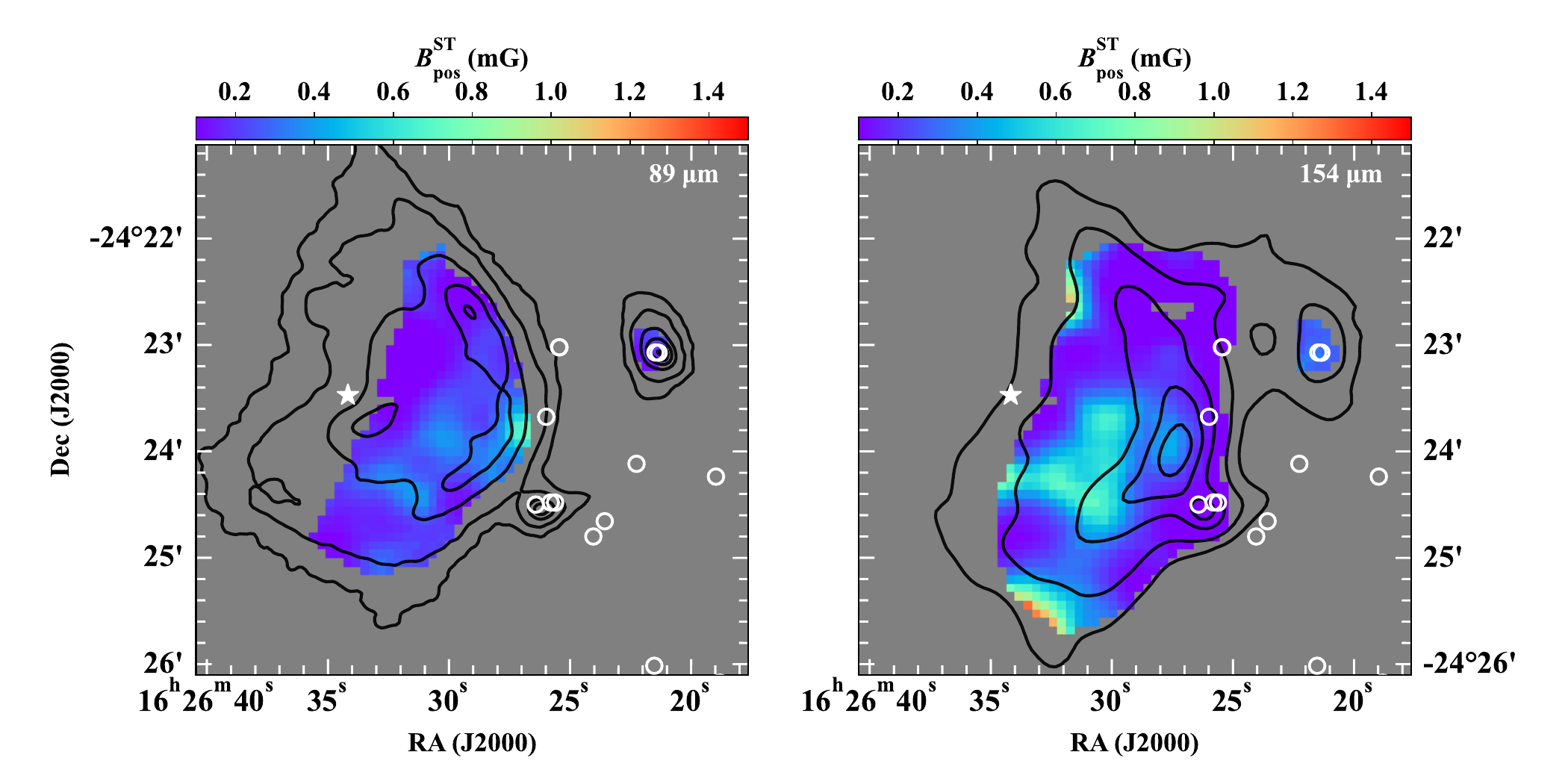} 
\caption{Maps of the B-field strength projected in the plane-of-the-sky in \opha~obtained by using the ST method, $B_\mathrm{pos}^{\mathrm{ST}}$, in band C (left panel) and band D (right panel). Other notations are similar as in Fig.~\ref{fig:Bpos_map}. }
\label{fig:Bpos_map_ST}
\end{figure*}

Figure~\ref{fig:bpos_compare} shows the histogram distribution of the estimated $B^\mathrm{DCF}_\mathrm{pos}$ obtained with HAWC$+$ data at 89 and 154~\um. Figure~\ref{fig:Bpos_map_ST} shows the maps of B-field strength estimated using the ST method ($B^\mathrm{ST}_\mathrm{pos}$) at 89 and 154~\um. 

% --------------------
% -------------------

% ------------------------------------------------
\section{Velocity dispersion traced by other molecular 
lines}\label{app:velo_disper_comparison}

To test whether the FWHM map of HCO$^{+}$(4$-$3) line is a good choice, we further use the N$_2$H$^{+}$(3$-$2) molecular data observed by APEX to estimate the total velocity dispersion, $\sigma_\varv$. The majority of the \opha~cloud exhibits the N$_2$H$^{+}$(3$-$2) emission showing 3 single Gaussian profiles; the middle peak is much stronger compared with the other two components. This strongest hyperfine line indicates the central velocity which is in the range of 2.54--3.97~km~s$^{-1}$, with an average velocity of 3.42~km~s$^{-1}$, in agreement with the systemic velocity of \opha~of $\sim$3.44~km~s$^{-1}$ estimated by \cite{andre07}. To each pixel we fit the spectrum with the sum of 3 single Gaussian profiles and then use the line width of the primary peak as the $\sigma_\varv$ toward the computing pixel. We construct the map of the FWHM velocity dispersion, following the procedure presented in Sect.~\ref{ssec:velocity-dispersion} for the case of the HCO$^{+}$(4$-$3) molecular line. 

\begin{figure}\centering
\includegraphics[width=0.46\textwidth]{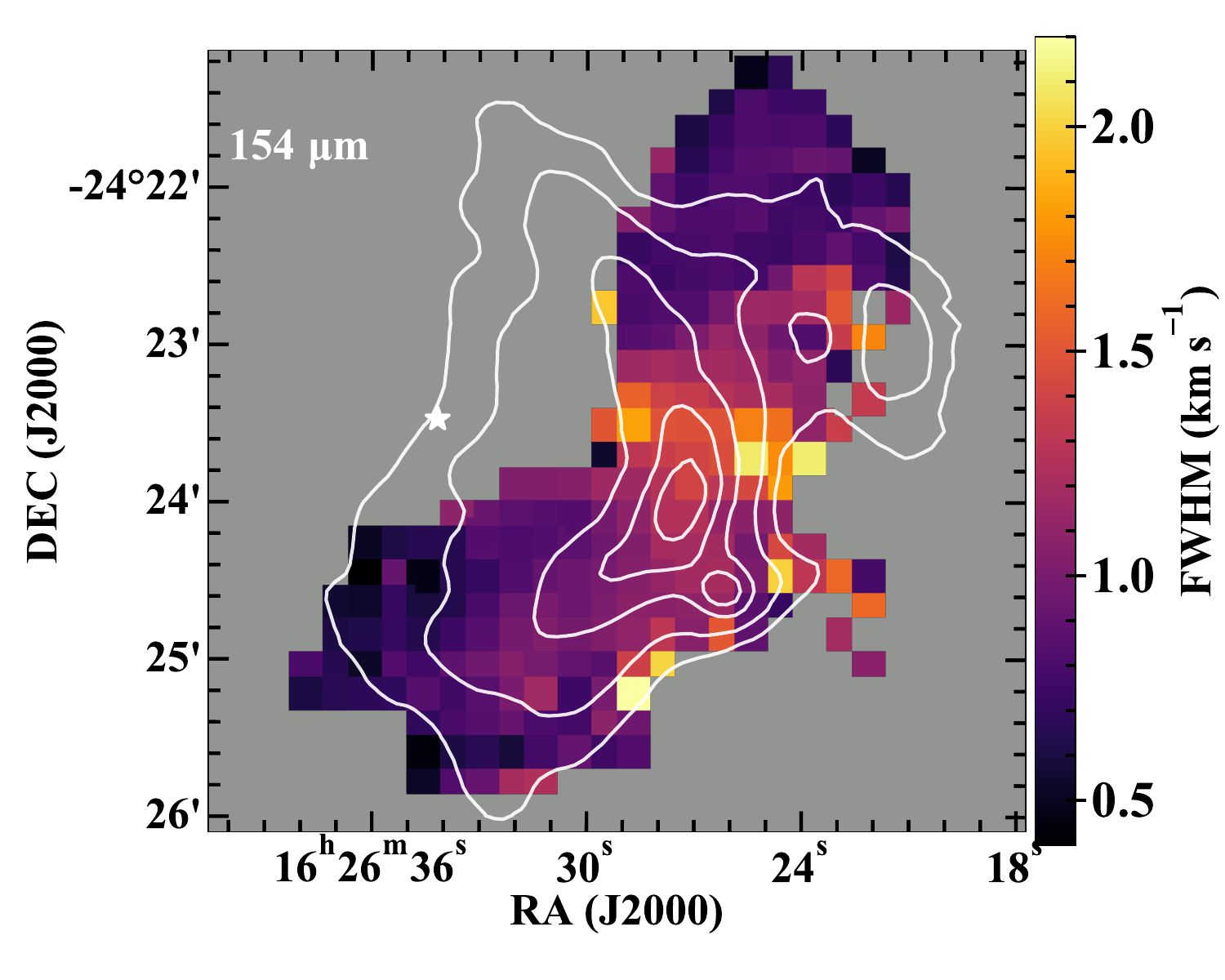}
\caption{Distribution of the non-thermal velocity FWHM of the N$_2$H$^{+}$(3$-$2) molecular line toward \opha. White star marks the position of the Oph-S1 star. White contours show the continuum emission (Stokes-$I$) at 154~\um, with contours levels at 0.12, 0.2, 0.4, 0.6, and 0.8 Jy/arcsec$^{2}$.}\label{fig:delta_velon2hp}
\end{figure}

Figure~\ref{fig:delta_velon2hp} shows the map of $\Delta\varv_\mathrm{NT}$ using the N$_2$H$^+$(3$-$2) molecular line toward \opha. The $\Delta\varv_\mathrm{NT}$ values ranges from 0.18 to 1.73~km~s$^{-1}$ with a median value of 0.95~km~s$^{-1}$, in good agreement with those obtained by HCO$^{+}$(4$-$3) line. We note, however, that the spatial extent of the HCO$^{+}$(4$-$3) emission line is largeer than that of the N$_2$H$^+$(3$-$2). Thus, we use the map of FWHM of the HCO$^{+}$(4$-$3) line to calculate the $B_{\mathrm{pos}}^{\mathrm{DCF}}$~strengths in \opha. Similarly, \cite{White15} reported a mean FWHM velocity dispersion of 1.5~km~s$^{-1}$ using the C$^{18}$O(3$-$2) data from JCMT/HARP in overall entire $\rho$ Ophiuchus cloud. Using the N$_2$H$^+$(1$-$0) data from IRAM, \cite{andre07} estimated an average value of the 1D nonthermal velocity dispersion toward \opha, $\sigma_\mathrm{NT}$=0.2~km~s$^{-1}$, corresponding to $\Delta\varv_\mathrm{NT}\sim$0.47~km~s$^{-1}$. Observations of the NH$_3$ hyperfine emission from the Green Bank Telescope also returned a similar 1D velocity dispersion with the average value of 0.3~km~s$^{-1}$, corresponding to an average of FWHM of 0.65~km~s$^{-1}$ \citep{Chen19}.

%=================================================
%=================================================
\FloatBarrier
\section{Mass-to-flux ratio, Alfv\'enic Mach number, beta plasma, and $B-n_\mathrm{H}$~relation results obtained by using $B_{\mathrm{pos}}^{\mathrm{ST}}$}

\begin{figure}[h!]\centering
\includegraphics[width=0.44\textwidth]{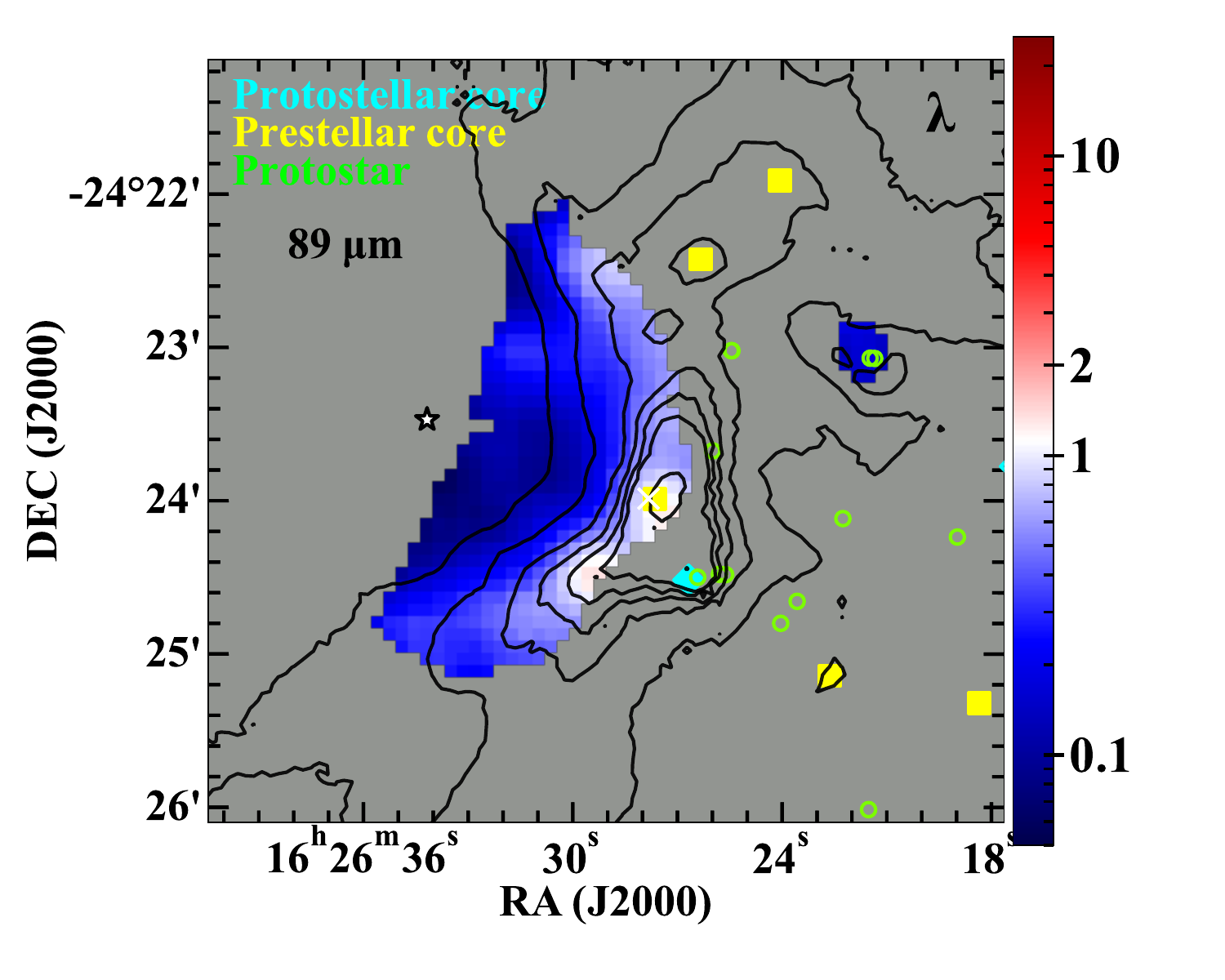}\\
\includegraphics[width=0.44\textwidth]{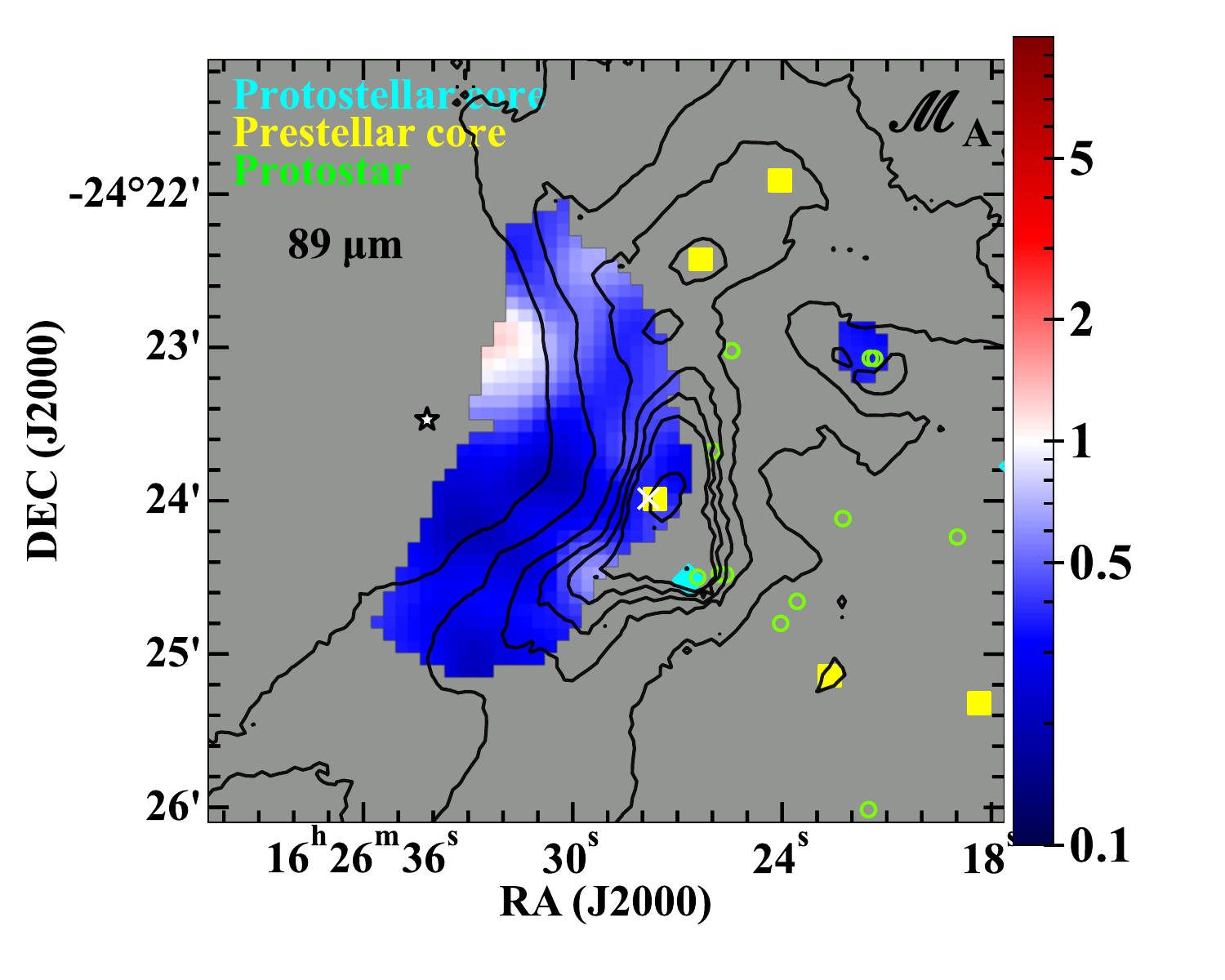}\\ 
\includegraphics[width=0.44\textwidth]{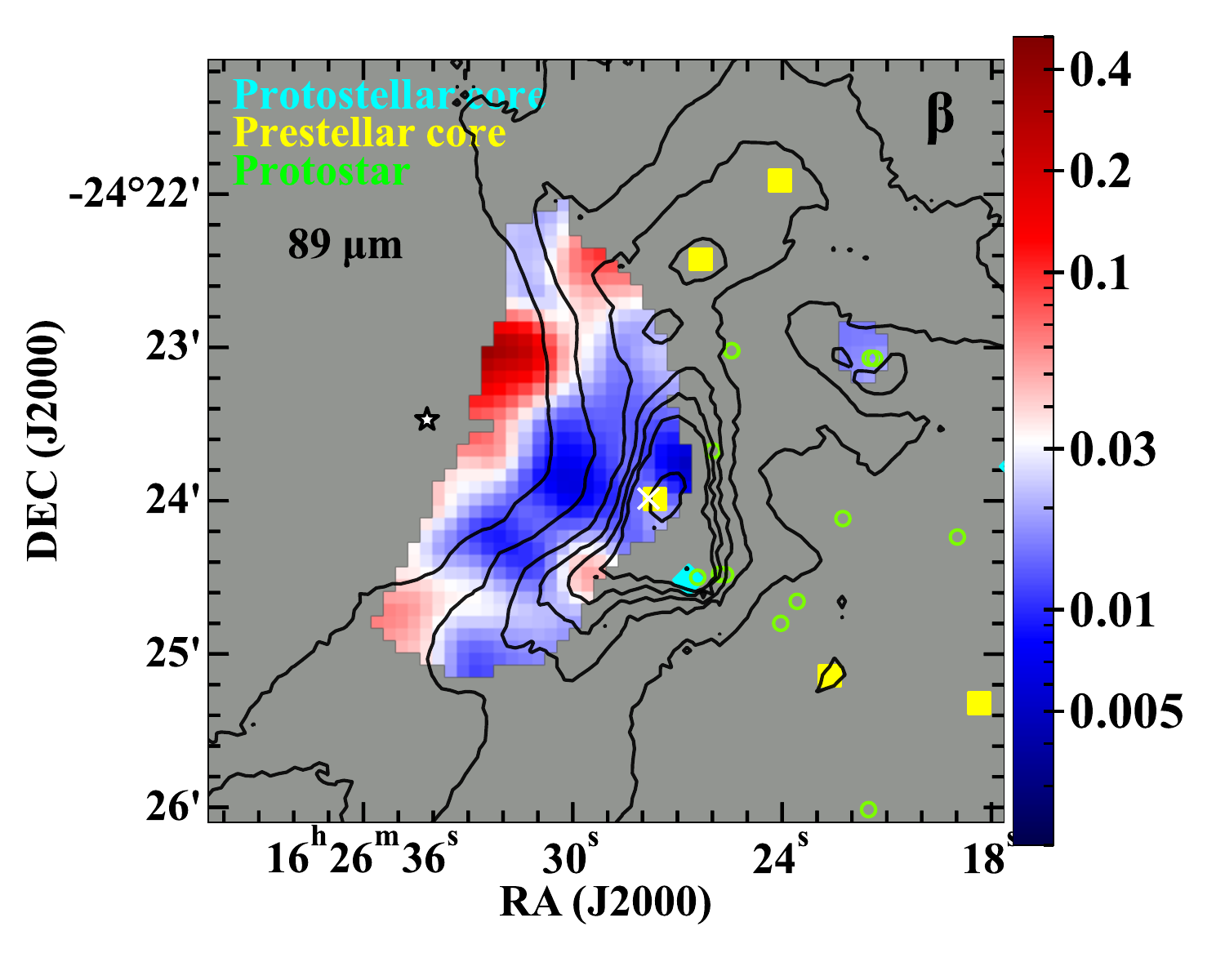} 
\caption{Similar to Fig.~\ref{fig:lambda_alfven_beta_pressure_maps} but using $B_\mathrm{pos}^\mathrm{DCF}$ in band C. }
\label{fig:lambda_alfven_beta_pressure_maps_C} 
\end{figure}

\begin{figure*}[h!]\centering
\includegraphics[width=0.44\textwidth]{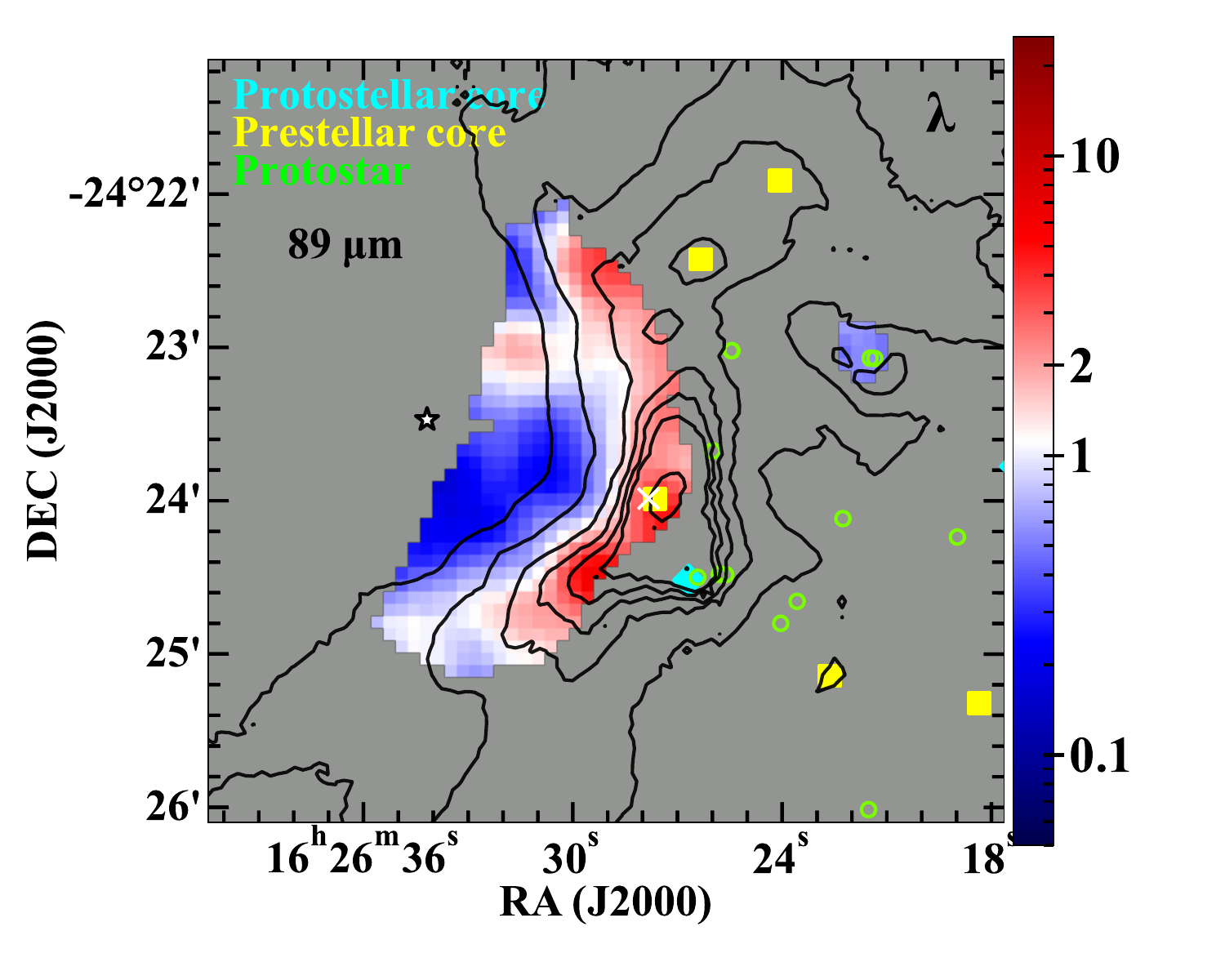}
\includegraphics[width=0.44\textwidth]{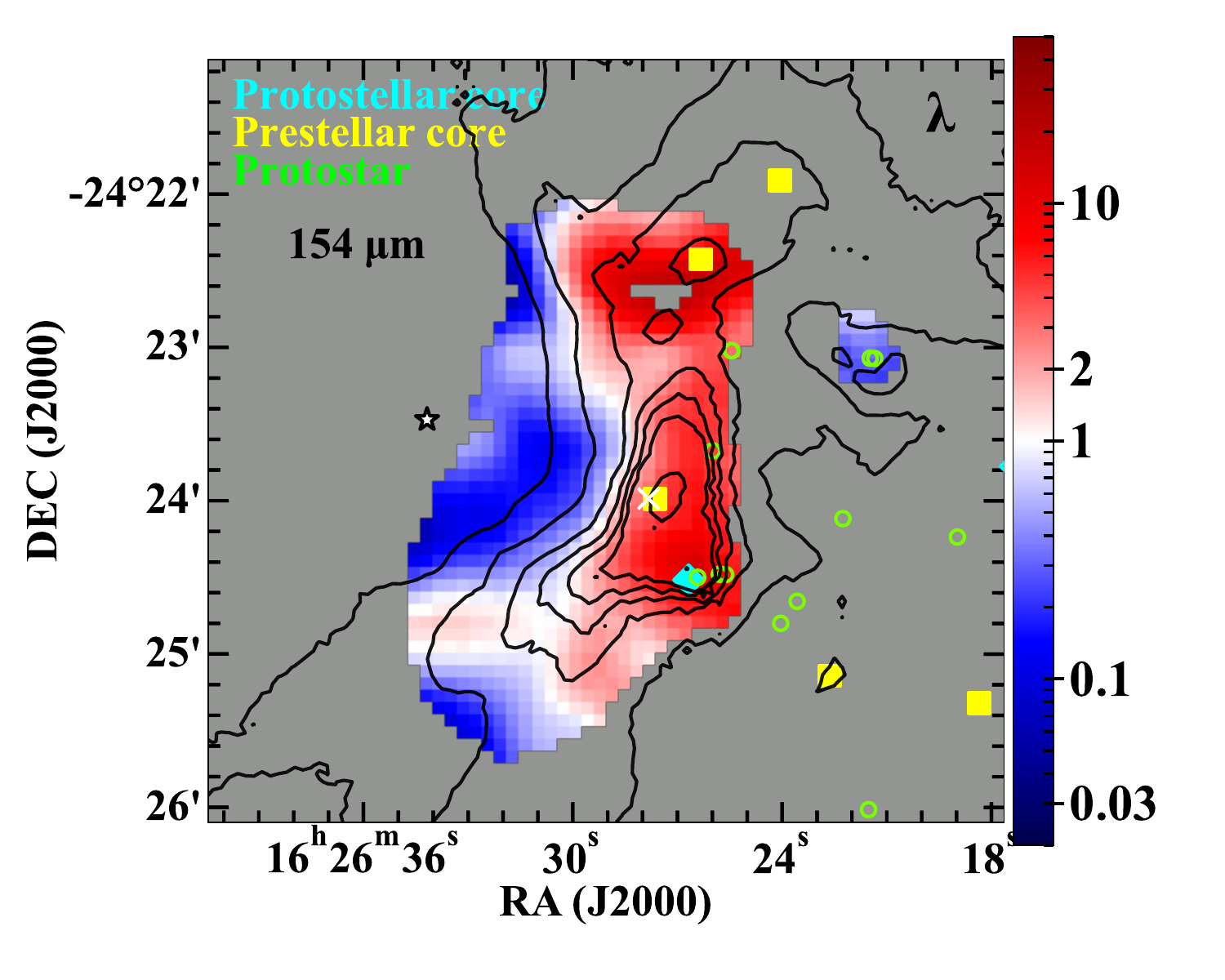}\\ 
\includegraphics[width=0.44\textwidth]{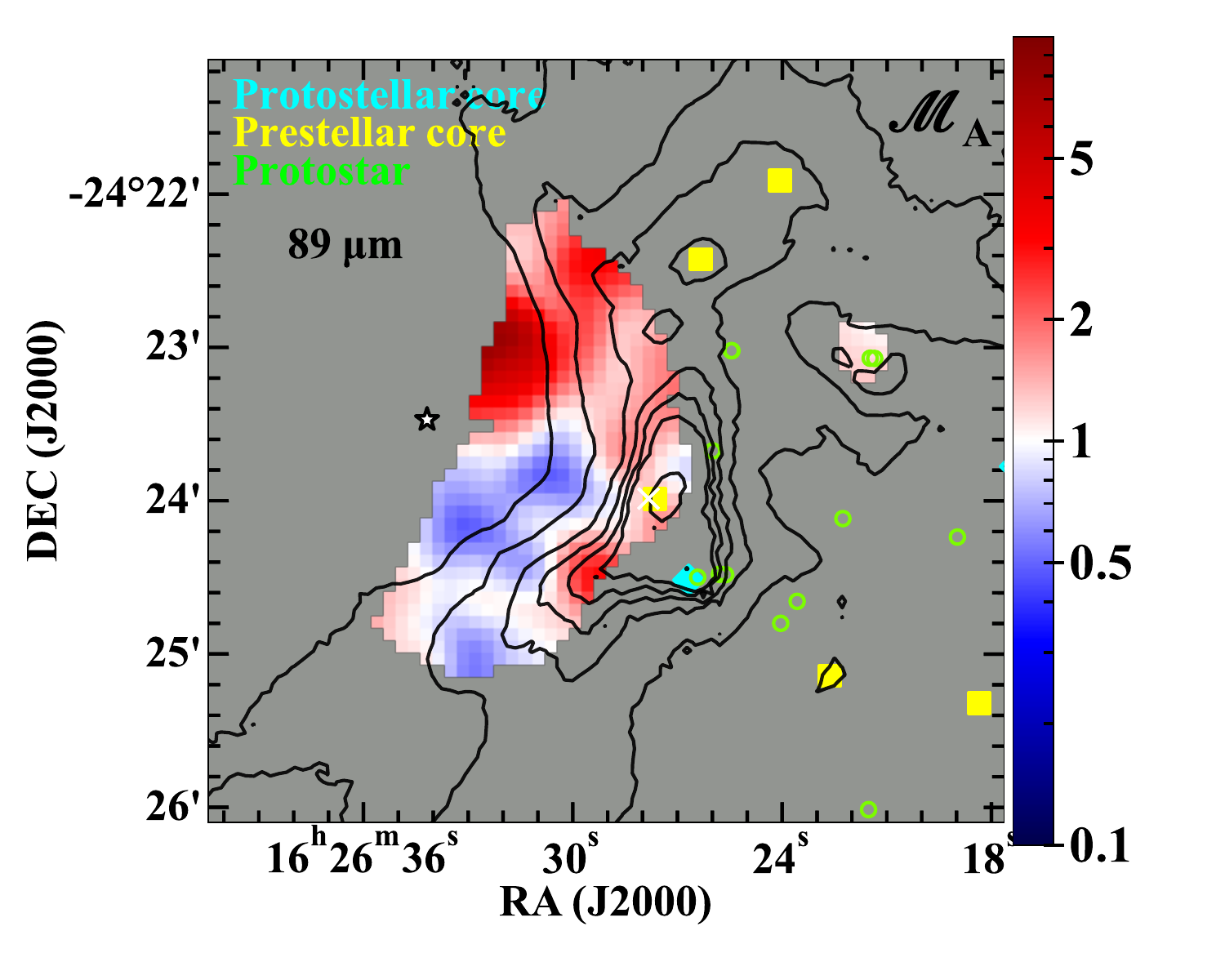}
\includegraphics[width=0.44\textwidth]{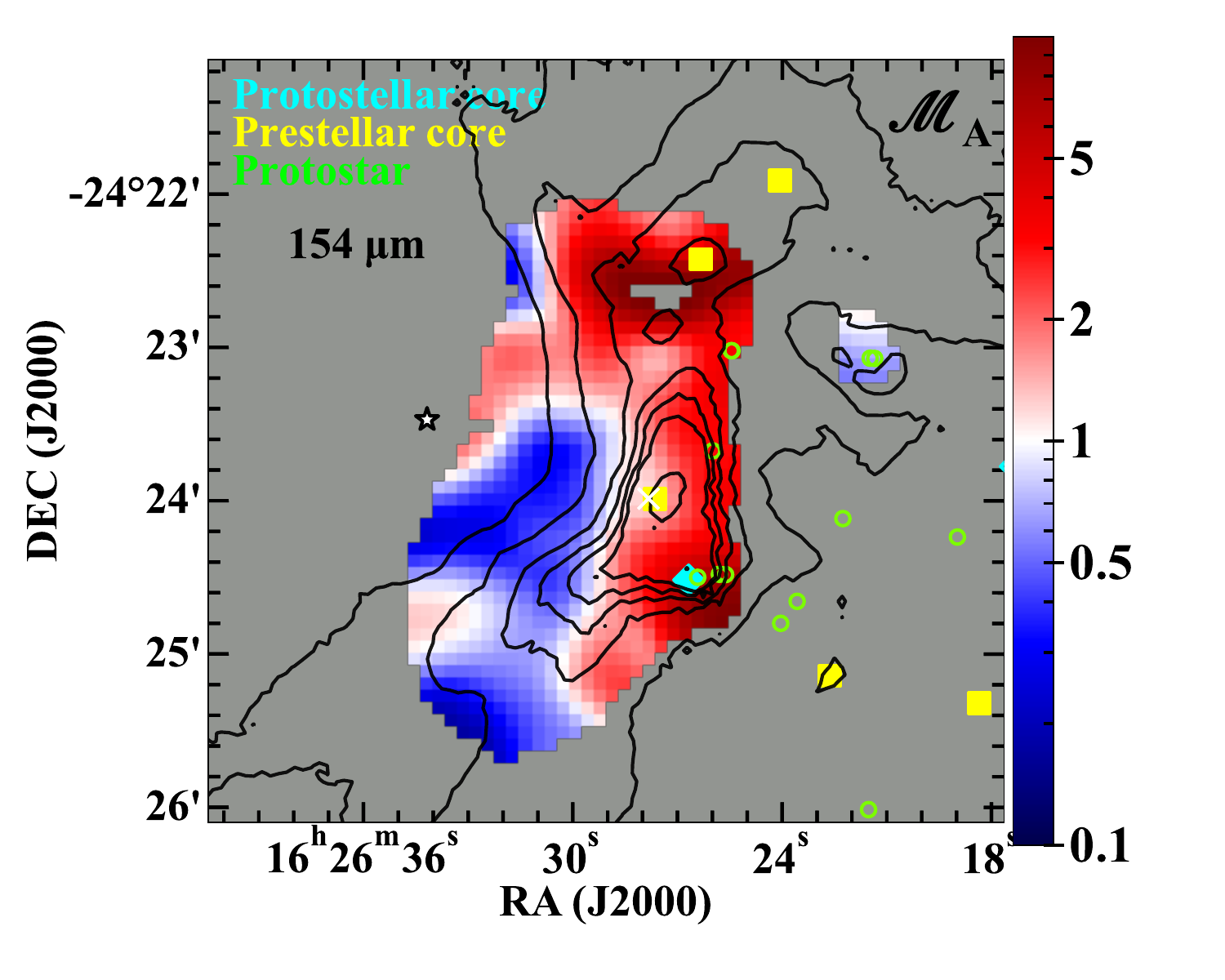}\\ 
\includegraphics[width=0.44\textwidth]{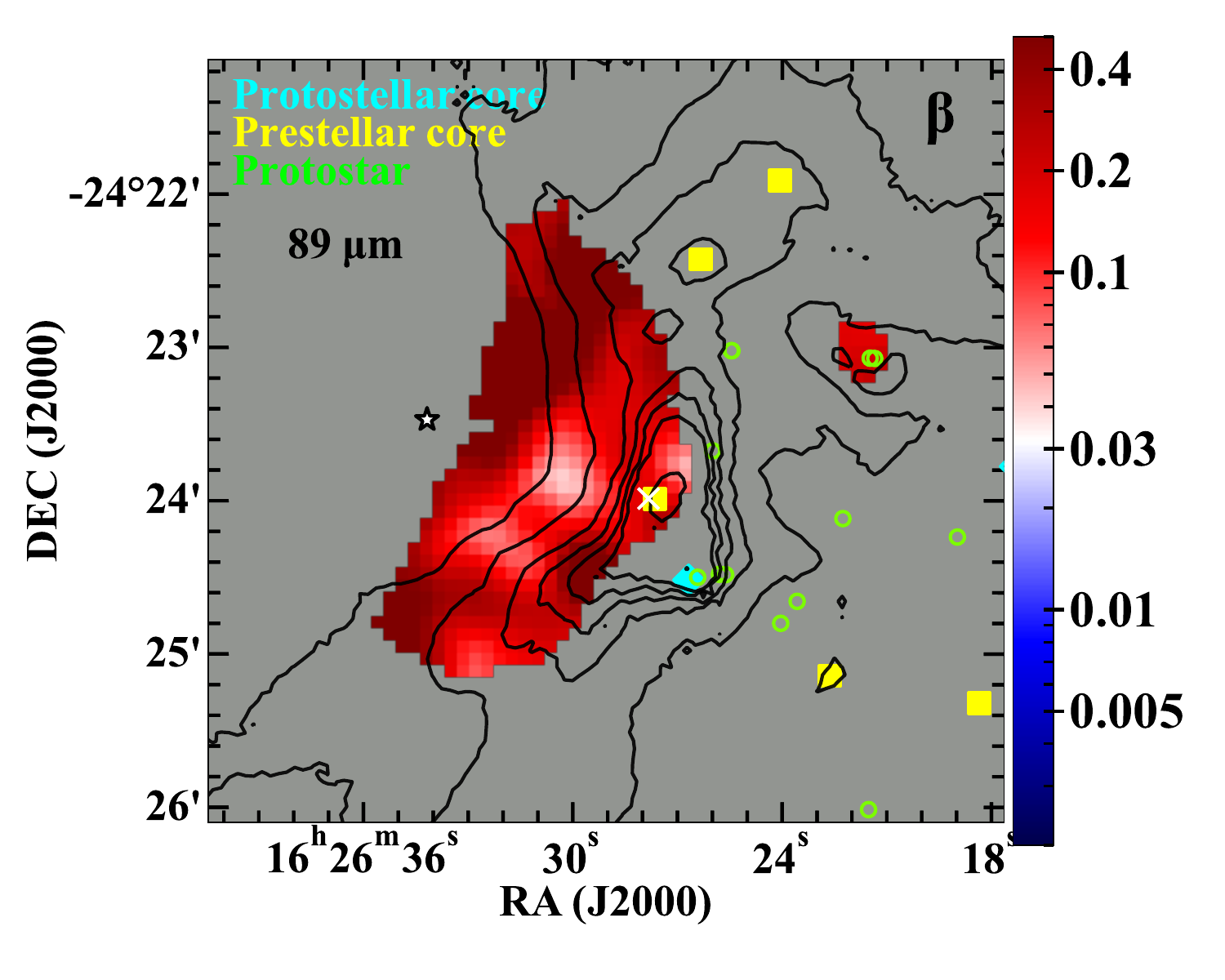} 
\includegraphics[width=0.44\textwidth]{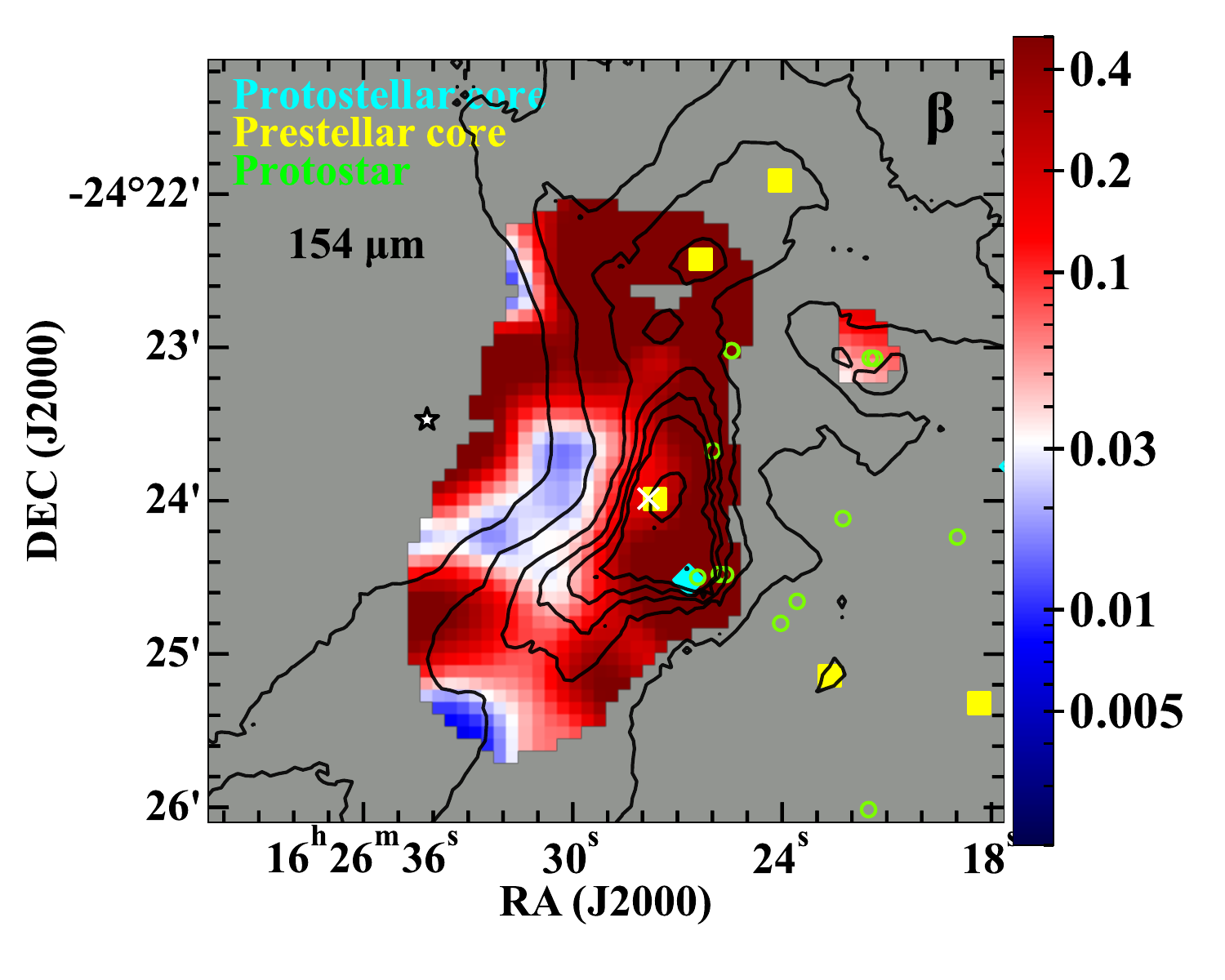} 
\caption{Similar to Fig.~\ref{fig:lambda_alfven_beta_pressure_maps} but using $B_\mathrm{pos}^{\mathrm{ST}}$ in band C (left panels) and D (right panels). }
\label{fig:lambda_alfven_beta_pressure_maps_ST}
\end{figure*}

Figure~\ref{fig:lambda_alfven_beta_pressure_maps_C} shows the maps of mass-to-flux ratio (top panel), Alfv\'enic Mach number (middle panel), and beta plasma (bottom panel) toward \opha~using $B_\mathrm{pos}^{\mathrm{DCF}}$ in band C. Figure~\ref{fig:lambda_alfven_beta_pressure_maps_ST} show maps of mass-to-flux ratio (top), Alfv\'enic Mach number (middle), and beta plasma (bottom) toward \opha~using $B_\mathrm{pos}^{\mathrm{ST}}$ in band C (left panels) and D (right panels). 

\begin{figure}[tp!]\centering
\includegraphics[width = 0.475\textwidth]{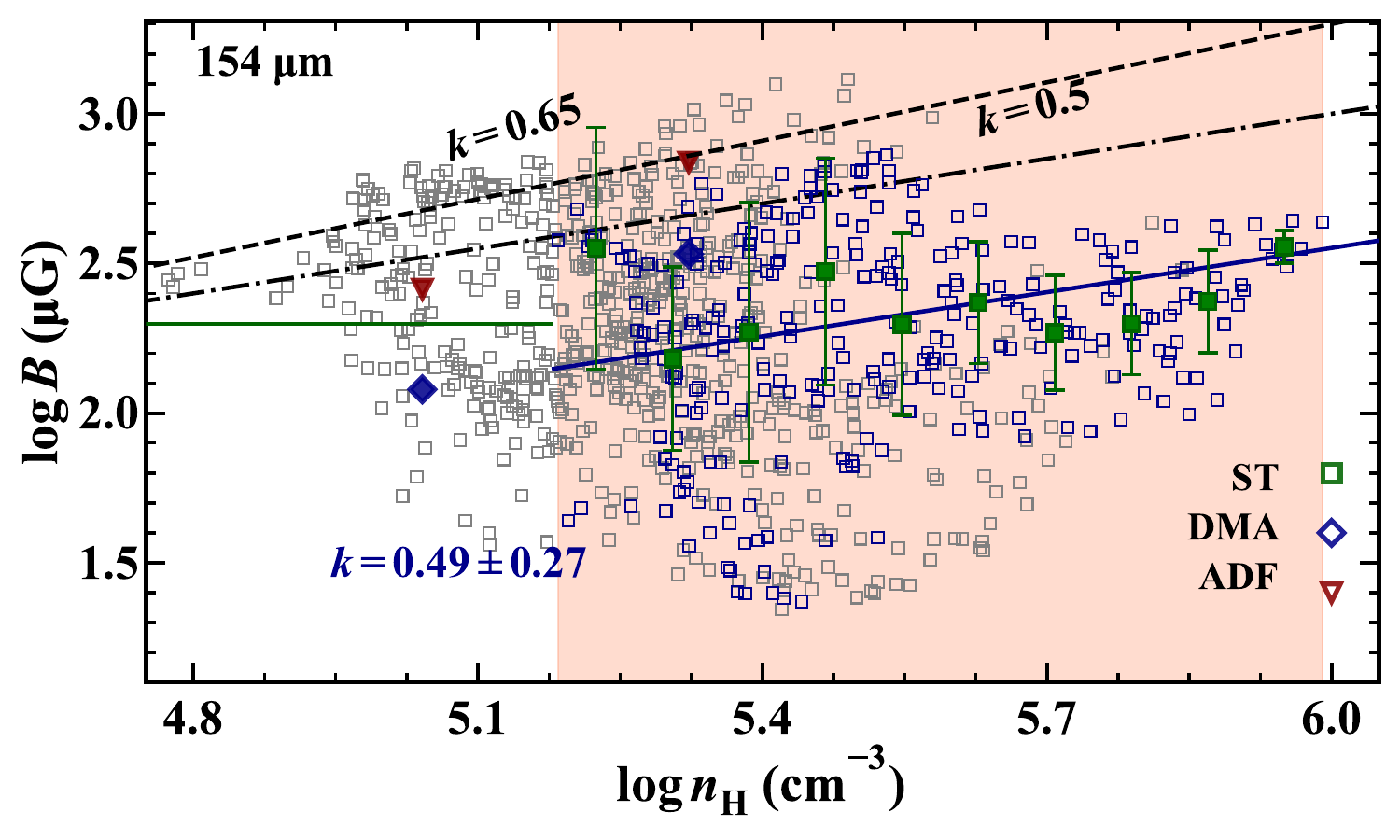}
\caption{Similar to Fig.~\ref{fig:Bpos_vsnH2} but using B-field strength estimated with the ST method in band D. Small squares represent $B$ strength estimated with the ST method at band D. Blue small squares further show the data at the denser region ($\log n_\mathrm{H} \ge 5.2$) with Stokes$-I \ge 0.4~\mathrm{Jy/arcsec^2}$. 
Green bigger squares represent median value of $ \log B$ binned in each interval of $\log n\mathrm{_H}$ of $\sim$0.08 in the denser region. Mean values of $\log B$ estimated with ADF (red triangles) and DMA (blue diamonds) methods toward subregions (a) and (e) are overplotted. The solid green line shows the median data at lower densities region where $\log n_\mathrm{H} \ge $5.2. 
The solid blue line shows the power-law fit toward the binned data of $B$ using the DCF method toward denser region (covered by the orange shaded box, resulting $k=0.49\pm0.27$. The black dash-dotted line indicates the critical power-law $B\propto n_\mathrm{H}^{0.5}$ for strong B-fields, predicted from the theoretical models \citep{Mouscho99}. The black dashed line indicates the relation found by \cite{Crutcher10} using the Zeeman observational data. }
\label{fig:Bpos_vsnH2_ST}
\end{figure}

Figure \ref{fig:Bpos_vsnH2_ST} shows the relation between $\log B$ and $\log n_\mathrm{H}$ using $B_{\mathrm{pos}}^{\mathrm{ST}}$ in band D. 

\FloatBarrier

% =========

% ------------------------------------------------
\section{Main structure of the ridge identification}\label{app:ridge}
\begin{figure}[tbp!]
\centering
\includegraphics[width=0.46\textwidth]{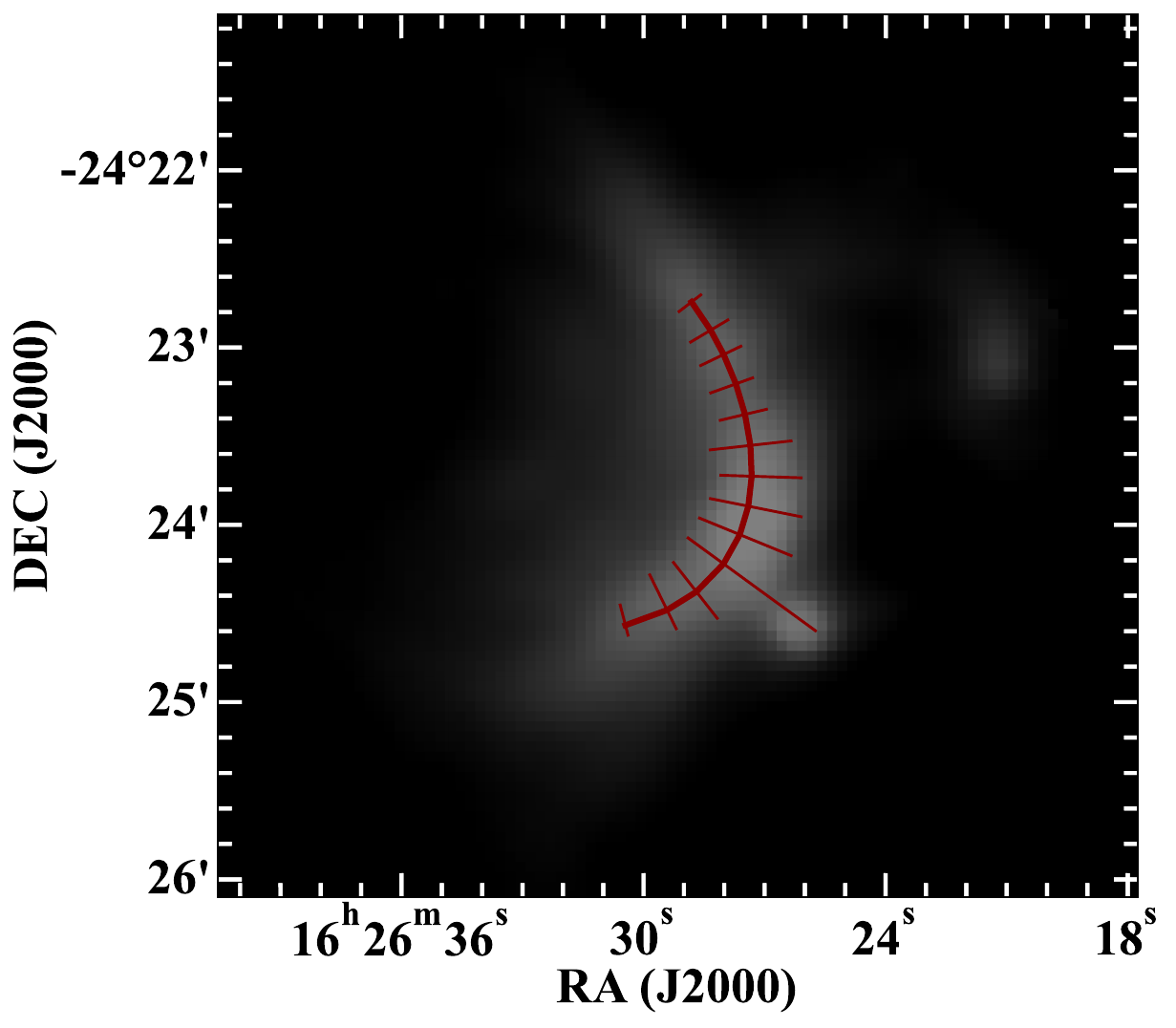}
\caption{Schematic view of the ridge in \opha. The background color map shows the Stokes-$I$ intensity at 154~\um. The thick red line indicates the main shape of the ridge. Thin red lines show the lines perpendicular to the tangent line at each position on the ridge. }
\label{fig:ridge}
\end{figure}

We identified the elongated shape of the ridge in the densest region of \opha~using the Python package \texttt{RadFil}. We first applied a threshold of Stokes-$I\geq$0.5~Jy/arcsec$^2$ in the map of Stokes-$I$ at 154~\um~to build a mask for the main shape of the densest region (the ridge). The spine of the ridge was identified by using the Python package \texttt{FilFinder}\footnote{\url{https://github.com/e-koch/FilFinder}} from \cite{koch15}. We then smoothed the ridge spine and define a tangent line at every position on the ridge, where the orientation of the ridge ($\theta_\mathrm{ridge}$) at every position was identified. Figure~\ref{fig:ridge} shows the schematic view of the ridge and the perpendicular cuts at each position on the ridge. The position angle of the tangent line at each position located on the ridge is defined as the orientation of the ridge at that position, measured east of north, and ranges from 0\degr~to 180\degr. 
%=================================================
% =================================================
\end{appendix}
\end{document}